\documentclass{ws-ijmpa}
\usepackage[super,compress]{cite}
\usepackage{graphicx}
\usepackage{epsfig}
\usepackage{dcolumn}
\usepackage{bm}
\usepackage{psfig}
\usepackage{color}
\usepackage{overpic}
\usepackage{url}

\newcommand{\xyz}{\rm XYZ}
\newcommand{\xx}{X(3872)}
\newcommand{\y}{Y(4260)}

\newcommand{\zc}{Z_c(3900)}
\newcommand{\zcp}{Z_c(4020)}

\newcommand{\BR}{{\cal B}}

\newcommand{\pip}{\pi^+}
\newcommand{\pim}{\pi^-}
\newcommand{\piz}{\pi^0}
\newcommand{\etap}{\eta^{\prime}}
\newcommand{\hc}{h_c}
\newcommand{\pphc}{\pi^+\pi^- h_c}
\newcommand{\etac}{\eta_c}

\newcommand{\dstrbar}{\bar{D}^{*}}
\newcommand{\psp}{\psi(2S)}
\newcommand{\psip}{\psi(2S)}
\newcommand{\pspp}{\psi(3770)}
\newcommand{\jpsi}{J/\psi}

\newcommand{\EE}{e^+e^-}
\newcommand{\MM}{\mu^+\mu^-}
\newcommand{\LL}{\ell^+\ell^-}

\newcommand{\pp}{\pi^+\pi^-}
\newcommand{\ppp}{\pi^+\pi^-\pi^0}
\newcommand{\kk}{K^+K^-}
\newcommand{\ks}{K_{S}^{0}}

\newcommand{\ppjpsi}{\pi^+\pi^- J/\psi}

\newcommand{\ddpi}{D^0D^{*-}\pi^+ + c.c.}
\newcommand{\beq}{\begin{equation}}
\newcommand{\eeq}{\end{equation}}
\newcommand{\bitm}{\begin{itemize}}
\newcommand{\eitm}{\end{itemize}}

\def\Journal#1#2#3#4{{#1} {\bf #2}, #3 (#4)}

\def\PRL{Phys. Rev. Lett.}
\def\PRD{Phys. Rev. D}

\begin{document}
\markboth{Chang-Zheng Yuan}{The $\xyz$ states revisited}

%
\catchline{}{}{}{}{}
%

\title{The $\xyz$ states revisited}

\author{Chang-Zheng Yuan}

\address{
$^1$ Institute of High Energy Physics, Chinese Academy of
Sciences, \\
19B Yuquan Road, Beijing 100049, China \\
$^2$ University of Chinese Academy of Sciences, \\
19A Yuquan Road, Beijing 100049, China\\
yuancz@ihep.ac.cn}

\maketitle

\begin{abstract}

The BESIII and the LHCb became the leading experiments in the
study of the exotic states after the Belle and BaBar experiments
finished their data taking in the first decade of this century. We
review the progress in the study of the $\xyz$ states at BESIII
and LHCb experiments with their unique data samples in $\EE$
annihilation at center-of-mass energies of 3.8--4.6~GeV and in
$pp$ collision at center-of-mass energies 7, 8, and 13~TeV,
respectively. With these data samples, we have deepened our
understanding of the most famous charmonium-like states $\xx$,
$\y$, $\zc$, and $Z_c(4430)$, as well as other similar states like
the $Y(4140)$ and $X(3823)$. We review the progress in the study
of these states, and also discuss perspectives at future
experiments.


\end{abstract}

\ccode{PACS numbers: 14.40.Rt, 14.40.Pq, 13.25.Gv, 13.20.Gd,
13.66.Bc}

\tableofcontents

\section{\boldmath Introduction}

In the conventional quark model, mesons are composed of one quark
and one anti-quark, while baryons are composed of three quarks.
Although this picture is very simple, it describes almost all the
hadrons observed so far~\cite{pdg}. However, exotic hadronic
states with other configurations have been proposed and searched
for since long time ago~\cite{klempt}.

Many charmonium and charmonium-like states were discovered at two
$B$-factories BaBar~\cite{babar} and Belle~\cite{belle} in the
first decade of this century~\cite{PBFB}. Whereas some of these
are good candidates of charmonium states, as predicted in
different models, many other states have exotic properties, which
may indicate that exotic states, such as multi-quark state, meson
molecule, hybrid, or hadron-quarkonium, have been
observed~\cite{reviews}. Experimentally, these states are also
called $\xyz$ states, to indicate their nature is still unclear.

BaBar~\cite{babar} and Belle~\cite{belle} experiments finished
their data taking in 2008 and 2010 with total integrated
luminosity of 557 and 1040~fb$^{-1}$, respectively. The data are
still used for various physics analyses until now. In 2008, two
new experiments, the BESIII~\cite{bes3}, a $\tau$-charm factory
experiment at the BEPCII $\EE$ collider, and the
LHCb~\cite{lhcb_detector}, a $B$-factory experiment at the LHC,
started data taking, and contributed to the study of the $\xyz$
particles ever since.

In this article, we review the study of the $\xyz$ particles from
the BESIII~\cite{bes3} and the LHCb~\cite{lhcb_detector}
experiments. We first introduce the two experiments, the data
samples, and the advantages and disadvantages for the study of the
hadron spectroscopy at these experiments, and then focus on the
measurements of the three mostly studied $\xyz$ states, i.e., the
$\xx$, the $\y$, and the $\zc$. We also discuss other $\xyz$
states, including the $\zcp$, the $Z_c(4430)$, the $X(3823)$, and
the $Y(4140)$ family. At the end of the article, we give
perspectives on the study at these two experiments, and also point
out possible studies at next generation experiments.

\section{The BESIII and LHCb experiments and data samples}

\subsection{The BESIII experiment}

The BESIII experiment~\cite{bes3} at the BEPCII storage ring
started its first collisions in the tau-charm energy region in
2008. The BESIII detector has an effective geometrical acceptance
of 93\% of $4\pi$. It contains a small cell helium-based
multi-layered drift chamber (MDC) which provides momentum
measurements of charged particles; a time-of-flight system (TOF)
based on plastic scintillator which helps to identify charged
particles; an electromagnetic calorimeter (EMC) made of CsI(Tl)
crystals which is used to measure the energies of photons and
provide trigger signals; and a muon system (MUC) made of Resistive
Plate Chambers (RPC). The momentum resolution of the charged
particles is $0.5$\% at 1~GeV/$c$ in a 1~Tesla magnetic field; the
energy loss ($dE/dx$) measurement provided by the MDC has a
resolution better than 6\% for electrons from Bhabha scattering;
the photon energy resolution can reach $2.5$\% ($5$\%) at 1~GeV in
the barrel (endcaps) of the EMC; and the time resolution of TOF is
$80$~ps in the barrel and $110$~ps in the endcaps. In 2015, the
endcap TOF was replaced with MRPC, and the time resolution is
improved to be 60~ps~\cite{bes3_MRPC}.

After a few years running at energies for its well-defined physics
programs~\cite{BESIII_YB}, i.e., at $\jpsi$ and $\psp$ peaks in
2009 and the $\pspp$ peak in 2010 and 2011, the BESIII experiment
started to collect data for the study of the $\xyz$ particles,
which were not described in the Yellow Book~\cite{BESIII_YB}.
BESIII took its first data sample at the $\psi(4040)$ resonance in
May 2011, with the aim of searching for the $X(3872)$ and the
excited $P$-wave charmonium spin-triplet states in the
$\psi(4040)$ radiative transitions. This sample is about
0.5~fb$^{-1}$, which is limited by the one-month running time left
after the $\pspp$ data taking in the 2010--2011 run.

The upgrade of BEPCII's LINAC in summer 2012 increased the highest
beam energy from 2.1 to 2.3~GeV, making it possible to collect
data at higher center-of-mass (c.m.) energies (up to 4.6~GeV). In
one month's data of 525~pb$^{-1}$ (from December 14, 2012 to
January 14, 2013) at c.m. energy of $4.26$~GeV, the charged
charmonium-like state $\zc$ was discovered~\cite{zc3900}, this
results in changes to the data collection plan for the 2012--2013
run. More data were accumulated at c.m. energies of 4.26~GeV and
then 4.23~GeV where a higher $\EE\to \pp\jpsi$ production rate is
observed. Data at the $Y(4360)$ peak were also obtained in spring
2013, and data at even higher energies (4.42 and 4.6~GeV) were
recorded in 2014 after a fine scan of the total hadronic cross
sections between 3.8 and 4.6~GeV at more than 100 energy points
(``$R$-scan data" hereafter), with a total integrated luminosity
of about 800~pb$^{-1}$. The data taking in 2015-2016 was dedicated
to $D_s$ decays at c.m. energy $\sqrt{s}=4.178$~GeV which can also
be used for $\xyz$ study, followed by more data points between
4.19 and 4.28~GeV dedicated to $\xyz$ related analyses in
2016-2017 running year.

The data samples for the $\xyz$ study (``$\xyz$ data" hereafter)
are presented in Table~\ref{ecm_lum_xyz}, which lists the nominal
c.m. energy, measured c.m. energy (when it is available), and
integrated luminosity at each energy point. These data were used
for all the analyses presented in this article. The c.m. energy
and the luminosity of the $R$-scan data can be found in
Ref.~\refcite{lum_Rscan} and are not listed here.

\begin{table}[htbp]
\tbl{The measured c.m. energy~\cite{ecm_gaoq}, integrated
luminosity~\cite{lum_songwm} of each data sample collected for the
study of the $\xyz$ states. The uncertainties on the integrated
luminosities are statistical only; a 1\% systematic uncertainty
common to all the data points is not listed. ``--" means not
available yet and numbers without error are rough estimation.}
{\begin{tabular}{@{}ccr@{}} \toprule
Data sample & c.m. energy~(MeV)  & ${\cal L}$ ($\rm pb^{-1}$) \\
\colrule
3810       &   3807.65$\pm$0.10$\pm$0.58            &50.54$\pm$0.03     \\
3900       &   3896.24$\pm$0.11$\pm$0.72            &52.61$\pm$0.03     \\
4009       &   4007.62$\pm$0.05$\pm$0.66            &481.96$\pm$0.01    \\
4090       &   4085.45$\pm$0.14$\pm$0.66            &52.63$\pm$0.03     \\
4180       &             4178                       &  $\sim 3190$      \\
4190       &   4188.59$\pm$0.15$\pm$0.68            &43.09$\pm$0.03     \\
4190       &         --      &   $\sim 500$ \\
4200       &         --      &   $\sim 500$ \\
4210       &   4207.73$\pm$0.14$\pm$0.61            &54.55$\pm$0.03     \\
4210       &         --      &   $\sim 500$ \\
4220       &   4217.13$\pm$0.14$\pm$0.67            &54.13$\pm$0.03     \\
4220       &         --      &   $\sim 500$ \\
4230       &   4226.26$\pm$0.04$\pm$0.65            &1091.74$\pm$0.15  \\
4237       &         --      &   $\sim 500$ \\
4245       &   4241.66$\pm$0.12$\pm$0.73            &55.59$\pm$0.04    \\
4246       &         --      &   $\sim 500$ \\
4260       &   4257.97$\pm$0.04$\pm$0.66            &825.67$\pm$0.13   \\
4270       &         --      &   $\sim 500$ \\
4280       &         --      &   $\sim 200$ \\
4310       &   4307.89$\pm$0.17$\pm$0.63            &44.90$\pm$0.03   \\
4360       &   4358.26$\pm$0.05$\pm$0.62            &539.84$\pm$0.10  \\
4390       &   4387.40$\pm$0.17$\pm$0.65            &55.18$\pm$0.04   \\
4420       &   4415.58$\pm$0.04$\pm$0.72            &1073.56$\pm$0.14  \\
4470       &   4467.06$\pm$0.11$\pm$0.73            &109.94$\pm$0.04   \\
4530       &   4527.14$\pm$0.11$\pm$0.72            &109.98$\pm$0.04   \\
4575       &   4574.50$\pm$0.18$\pm$0.70            &47.67$\pm$0.03    \\
4600       &   4599.53$\pm$0.07$\pm$0.74            &566.93$\pm$0.11   \\
 \botrule
\end{tabular} \label{ecm_lum_xyz} }
\end{table}

Compared with the $B$ factories BaBar and Belle, BESIII has its
advantages in the study of the $\xyz$ states, especially in the
study of the vector $Y$ states. BESIII collects $\EE$ annihilation
data at c.m. energies correspond to the production of the $Y$
states directly, while the $B$ factories use data produced via
initial state radiation (ISR), so the BESIII has a much higher
detection efficiency and can take more data at any energy of
interest (for example, the efficiency is 46\% at
BESIII~\cite{zc3900} and about 10\% at Belle~\cite{belley} for
selecting $\y\to \pp\jpsi\to \pp\LL$ ($\ell=e$, or $\mu$) events).
This makes the study of the $Z_c$ states from the $Y$ decays also
more efficient at BESIII than at the $B$ factories. However, $B$
factories can measure the cross sections in a wide energy range
since all the events are produced at the same time, while BESIII
needs to tune the c.m. energy point by point to collect data, thus
can only cover limited energy range.

Needless to say that the $B$ factories can study the $\xyz$ states
with $B$ decays, two-photon fusion, as well as double-charmonium
production and $\Upsilon(nS)$ ($n=1$, $2$, $3$) decays, while
BESIII is limited to $\EE$ annihilation.

\subsection{The LHCb experiment}

LHCb is a dedicated heavy flavor physics experiment at the LHC but
has unexpected potential in the study of hadron spectroscopy. The
detector is a single-arm spectrometer with a forward angular
coverage from approximately 15~mrad to 300 (250)~mrad in the
bending (non-bending) plane~\cite{lhcb_detector}.

The spectrometer magnet is a warm dipole magnet providing an
integrated field of about 4~Tm, which deflects charged particles
in the horizontal plane. The tracking system consists of the
VErtex LOcator (VELO), and four planar tracking stations: the
Tracker Turicensis (TT) upstream of the dipole magnet, and
tracking stations T1--T3 downstream of the magnet. Charged
particles require a minimum momentum of $1.5$~GeV/$c$ to reach the
tracking stations, T1--T3. Charged hadron identification in the
momentum range from 2 to 100~GeV/$c$ is achieved by two Ring
Imaging Cherenkov detectors (RICH1 and RICH2). The calorimeter
system is composed of a Scintillating Pad Detector (SPD), a
Preshower (PS), a shashlik type electromagnetic calorimeter (ECAL)
and a hadronic calorimeter (HCAL). The muon detection system
provides muon identification and contributes to the L0 trigger of
the experiment. The minimum momentum that a muon must have to
traverse the five stations is approximately $6$~GeV/$c$.

The integrated luminosity recorded by the LHCb detector is
1.1~fb$^{-1}$ at 7~TeV in 2011, 2.1~fb$^{-1}$ at 8~TeV in 2012,
and about 4~fb$^{-1}$ at 13~TeV from 2015 up to now, and the plan
is to accumulate up to 5~fb$^{-1}$ at 13~TeV by the end of
2018~\cite{tomasz_charm2018}.

Due to the large production cross sections, more $B$ mesons and
other particles are produced at LHCb than at other experiments
like BaBar and Belle. However, to avoid background from direct
$pp$ collision, many of the studies were performed with a mother
particle which has a long decay length. So far, most of the
results related to the $\xyz$ particles are from $B$ decays. The
large $B$ sample makes the determination of the quantum numbers
and the decay dynamics of the $\xyz$ states possible via partial
wave analysis (PWA) of the $B$ decays.

\section{The iso-triplet charmonium-like states: $Z_c$s}

Searching for the charged charmonium-like state is one of the most
promising ways of studying the exotic hadrons, since such a state
must contain at least four quarks and thus could not be a
conventional meson. The searches were performed in the combination
of one charged pion and a charmonium state, like $\jpsi$, $\psp$,
and $\hc$, since they are narrow and the reconstruction in the
experiment is relatively easy.

The first reported charged charmonium-like state, $Z_c(4430)^-$,
was in the $\pi^-\psp$ invariant mass distribution in $B\to K
\pi^-\psp$ decays in the Belle
experiment~\cite{Belle_zc4430,Belle_zc4430pwa}, it was confirmed
by the LHCb experiment seven years later~\cite{LHCb_zc4430}. The
$\zc^-$ was observed in $\pi^-\jpsi$ invariant mass distribution
in the study of $\EE\to \ppjpsi$ at BESIII~\cite{zc3900} and
Belle~\cite{belley_new} experiments, and the $\zcp^-$ was observed
in $\pi^-\hc$ system in $\EE\to \pphc$~\cite{zc4020} only at
BESIII. There is also evidence for $Z_c$ structures in the
$\pi\psp$ system at Belle~\cite{belle_y4660_new} and
BESIII~\cite{bes3_pppsp} in $\EE\to \pp\psp$.

These states seem to indicate that a new class of hadrons has been
observed. As there are at least four quarks within all these $Z_c$
states, they have been interpreted either as compact tetraquark
states, molecular states of two charmed mesons ($\bar{D}D^*+c.c.$,
$\bar{D^*}D^*$, $\bar{D}D_1+c.c.$, $\bar{D^*}D_1+c.c.$, etc.),
hadro-quarkonium states, or other configurations~\cite{reviews}.

There are other $Z_c$ states observed in different processes, such
as the two structures at masses 4050 and 4250~MeV/$c^2$ in
$\pi^-\chi_{c1}$ system in $B\to K \pi^-\chi_{c1}$
decays~\cite{belle_zc12}; and the $Z_c(4200)^-$ in the
$\pi^-\jpsi$ invariant mass distribution in $B\to K \pi^-\jpsi$
decays~\cite{belle_zc4200} from Belle experiment. We will focus on
the $Z_c(4430)^-$, $\zc^-$, and $\zcp^-$ in this article as new
information is available recently, we also report new observation
in $\pi^-\psp$ system from BESIII experiment~\cite{bes3_pppsp}.

\subsection{The $\zc$}

\subsubsection{Observation of the $\zc$}

\begin{figure}
\begin{center}
\includegraphics[height=4.3cm]{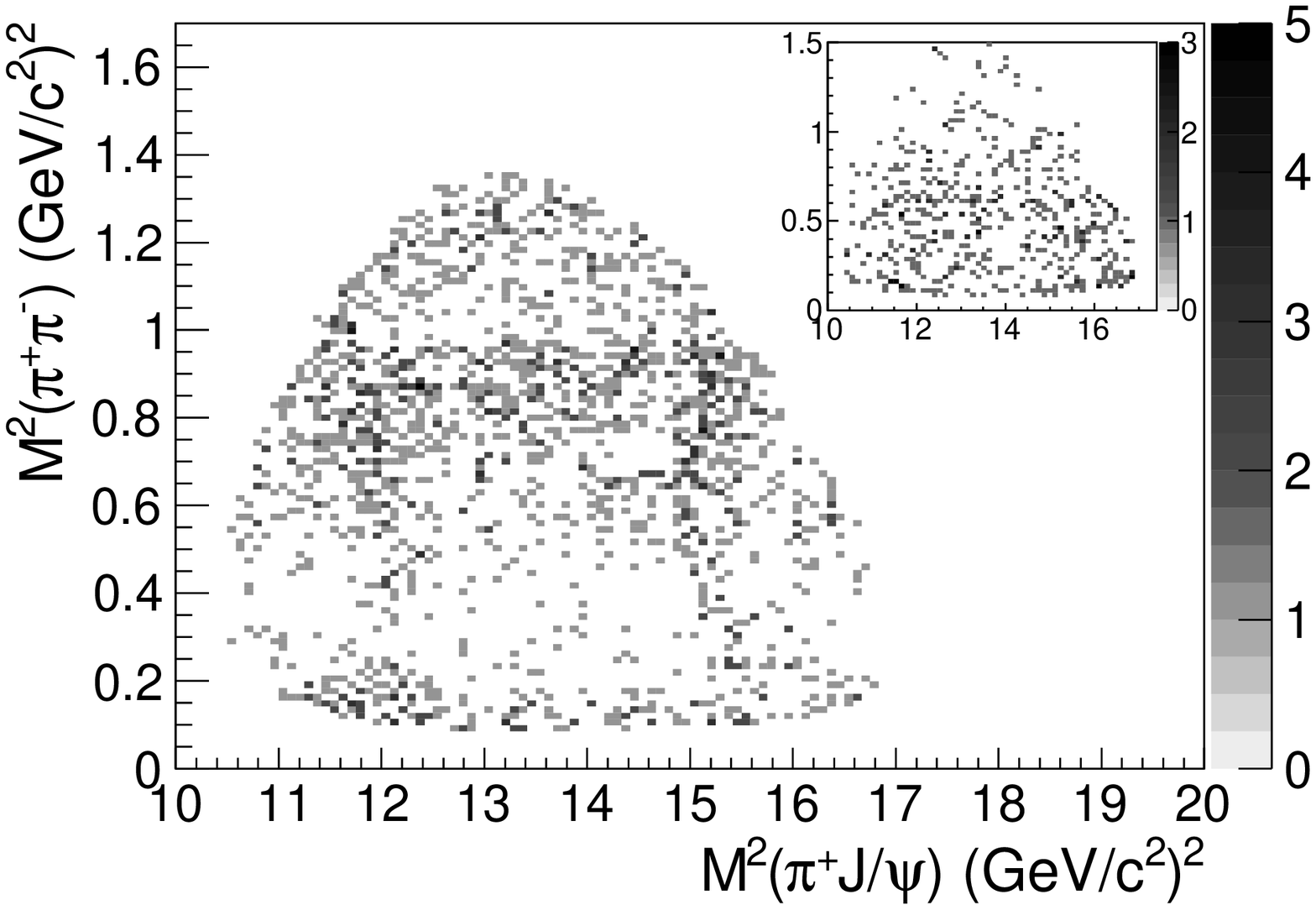}
\includegraphics[height=4.3cm]{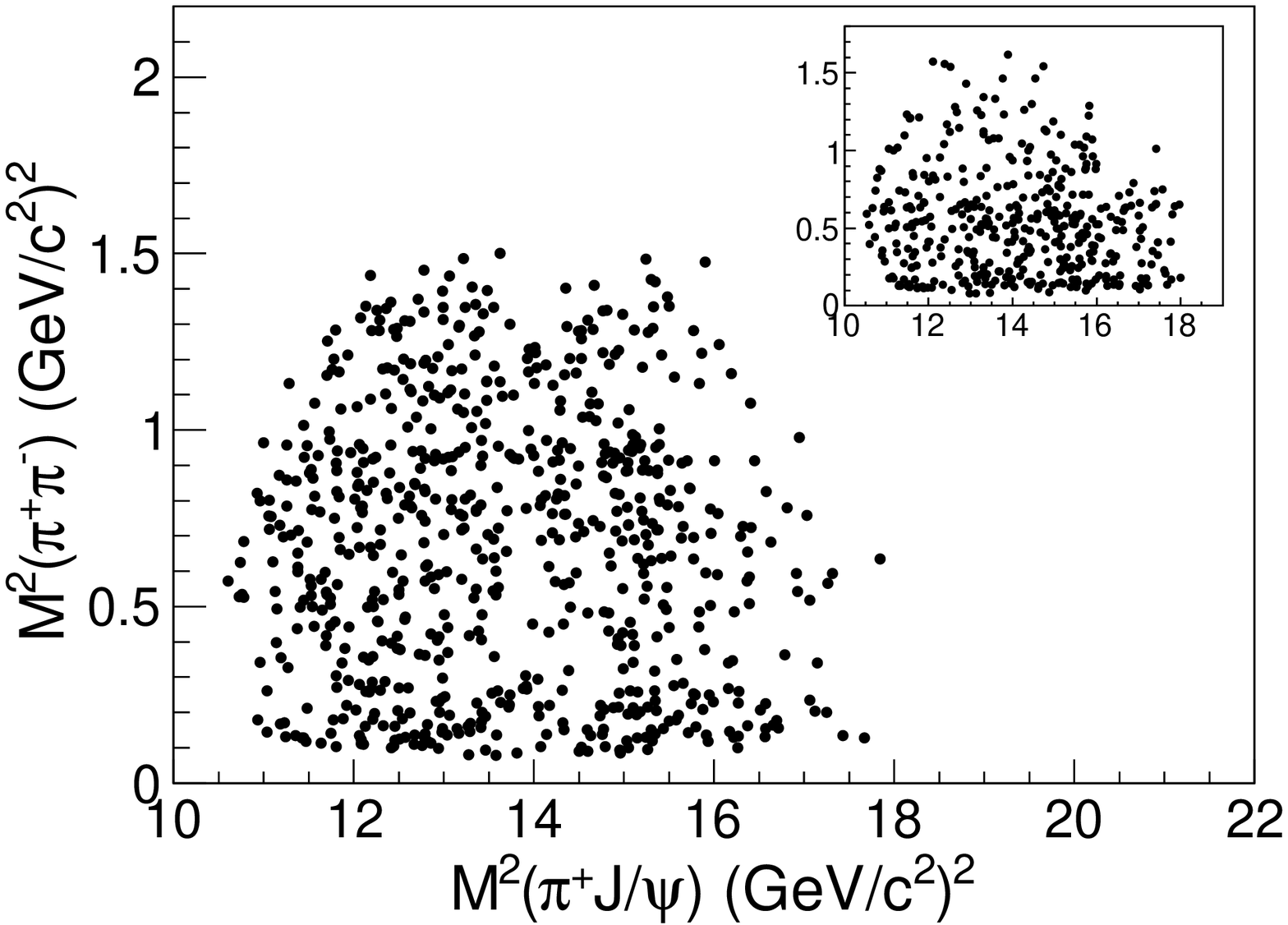}
\caption{Dalitz plots for selected $\EE\to \ppjpsi$ events in the
$\jpsi$ signal region at BESIII (left panel) and at Belle (right
panel). The insets show background events from the $\jpsi$ mass
sidebands~(not normalized).} \label{dalitz_zc}
\end{center}
\end{figure}

The BESIII experiment studied the $\EE\to \ppjpsi$ process at a
c.m. energy of 4.26~GeV using a 525~pb$^{-1}$ data
sample~\cite{zc3900}, with $\jpsi$ decays into a pair of $\EE$ or
$\MM$. $595\pm 28$ signal events in the $\EE$ mode and $882\pm 33$
signal events in the $\MM$ mode are reconstructed and the cross
section is measured to be $(62.9\pm 1.9 \pm 3.7)$~pb, which agrees
with the existing results from the BaBar~\cite{babar_y4260_new}
and Belle~\cite{belley} experiments. The $\jpsi$ signal is
selected by requiring the invariant mass of the lepton pair is
consistent with the $\jpsi$, and a sample of 1595 $\ppjpsi$ events
with a purity of 90\% is obtained. The intermediate states are
studied by examining the Dalitz plot (shown in
Fig.~\ref{dalitz_zc}) of the selected candidate events.

\begin{figure}[htbp]
\centering
 \includegraphics[height=4.3cm]{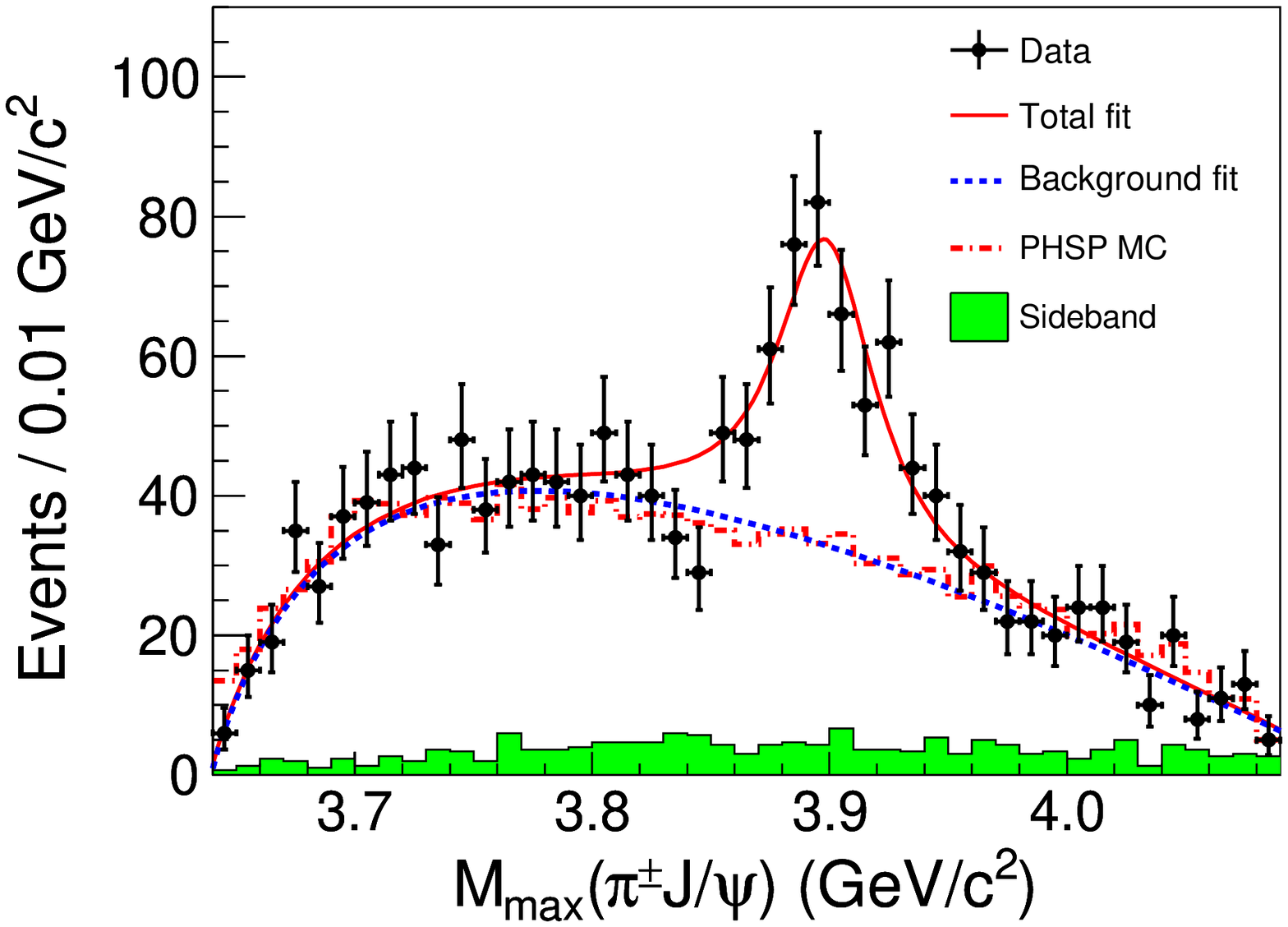}
 \includegraphics[height=4.3cm]{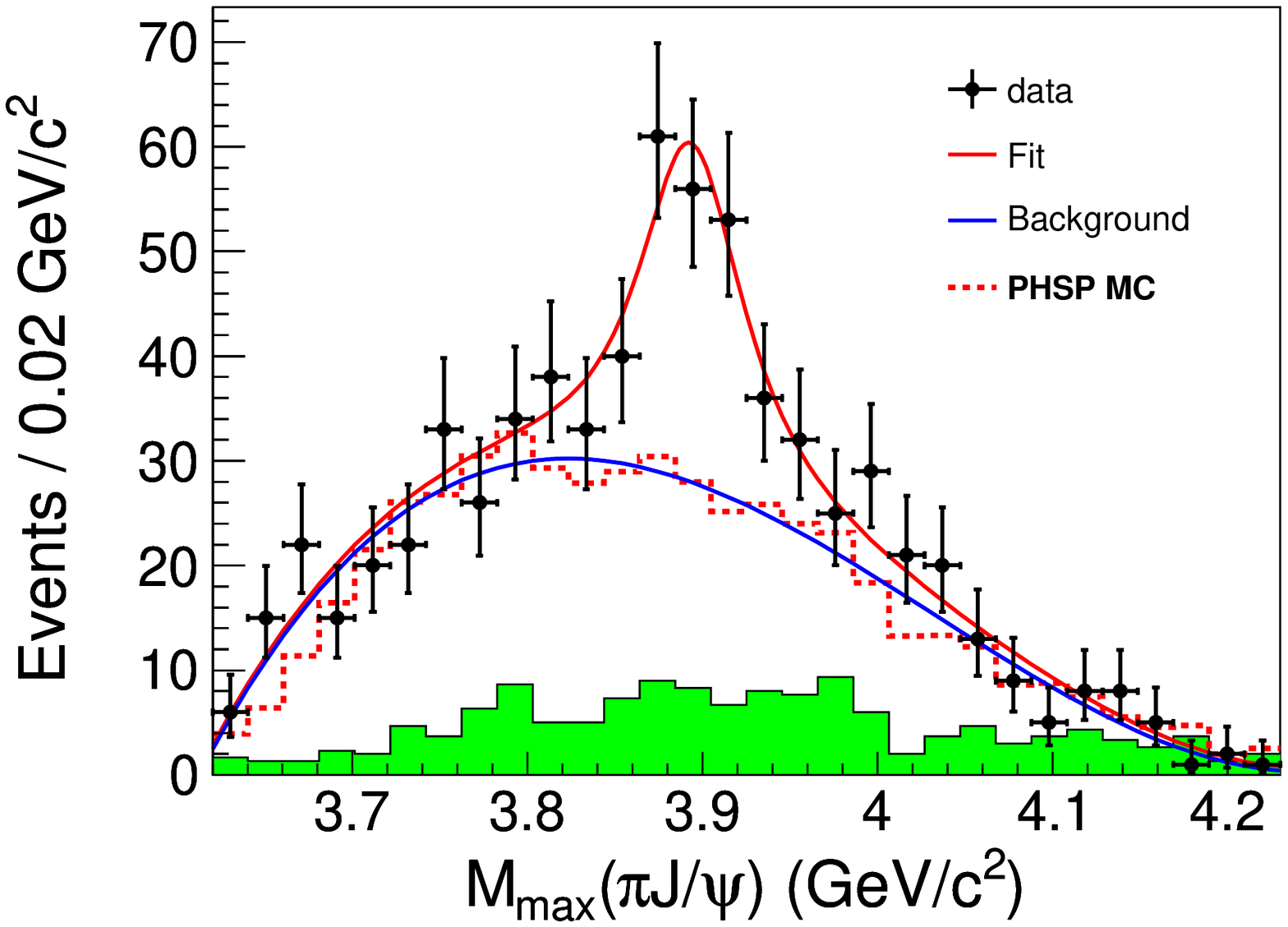}
 \caption{Unbinned maximum likelihood fit to the distribution of
the $M_{\mathrm{max}}(\pi J/\psi)$ (left panel from BESIII and
right panel from Belle). Points with error bars are data, the
curves are the best fit, the dashed histograms are the phase space
distributions and the shaded histograms are the non-$\ppjpsi$
background estimated from the normalized $\jpsi$ sidebands.}
\label{projfit}
\end{figure}

Besides the known $\sigma$ and $f_0(980)$ structures in the $\pp$
system, a structure at around 3.9~GeV/$c^2$ was observed in the
$\pi^\pm \jpsi$ invariant mass distribution with a statistical
significance larger than $8\sigma$, which is referred to as the
$\zc$. A fit to the $\pi^\pm\jpsi$ invariant mass spectrum with a
constant width Breit-Wigner (BW) function (see
Fig.~\ref{projfit}), neglecting interference with other
amplitudes, results in a mass of $(3899.0\pm 3.6\pm 4.9)~{\rm
MeV}/c^2$ and a width of $(46\pm 10\pm 20)$~MeV. The associated
production ratio is measured to be $\frac{\sigma(\EE\to \pi^\pm
\zc^\mp\to \ppjpsi))} {\sigma(\EE\to \ppjpsi)}=(21.5\pm 3.3\pm
7.5)\%$.

At the Belle experiment, the cross section of $\EE\to \ppjpsi$ is
measured from 3.8--5.5~GeV using the ISR method with a
967~fb$^{-1}$ data sample collected at or near the $\Upsilon(nS)$
($n = 1,\ 2,\ 4,\ 5$) resonances~\cite{belley_new}. Events in the
$\y$ signal region ($4.15  < M(\ppjpsi) < 4.45$~GeV/$c^2$) are
investigated, as is shown in Fig.~\ref{dalitz_zc}. The $\zc$ state
(referred to as $Z(3900)^+$ in the Belle paper) with a mass of
$(3894.5\pm 6.6\pm 4.5)~{\rm MeV}/c^2$ and a width of $(63\pm
24\pm 26)$~MeV is observed in the $\pi^\pm\jpsi$ mass spectrum
(see Fig.~\ref{projfit}) with a statistical significance larger
than $5.2\sigma$. The ratio of the production rates
$\frac{\sigma(\EE\to \pi^\pm \zc^\mp\to \ppjpsi))} {\sigma(\EE\to
\ppjpsi)}=(29.0\pm 8.9)\%$, where the error is statistical only.

The $\zc$ state was confirmed shortly after with CLEO-c data at a
c.m. energy of 4.17~GeV~\cite{seth_zc}, and the mass and width
agreed very well with the BESIII and Belle measurements. In
addition, a $3.5\sigma$ evidence for $\zc^{0}$ in the CLEO-c data
was also reported in $\EE\to \piz\piz\jpsi$
process~\cite{seth_zc}.

BESIII measured the cross sections of $e^+e^-\to \pi^0\pi^0
J/\psi$ with data at c.m. energy ranges from
4.19--4.42~GeV~\cite{zc03900}. A neutral state $\zc^0\to
\piz\jpsi$ with a significance of $10.4\sigma$ was observed, with
the mass and width measured to be $(3894.8\pm 2.3\pm
3.2)$~MeV/$c^2$ and $(29.6\pm 8.2\pm 8.2)$~MeV, respectively. This
state decays to $\pi^0 J/\psi$ and its mass is close to that of
$\zc^\pm$, so it is interpreted as the neutral partner of the
$\zc^\pm$. The measured production rate of $\EE\to \pi^0\zc^0$ is
about half of that for $\EE\to \pi^+\zc^-+c.c.$, which is
consistent with the isospin symmetry expectation. This determines
the $\zc$ is an isovector state.

\subsubsection{Observation of $\zc\to D\bar{D}^*+c.c.$}

The $\zc$ observed in the $\pi\jpsi$ final state is close to and
above the $D\bar{D}^*+c.c.$ mass threshold. With the same data
sample at $\sqrt{s}=4.26$~GeV, the BESIII experiment studied
$\EE\to \pi^\pm (D\bar{D}^*)^{\mp}$ and observed the open-charm
decay $\zc^\pm\to (D\bar{D}^*+c.c.)^\pm$~\cite{zc3885st}.

The $\EE\to \pi^{+} (D\dstrbar)^{-}+c.c.$ events are selected by a
so-called single-tag technique in which only the bachelor
$\pi^\pm$ and one final-state $D$ meson are detected, and the
$\dstrbar$ is inferred from energy-momentum conservation. In this
analysis, both isospin channels $\pip D^0 D^{*-}+c.c.$ and $\pip
D^- D^{*0}+c.c.$ are studied. The $D$ mesons are reconstructed in
the $D^0\to K^-\pi^+$ and $D^+\to K^-\pip\pip$ decay channels.

\begin{figure}[htbp]
\centering
  \includegraphics[width=6.0cm]{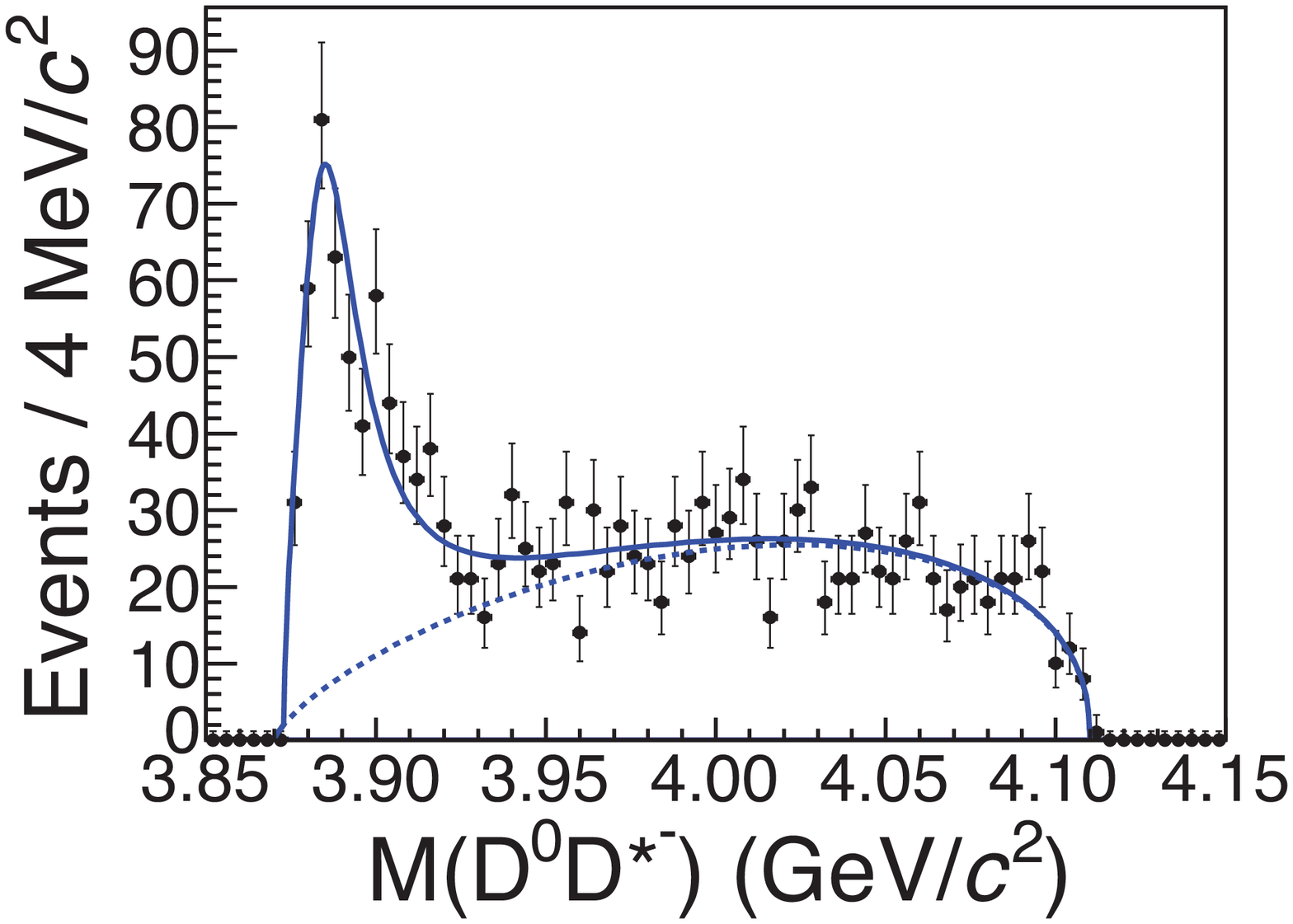}
  \includegraphics[width=6.0cm]{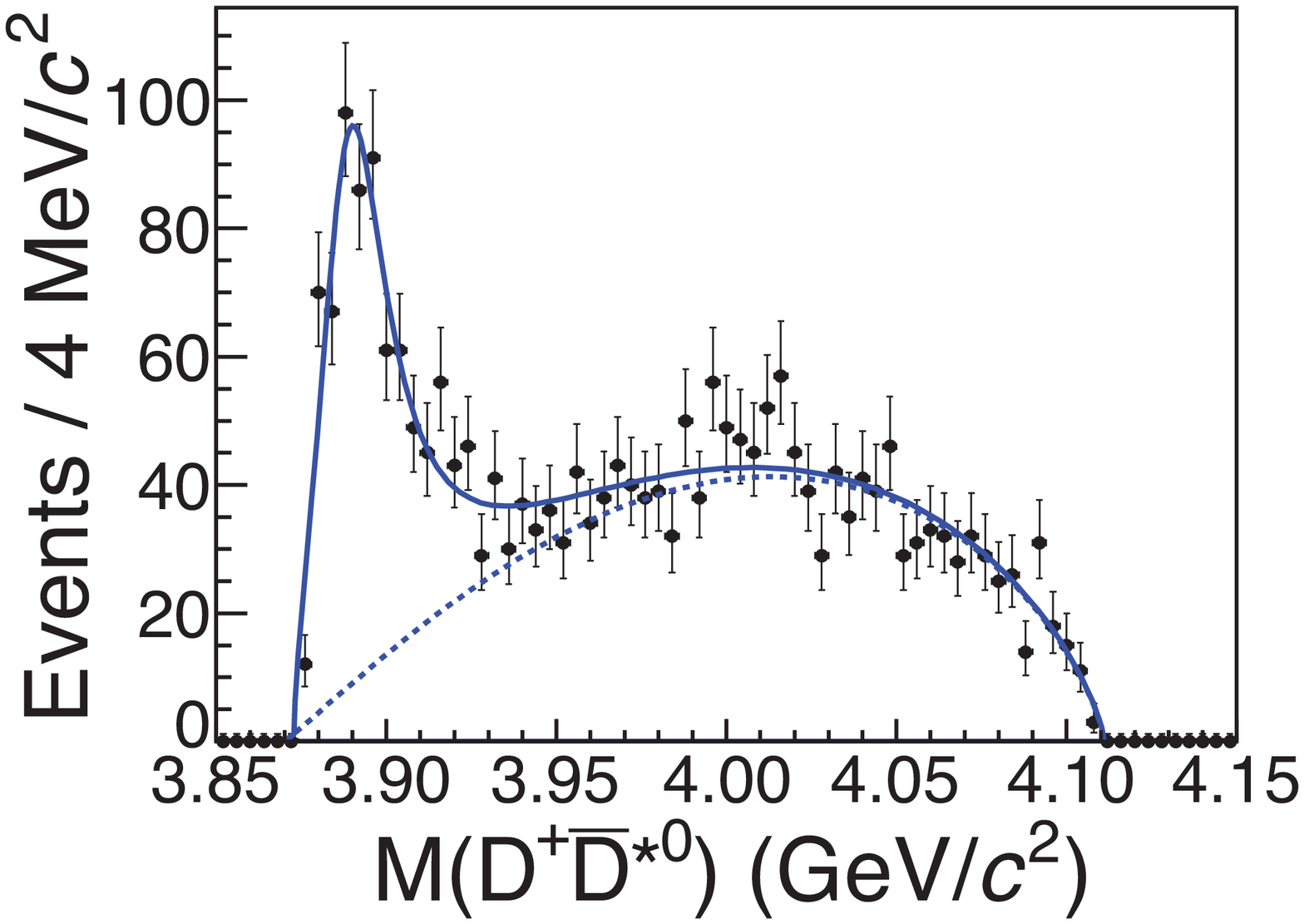}
\caption{The $M(D^0 D^{*-})$ (left) and $M(D^+\bar{D}^{*0})$
(right) distributions for selected events at $\sqrt{s}=4.26$~GeV
in the single-tag analysis. The curves show the best fits. }
\label{xuxp}
\end{figure}

A structure close to the threshold is observed in the
$(D\bar{D}^*)^{\pm}$ invariant mass distribution. When fitted to a
BW function with mass-dependent width, the pole mass and width are
determined to be $(3883.9 \pm 1.5 \pm 4.2)$~MeV/$c^2$ and
$(24.8\pm 3.3 \pm 11.0)$~MeV, respectively (see Fig.~\ref{xuxp}).
The production rate is measured to be $\sigma(\EE \to \pi^{\mp}
\zc^{\pm})\times \BR(\zc^{\pm}\to (D\bar{D}^*)^{\pm}) =(83.5\pm
6.6 \pm 22.0)$~pb.

The analysis is refined with double-tag method and more
luminosity~\cite{zc3885dt}. In this analysis, both $\EE\to
\pi^+D^0D^{*-}+c.c.$ and $\pi^+ D^-D^{*0}+c.c.$ are measured with
data samples at $\sqrt{s}$=4.23 and 4.26~GeV. The bachelor $\pi^+$
and the $D$ meson pair in the final state are reconstructed, with
the $\pi$ from $D^{*-}$ and $D^{*0}$ decays inferred using
energy-momentum conservation. The $D^0$ is reconstructed in four
decay modes ($K^-\pi^+$, $K^-\pi^+\pi^0$, $K^-\pi^+\pi^+\pi^-$,
and $K^-\pi^+\pi^+\pi^-\pi^0$), and the $D^-$ in six decay modes
($K^+\pi^-\pi^-$, $K^+\pi^-\pi^-\pi^0$, $K^0_S\pi^-$,
$K^0_S\pi^-\pi^0$, $K^0_S\pi^+\pi^-\pi^-$, and $K^+K^-\pi^-$). The
double $D$ tag technique allows the use of more $D$ decay modes
and the background level is greatly suppressed.

\begin{figure*}[htbp]
\centering
\epsfig{width=0.45\textwidth,clip=true,file=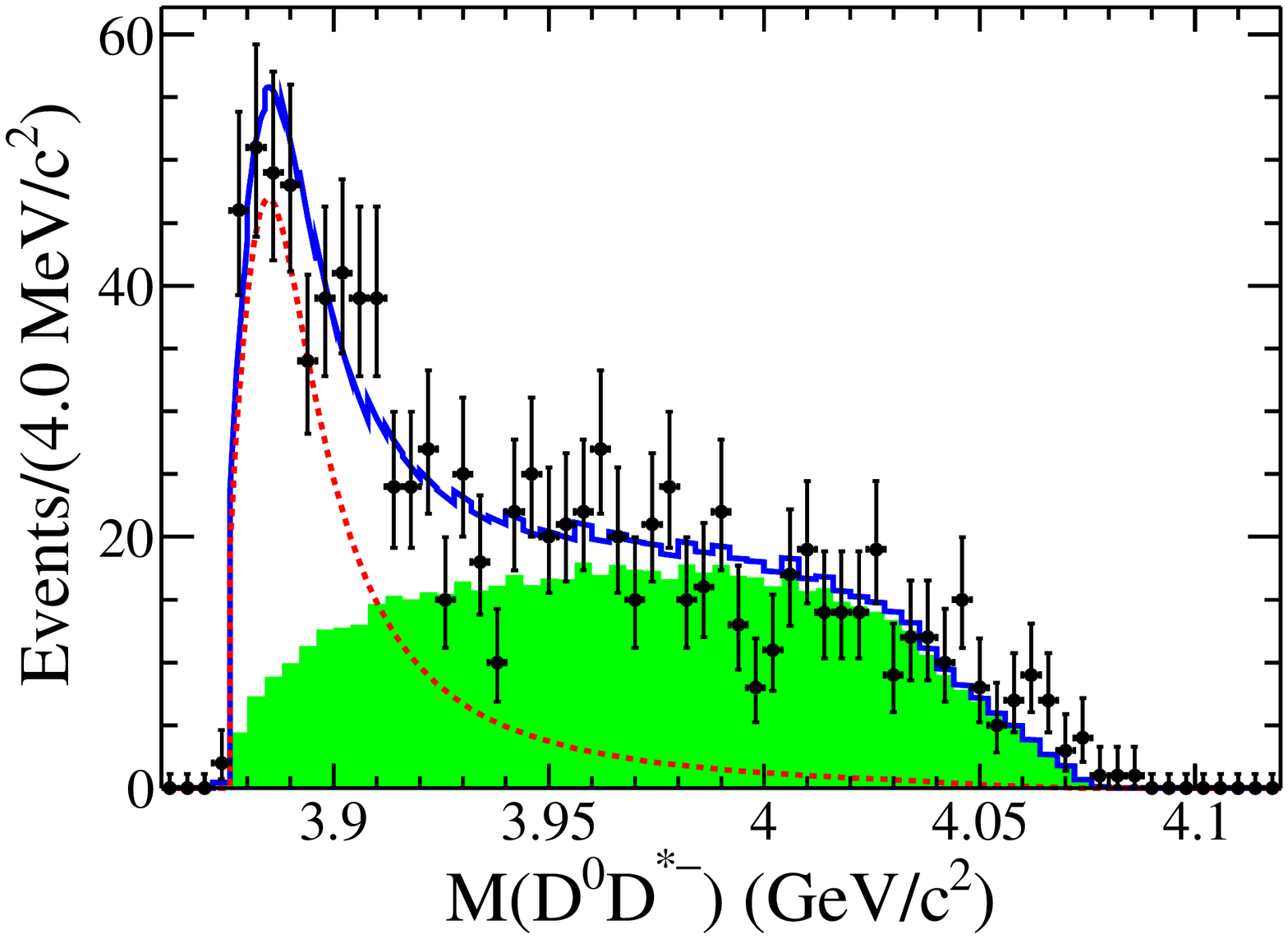}
\epsfig{width=0.45\textwidth,clip=true,file=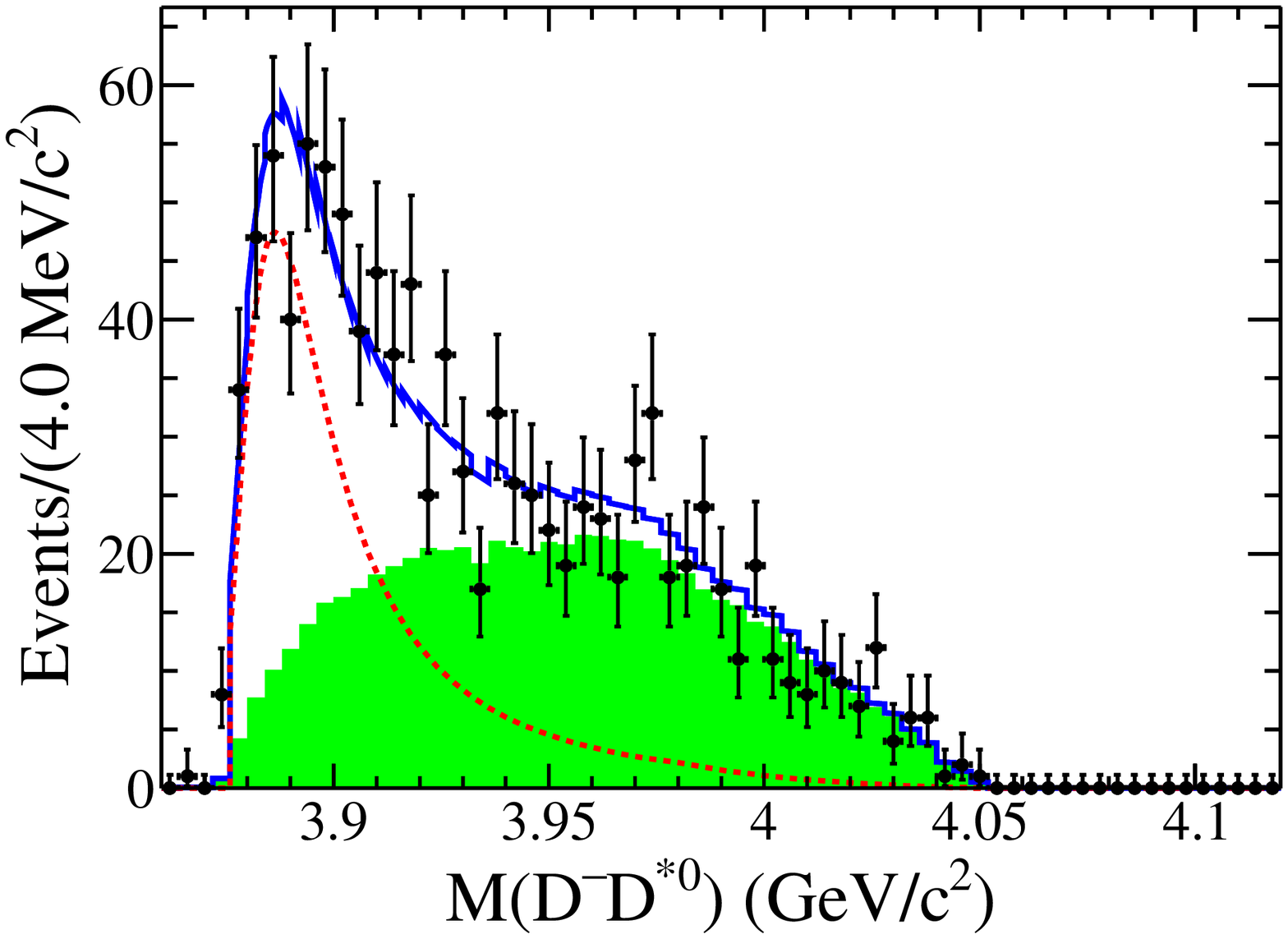}\\
\epsfig{width=0.45\textwidth,clip=true,file=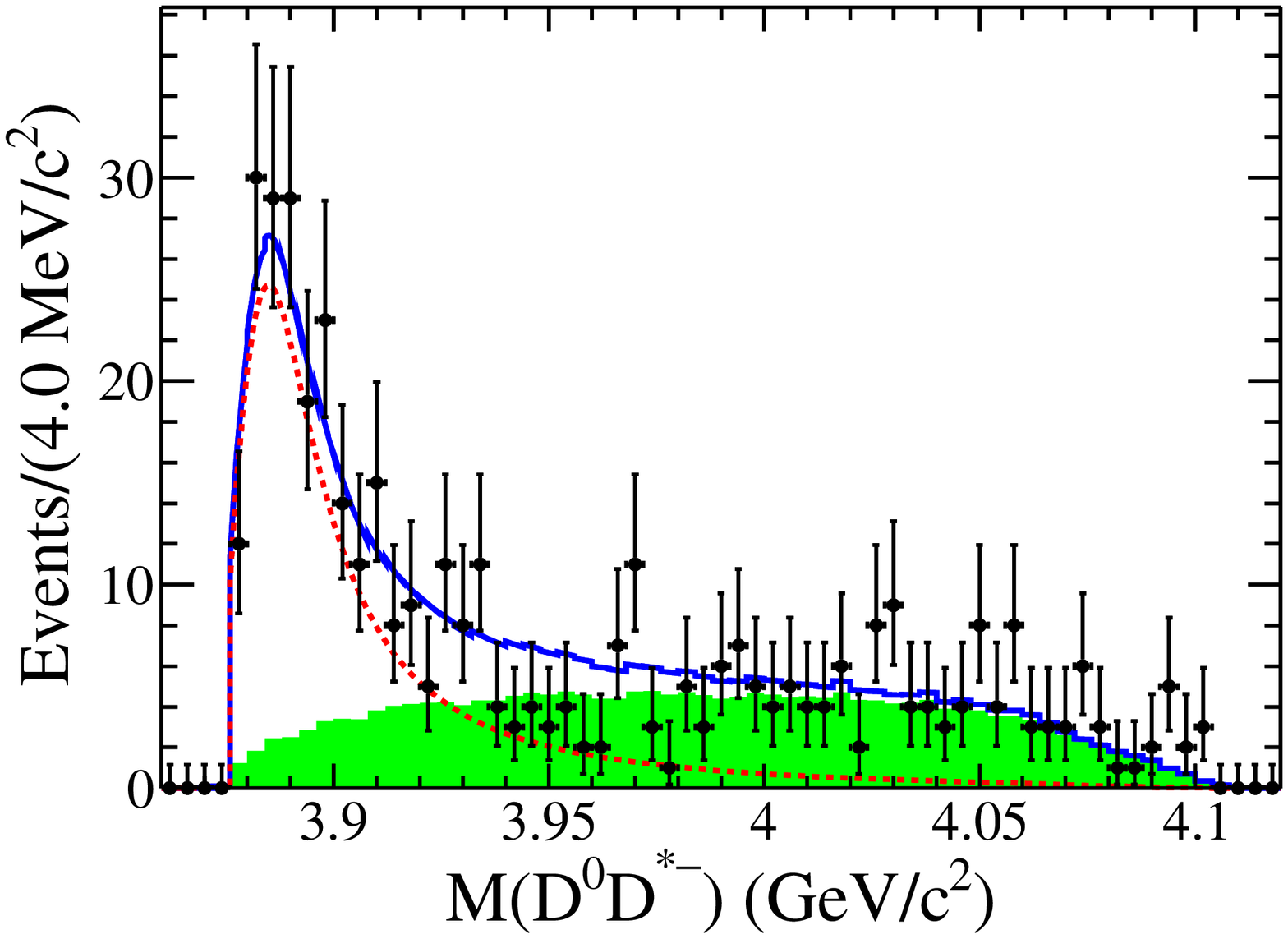}
\epsfig{width=0.45\textwidth,clip=true,file=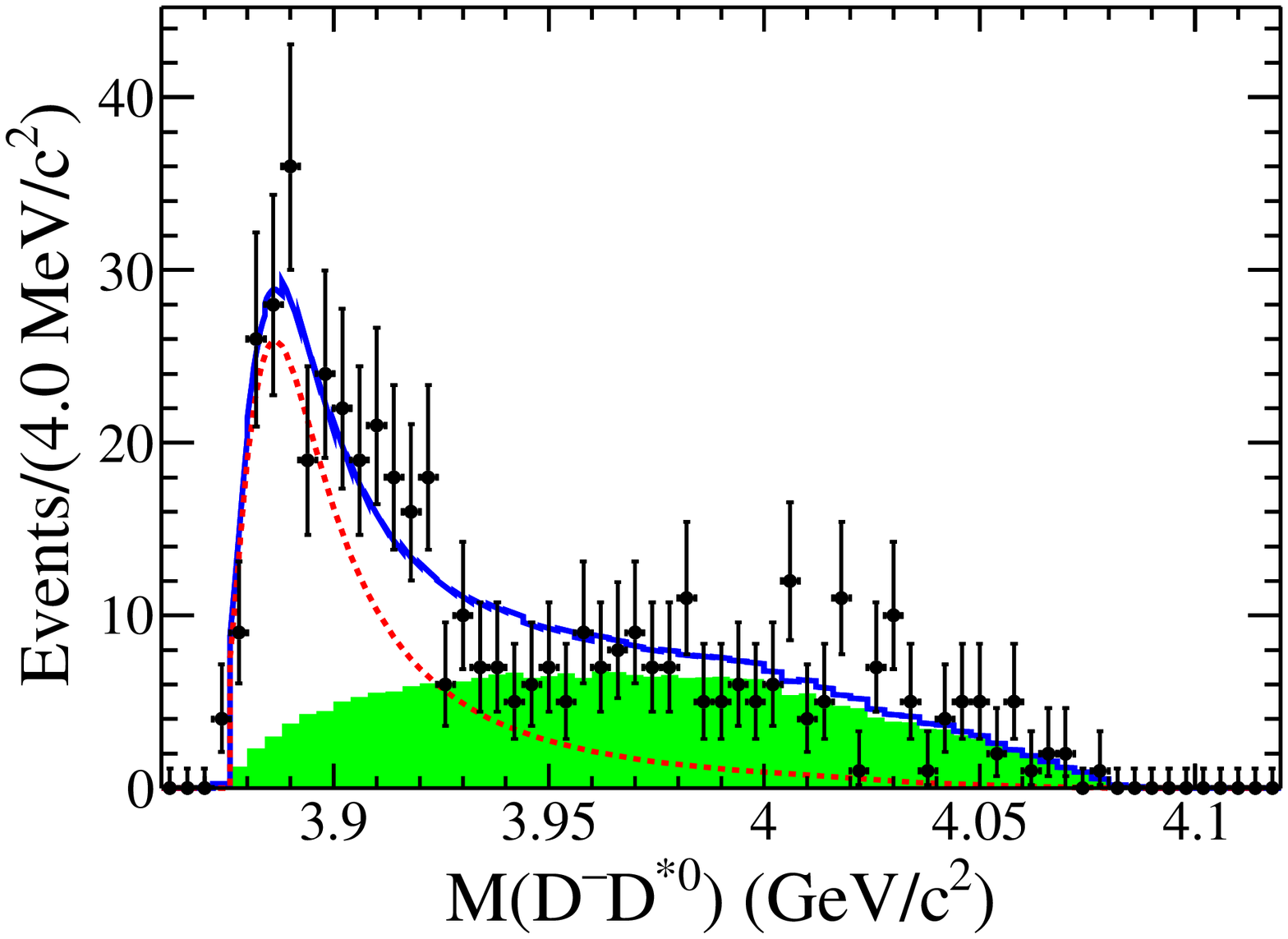}
\caption{Simultaneous fits to the $M(D\bar{D}^*)$ distributions in
the double-tag analysis. The top row is from data at
$\sqrt{s}$=4.23~GeV and the bottom row at $\sqrt{s}$=4.26~GeV. The
left column is for $\EE\to \pi^+D^0D^{*-}+c.c.$ and the right
column for $\EE\to \pi^+ D^-D^{*0}+c.c.$. The dots with error bars
are data and the lines show the fit to the data. The solid lines
(blue) describe the total fits, the dashed lines (red) describe
the signal shapes and the green areas describe the background
shapes.} \label{sig_fit}
\end{figure*}

The $M(D\bar{D}^*)$ distributions for the two processes at
$\sqrt{s}$=4.23 and 4.26~GeV are fitted simultaneously (shown in
Fig.~\ref{sig_fit}) with a BW function for the $\zc$ signal and a
phase space distribution for the background, added incoherently.
The mass and width of $\zc$ are fitted to be $(3890.3\pm
0.8)$~MeV/$c^2$ and $(31.5\pm 3.3)$~MeV, respectively, correspond
to the pole mass and pole width of $(3881.7\pm 1.6\pm
1.6)$~MeV/$c^2$ and $(26.6\pm 2.0\pm 2.1)$~MeV, respectively.

The production rate is measured as $\sigma(\EE\to
\pi^{\mp}\zc^{\pm})\times \BR(\zc^{\pm}\to (DD^*)^{\pm}) =
(141.6\pm 7.9\pm 12.3)$~pb at $\sqrt{s}$=4.23~GeV, and $(108.4\pm
6.9\pm 8.8)$~pb at $\sqrt{s}$=4.26~GeV.

The pole position of the $\zc$ and the production rate are
consistent with but more precise than those from single-tag
analysis. The double-tag analysis only has $\sim$9\% events in
common with the single-tag analysis, so the two analyses are
almost statistically independent and can be combined into a
weighted average. The combined pole mass and width are $(3882.2\pm
1.1\pm 1.5)$~MeV/$c^2$ and $(26.5\pm 1.7\pm 2.1)$~MeV,
respectively. The combined production rate $\sigma(\EE\to
\pi^{\mp}\zc^{\pm})\times \BR(\zc^{\pm}\to (DD^*)^{\pm})$ is
$(104.4\pm 4.8\pm 8.4)$~pb at $\sqrt{s}$=4.26~GeV.

In an analysis of $\EE\to \pi^0 (D\bar{D}^*)^0$, the $\zc^0\to
(D\bar{D}^*)^0$ is also observed~\cite{zc3885_0} and agree with
the expectation from isospin symmetry.

\subsubsection{The quantum numbers}
\label{Sec:jpc}

In both the single-tag and double-tag analyses of $\zc\to
D\bar{D}^*+c.c.$~\cite{zc3885st,zc3885dt}, by checking the angular
distribution of the $\pi$ accompanying the $\zc$, BESIII finds
that the spin-parity $J^P=1^+$ of the $\zc$ is favored over
$J^P=0^-$ and $1^-$ ($J^P=0^+$ is not allowed due to spin-parity
conservation in $\zc\to \pi\jpsi$), but $J>1$ cannot be ruled out
by simply checking one angular distribution.

BESIII determines the spin-parity of the $\zc$ based on a PWA of
$\EE\to \pp\jpsi$ events at $\sqrt{s}=4.23$ and
$4.26$~GeV~\cite{zc3900_jpc}. Following the event selection
reported in Ref.~\refcite{zc3900}, the numbers of selected
candidate events are 4154 at $\sqrt{s}=4.23$~GeV and 2447 at
$\sqrt{s}=4.26$~GeV, with 365 and 272 background events,
respectively, estimated by using the $\jpsi$ mass sidebands.

Amplitudes of the PWA are constructed with the helicity-covariant
method~\cite{chung}. The process $\EE\to \pp\jpsi$ is assumed to
proceed via the $\zc$ resonance, $i.e.$, $\EE\to \pi^\pm \zc^\mp$,
$\zc^\mp\to \pi^\mp\jpsi$, and via the non-$\zc$ decay $\EE\to
R\jpsi$, $R\to \pp$, with $R=\sigma$, $f_0(980)$, $f_2(1270)$, and
$f_0(1370)$. In the fit, the $\zc$ line shape is described with a
Flatt\'e-like formula taking into account the fact that the
$\zc^\pm$ decays are dominated by the final states
$(D\bar{D}^*)^\pm$~\cite{zc3885st,zc3885dt} and
$\pi^\pm\jpsi$~\cite{zc3900}. All processes are added coherently
to obtain the total amplitude.

\begin{figure*}[htbp]
\centering
\includegraphics[width=0.9\textwidth]{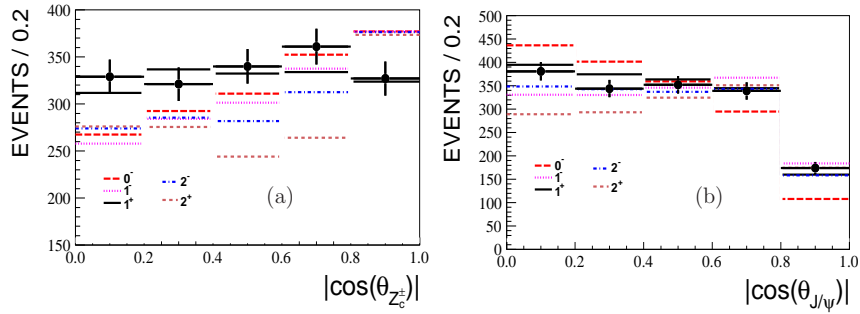}
\caption{\label{angdis} Polar angle distribution of $\zc^\pm$ in
$\EE\to \pi^+\zc^- + c.c.$ (a), and helicity angle distribution of
$\jpsi$ in the $\zc^\pm\to \pi^\pm\jpsi$ (b). The dots with error
bars are the $\zc$ enriched sample, and compared with the fits
with different $J^P$ hypotheses.}
\end{figure*}

\begin{figure*}[htbp]
\centering
\includegraphics[width=0.9\textwidth]{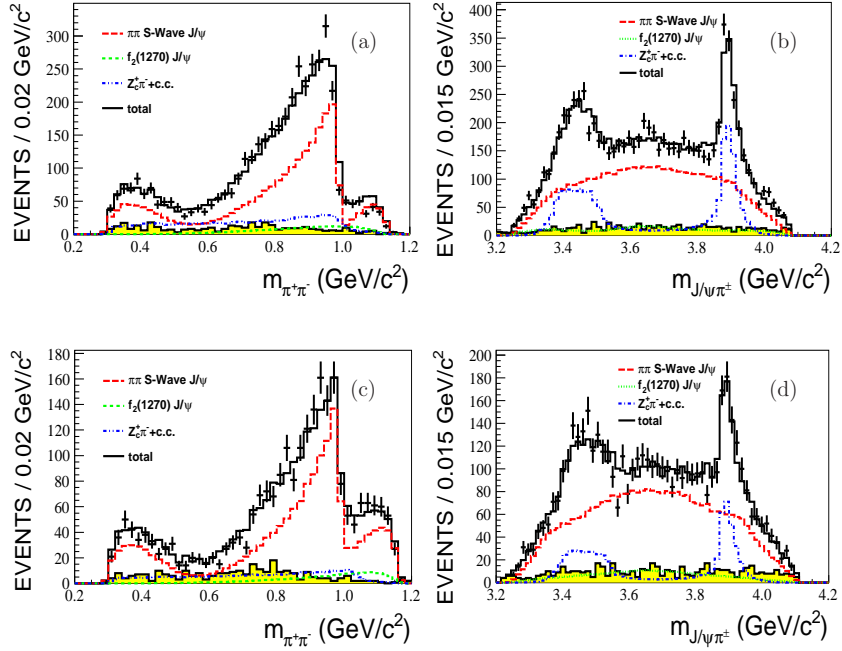}
\caption{\label{pwafitresult} Projections to $m_{\pp}$ (a,~c) and
$m_{\jpsi\pi^\pm}$ (b,~d) of the fit results with $J^P=1^+$ for
the $\zc$, at $\sqrt s=4.23$~GeV (a,~b) and $4.26$~GeV (c,~d). The
points with error bars are data, and the black histograms are the
total fit results. The shaded histogram denotes backgrounds. Plots
(b) and (d) are filled with two entries ($m_{\jpsi\pi^+}$ and
$m_{\jpsi\pi^-}$) per event.}
\end{figure*}

The fit indicates that the spin-parity $J^P=1^+$ of the $\zc$ are
favored by more than 7$\sigma$ over other quantum numbers ($0^-$,
$1^-$, $2^-$, and $2^+$), as can be seen in Fig.~\ref{angdis} for
a $\zc$ enriched sample ($m_{\jpsi\pi^\pm}\in
(3.86,3.92)$~GeV/$c^2$). Figure~\ref{pwafitresult} shows
projections of the fit results with $J^P=1^+$ for the $\zc$ state.
The pole mass of the $\zc$ is measured as $(3881.2\pm 4.2\pm
52.7)$~MeV/$c^2$ and pole width $(51.8\pm 4.6\pm 36.0)$~MeV. The
Born cross sections for $\EE\to \pi^+\zc^-+c.c.\to \ppjpsi$ are
measured to be $(21.8\pm 1.0\pm 4.4)$~pb at $\sqrt{s}=4.23$~GeV
and $(11.0\pm 1.2\pm 5.4)$~pb at $\sqrt{s}=4.26$~GeV.

\subsubsection{Evidence for $\zc\to \rho\eta_c$}

BESIII searches for $\EE\to \pp\piz\etac$ and intermediate states
decay into $\rho\etac$ with data collected at 4.23, 4.26, and
4.36~GeV~\cite{ycz_charm2018}. In this analysis, $\eta_c$ is
reconstructed with 9 hadronic final states: $p\bar{p}$, $2(K^+
K^-)$, $K^+ K^- \pi^+ \pi^-$,  $K^+ K^- \pi^0$, $p \bar{p} \pi^0$,
$\ks K^\pm \pi^\mp$, $\pi^+ \pi^- \eta$, $K^+ K^- \eta$, and
$\pi^+ \pi^- \pi^0 \pi^0$.

\begin{figure*}[htbp]
\centering
\epsfig{file=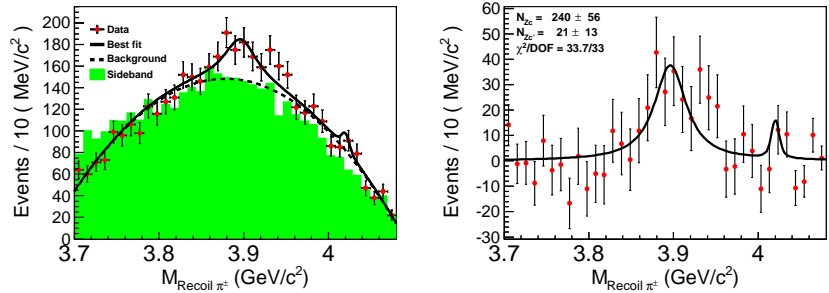,width=0.9\textwidth}
\caption{The $\pi^\pm$ recoil mass distribution in $\EE\to
\pi^{\pm}\rho^{\mp} \etac$ at $\sqrt{s}=4.23$~GeV and the fit with
$Z_c(3900/4020)^{\pm}$ signals (left panel); and the same plot
with background subtracted (right panel). Dots with error bars are
data, shaded histogram is from $\etac$ sidebands, normalized to
the number of backgrounds from the fit, the solid lines are total
fit and the dotted line is background. \label{fit_Zc_data_4230}}
\end{figure*}

Clear signal of $\EE\to \pp\piz\etac$ is observed at
$\sqrt{s}=4.23$~GeV. The $Z_c(3900/4020)^{\pm}\to \rho^\pm \etac$
signals are examined after requiring that the invariant mass of
the $\etac$ candidate is within the $\etac$ signal region and the
invariant mass of $\pi^{\pm}\pi^{0}$ is within the $\rho^\pm$
signal region. The recoil mass of the remaining $\pi^\mp$
(equivalent to the invariant mass of $\rho^\pm\etac$) is shown in
Fig.~\ref{fit_Zc_data_4230} for the data at $\sqrt{s}=4.23$~GeV,
the $\zc^\pm$ signal is found while there is no significant
$\zcp^\pm$ signal. The $\rho^\pm\etac$ invariant mass distribution
is fitted with the contributions from $\zc$ and $\zcp$ together
with a smooth background. $240\pm 56$ $\zc^\pm$ events is observed
with a statistical significance of $4.3\sigma$ ($3.9\sigma$
including the systematical uncertainty). The $\zc$ signals at
other c.m. energies and the $\zcp$ signals at all the c.m.
energies are not statistically significant.

The cross section is measured as
 \(
\sigma(\EE\to \pi^\mp \zc^\pm\to \pi^\mp\rho^\pm\etac)=(47 \pm 11
\pm 11)~{\rm pb}
 \)
at $\sqrt{s}=4.23$~GeV. This result is equal within errors to the
cross section of $\EE\to \ppp\etac$, which is $(46 \pm 12 \pm
10)\rm \,pb$. This indicates that the $\EE\to \ppp\etac$ process
is saturated by the $\EE\to \pi^\mp \zc^\pm\to
\pi^\mp\rho^\pm\etac$. No signal is observed at $\sqrt{s}=$4.26
and 4.36~GeV and the upper limits of the production cross sections
at the 90\% confidence level (C.L.) are determined.

Using the results from Ref.~\refcite{zc3900_jpc}, the ratio of the
branching fractions of different $\zc^\pm$ decays is calculated as
$R_{\zc}=\frac{\BR(\zc\to \rho\etac)}{\BR(\zc\to \pi\jpsi)}=2.1\pm
0.8$ at $\sqrt{s}=4.23$~GeV and less than 6.4 at
$\sqrt{s}=4.26$~GeV at the 90\% C.L. The theoretical predictions
for this ratio varies depending on model assumptions and ranges
from a few per cent to a few
hundreds~\cite{Esposito:2014hsa,Li:2014pfa,Ke:2013gia,zhusl_1,zhusl_2,Voloshin_rhoetac}.

\subsubsection{Hint for $\zc\to \pi \hc$}
\label{Sec:zctopihc}

\begin{figure}[htbp]
\begin{center}
\includegraphics[width=0.45\textwidth]{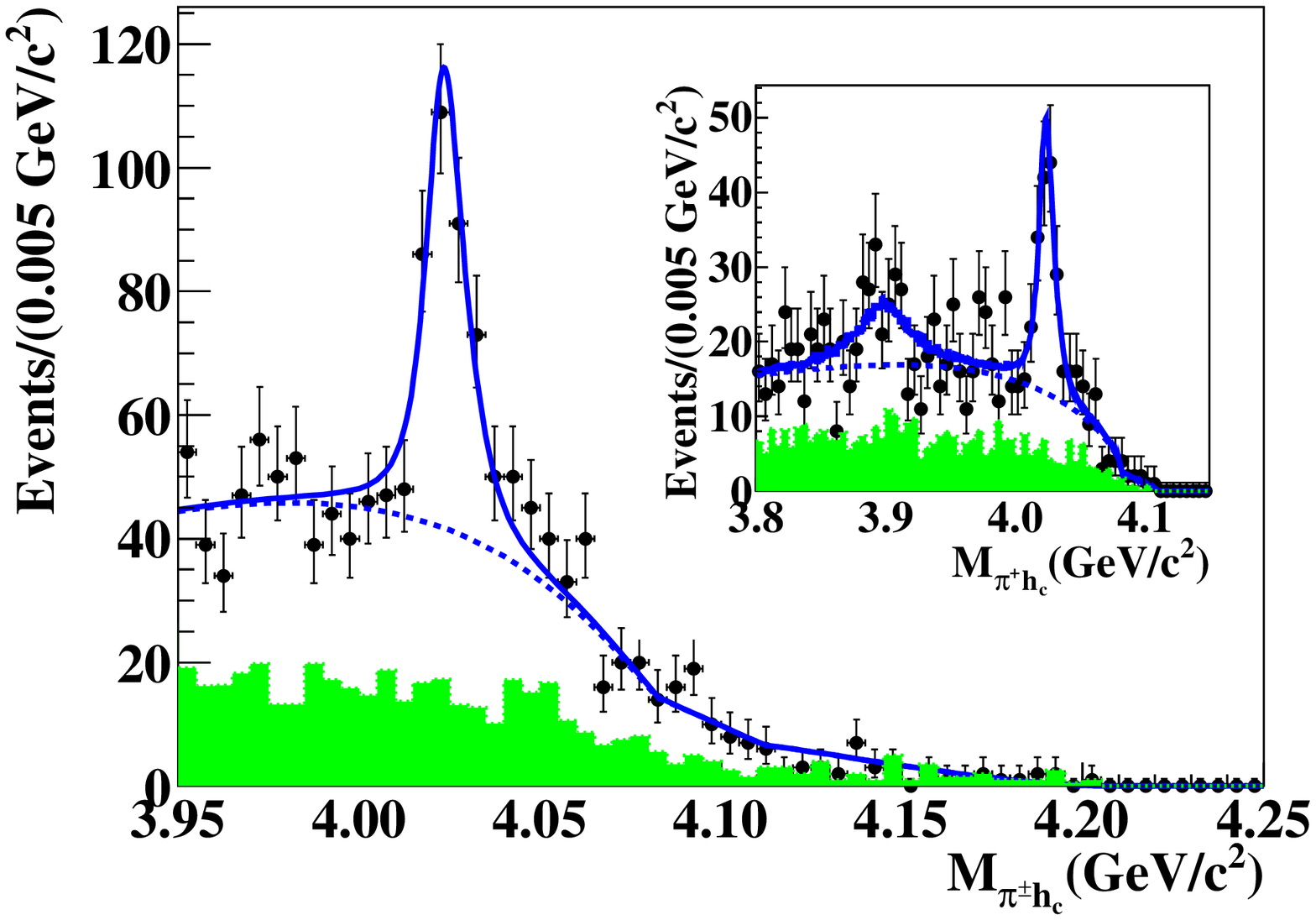}
\includegraphics[width=0.45\textwidth]{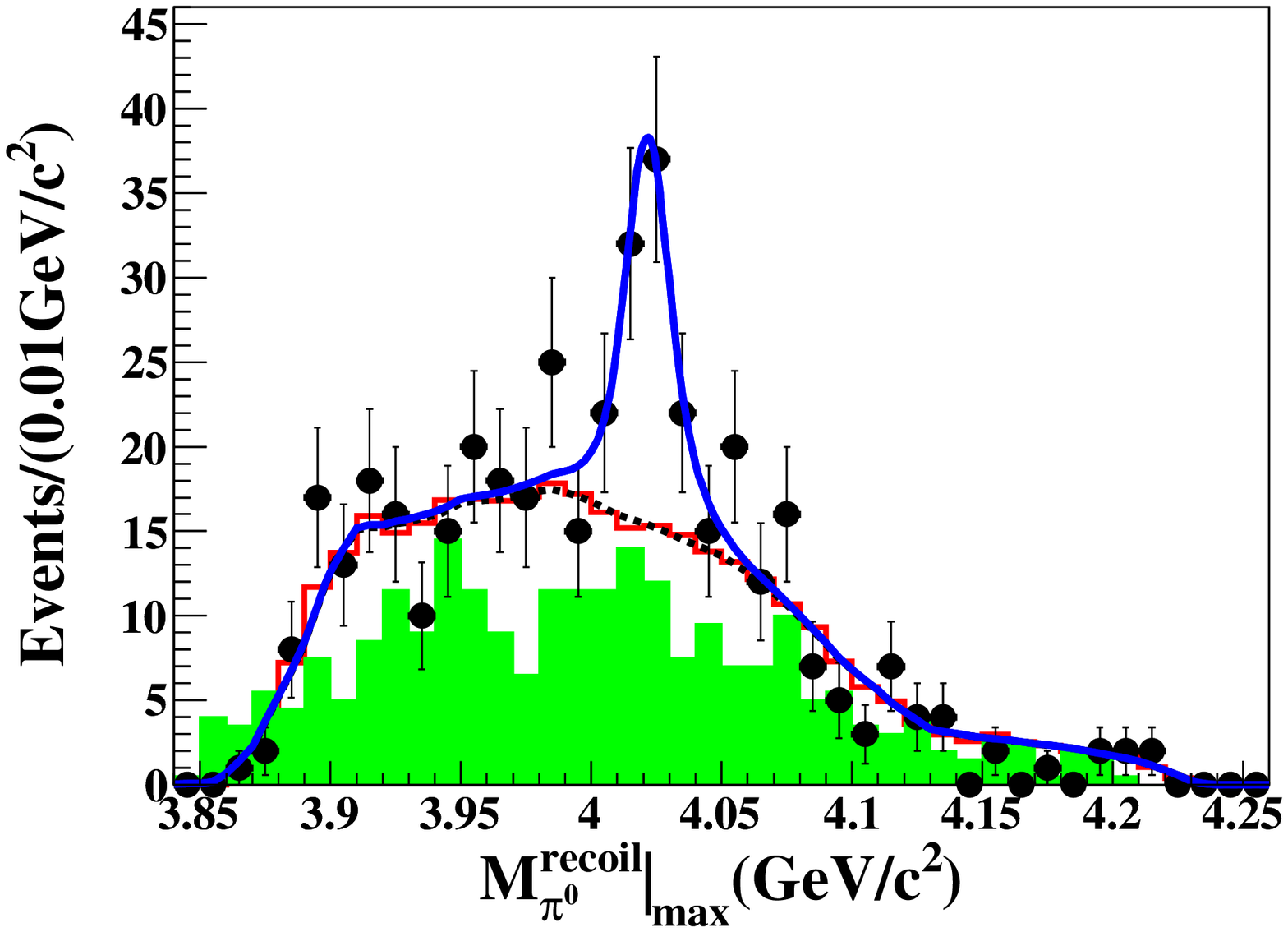}
\caption{Sum of the simultaneous fits to the $M(\pi^\pm h_c)$
(left panel) and $M(\piz h_c)$ (right panel) distributions from
$\EE\to \pphc$ and $\piz\piz\hc$, respectively, at 4.23, 4.26, and
4.36~GeV in the BESIII data; the inset in the left panel shows the
sum of the simultaneous fit to the $M({\pi^\pm h_c})$
distributions at 4.23 and 4.26~GeV with $\zc$ and $\zcp$. Dots
with error bars are data; shaded histograms are normalized
sideband background; the solid curves show the total fit, and the
dotted curves the backgrounds from the fit.} \label{1Dfit}
\end{center}
\end{figure}

BESIII measured cross sections of $\EE\to \pphc$ at c.m. energies
of 3.90--4.42~GeV with $\hc\to \gamma\etac$ and $\etac$ decays
into 16 hadronic final states~\cite{zc4020}. Intermediate states
were studied by examining the Dalitz plot of the selected $\pphc$
candidate events. While a new resonance $\zcp$ (see
Sec.~\ref{Sec:zcp}) is observed, there is only a faint signal at
around 3.9~GeV/$c^2$ (inset of Fig.~\ref{1Dfit}~(left)), which
could be $\zc$.

An unbinned maximum likelihood fit was applied to the
$M(\pi^\pm\hc)$ distribution assuming there are both $\zcp$ and
$\zc$ contributions with the mass and width of the latter fixed to
the BESIII measurements~\cite{zc3900}. The fit results in a
statistical significance of 2.1$\sigma$ (see the inset of
Fig.~\ref{1Dfit}~(left)) for the $\zc$. At the 90\% C.L., the
upper limits on the production cross sections are set to
$\sigma(\EE\to \pi^\pm \zc^\mp\to \pphc) <13$~pb at 4.23~GeV and
$<11$~pb at 4.26~GeV. These are lower than those of $\zc\to
\pi^\pm\jpsi$~\cite{zc3900_jpc}.

\subsubsection{Summary on the $\zc$}

From the above studies, we conclude that the $\zc$ is an isovector
state with positive $G$-parity and spin-parity $J^P=1^+$. It
decays into $\pi\jpsi$ and $D\dstrbar$, and it may also decay into
$\rho\etac$ and $\pi\hc$ final states. The neutral $\zc$ has a
negative $C$-parity.

Although the mass and width are reported in many measurements, in
most of the cases the interference between the $\zc$ and other
amplitudes is not considered properly. The most reliable
measurement is from the PWA of the $\EE\to \ppjpsi$
mode~\cite{zc3900_jpc}, with the pole mass of $(3881.2\pm 4.2\pm
52.7)$~MeV/$c^2$ and pole width of $(51.8\pm 4.6\pm 36.0)$~MeV.
The errors are still large, dominated by the uncertainties in the
parametrization of the $\pp$ S-wave amplitudes. The assumption
that $\zc$ decays dominantly into $\pi\jpsi$ and $D\dstrbar$ in
PWA may also introduce bias, as the $\rho\etac$ mode has been
observed with a larger decay rate than $\pi\jpsi$, and there could
be other decay modes such as $\pi\hc$, $\pi\psp$ and so on.

The reported production cross section for $\EE\to \pi^+\zc^-+c.c.$
suffers from the same problems mentioned in the $\zc$ mass and
width determination. The only reliable measurement is the product
cross section of $\EE\to \pi^+\zc^-+c.c.\to \ppjpsi$ determined
from PWA which is $(21.8\pm 1.0\pm 4.4)$~pb at $\sqrt{s}=4.23$~GeV
and $(11.0\pm 1.2\pm 5.4)$~pb at
$\sqrt{s}=4.26$~GeV~\cite{zc3900_jpc}. It seems that the $\zc$ is
produced at 4.23~GeV twice as much as that at 4.26~GeV. It would
be very important to measure the $\zc$ production cross sections
at other c.m. energies, to check if the line shape of $\EE\to
\pi\zc$ is the same as that of $\EE\to \ppjpsi$. This will be an
important piece of information in understanding the nature of the
$\zc$ and the production mechanism.

Although BESIII also searched for the $\zc$ decays into isospin
violating mode $\eta\jpsi$~\cite{Zc_jpsieta_bes3} and light hadron
final state $\omega\pi$~\cite{Zc_omegapi_bes3}, it is not
surprising that the decays are not observed and the upper limits
of the decay rates are one order of magnitude or even smaller than
$\zc\to \pi\jpsi$, as expected naively.

\subsection{The $\zcp$}

\subsubsection{Observation of the $\zcp$}
\label{Sec:zcp}

\begin{figure}[htbp]
\begin{center}
\includegraphics[width=0.5\textwidth]{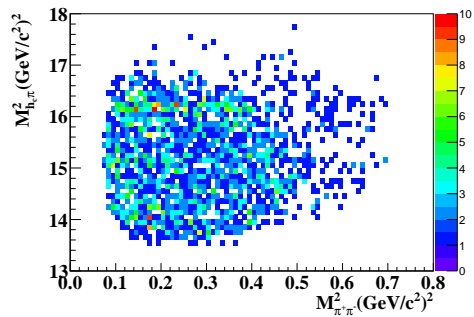}
\caption{Dalitz plot ($M^2_{\pi^+\hc}$ vs. $M^2_{\pp}$) for
selected $\EE\to \pphc$ events, summed over all energy points.}
\label{dalitz_pphc}
\end{center}
\end{figure}

As has been mentioned in Sec.~\ref{Sec:zctopihc}, BESIII measures
cross sections of $\EE\to \pphc$ at c.m. energies of
3.90--4.42~GeV~\cite{zc4020}. Intermediate states are studied by
examining the Dalitz plot of the selected $\pphc$ candidate
events. The $\hc$ signal is selected using $3.518 < M_{\gamma
\eta_c} < 3.538$~GeV/$c^2$, and $\pphc$ samples of 859 events at
4.23~GeV, 586 events at 4.26~GeV, and 469 events at 4.36~GeV are
obtained with purities of about $\sim$65\%. Although there are no
clear structures in the $\pp$ system, there is distinct evidence
for an exotic charmonium-like structure in the $\pi^\pm\hc$
system, as clearly shown in the Dalitz plot
(Fig.~\ref{dalitz_pphc}). Figure~\ref{1Dfit}~(left) shows the
projection of the $M(\pi^\pm\hc)$ (two entries per event)
distribution for the signal events, as well as the background
events estimated from normalized $\hc$ mass sidebands. There is a
significant peak at around 4.02~GeV/$c^2$ ($\zcp$), and there are
also some events at around 3.9~GeV/$c^2$ which could be $\zc$. The
individual datasets at $\sqrt{s}=4.23$, $4.26$, and $4.36$~GeV
show similar structures.

An unbinned maximum likelihood fit is applied to the
$M(\pi^\pm\hc)$ distribution summed over the 16 $\eta_c$ decay
modes. The data at 4.23, 4.26, and 4.36~GeV are fitted
simultaneously to the same signal function with common mass and
width. Figure~\ref{1Dfit}~(left) shows the fitted results. The
mass and width of $\zcp$ are measured to be $(4022.9\pm 0.8\pm
2.7)~{\rm MeV}/c^2$ and $(7.9\pm 2.7\pm 2.6)$~MeV, respectively.
The statistical significance of the $\zcp$ signal is found to be
greater than $8.9\sigma$.

The numbers of $\zcp$ events are determined to be $114\pm 25$,
$72\pm 17$, and $67\pm 15$ at 4.23, 4.26, and 4.36~GeV,
respectively. The cross sections are calculated to be
$\sigma(\EE\to \pi^\pm \zcp^\mp\to \pphc) = (8.7\pm 1.9\pm 2.8\pm
1.4)$~pb at 4.23~GeV, $(7.4\pm 1.7\pm 2.1\pm 1.2)$~pb at 4.26~GeV,
and $(10.3\pm 2.3\pm 3.1\pm 1.6)$~pb at 4.36~GeV, where the first
errors are statistical, the second ones systematic, and the third
ones from the uncertainty in $\BR(\hc\to \gamma\etac)$~\cite{pdg}.

BESIII also observes $\EE\to \piz\piz\hc$ at $\sqrt{s}=$4.23,
4.26, and 4.36~GeV for the first time~\cite{zc0_4020}. The
measured Born cross sections are about half of those for $\EE\to
\pphc$, which agree with expectations based on isospin symmetry
within systematic uncertainties. $\zcp^0$, the neutral isospin
partner of the $\zcp^\pm$ is observed in $\piz\hc$ invariant mass
distribution (Fig.~\ref{1Dfit}~(right)). This observation
indicates that there are no anomalously large isospin violations
in $\pp\hc$ and $\pi\zcp$ systems, and $\zcp$ is an isovector
state.

\subsubsection{Observation of  $\zcp\to D^{*}\bar{D}^{*}$}

\begin{figure}[htbp]
\centering
\includegraphics[height=4cm]{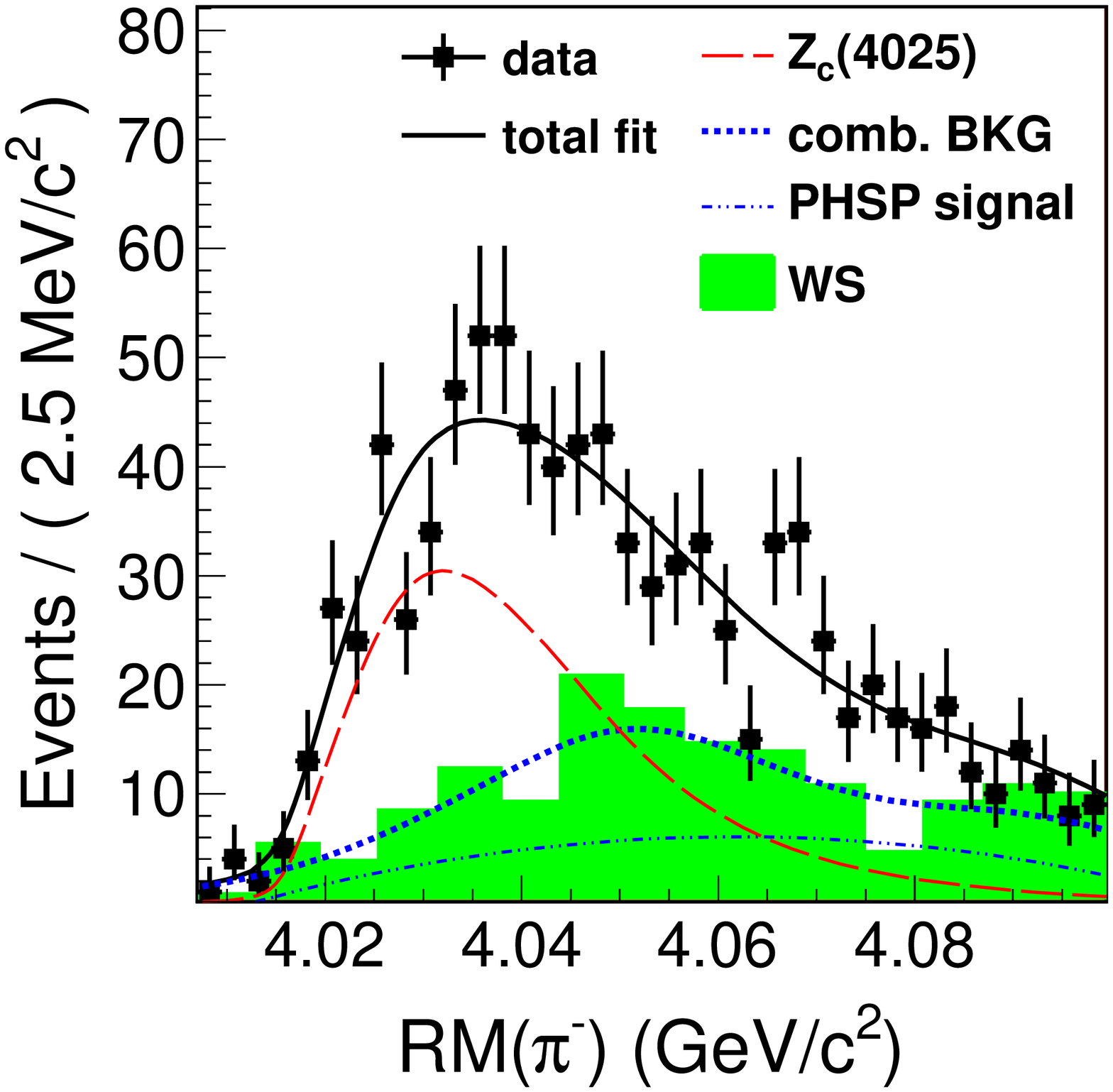}
\includegraphics[height=4cm]{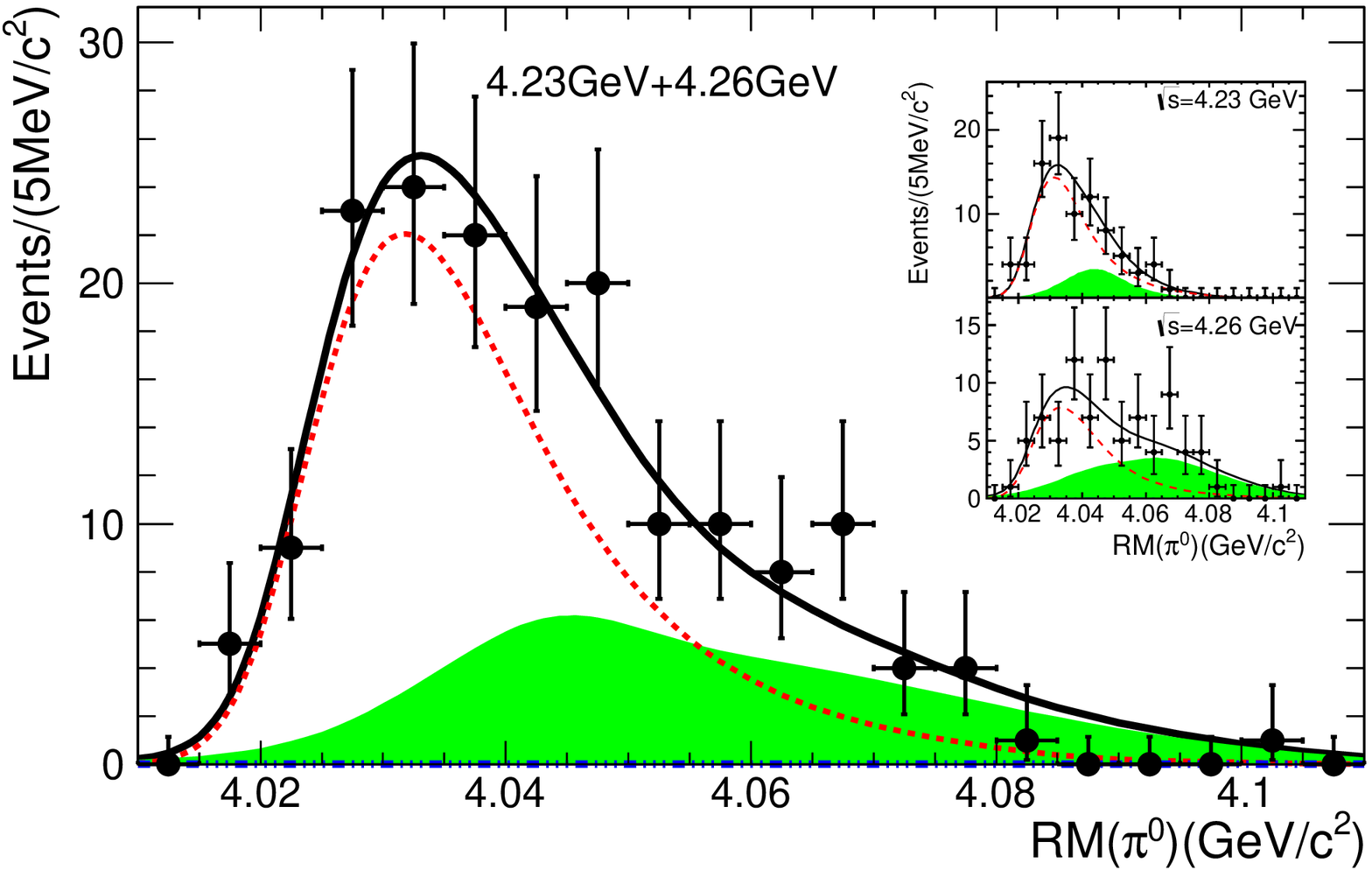}
\caption{Unbinned maximum likelihood fit to the $\pi^\mp$ recoil
mass spectrum (left) in $\EE\to (D^{*} \bar{D}^{*})^{\pm} \pi^\mp$
at $\sqrt{s}=4.26$~GeV, and to the $\piz$ recoil mass spectrum
(right) in $\EE\to (D^{*} \bar{D}^{*})^{0} \piz$ at
$\sqrt{s}=4.23$ and $4.26$~GeV at BESIII. } \label{fig:fit}
\end{figure}

The $\zcp$ is very close to and is above the $D^*\bar{D}^*$
threshold of 4.02~GeV/$c^2$, so it may couples to $D^*\bar{D}^*$
final state. The BESIII experiment studies the $\EE \to (D^{*}
\bar{D}^{*})^{\pm} \pi^\mp$ process using data at
$\sqrt{s}=$4.26~GeV~\cite{zc4025}. Based on a partial
reconstruction technique (only the bachelor $\pi^\mp$, one charged
$D$, and one $\piz$ from $D^*$ decays are reconstructed), the Born
cross section is measured to be $(137\pm 9\pm 15)$~pb. A structure
near the $(D^* \bar{D}^*)^{\pm}$ threshold in the $\pi^\mp$ recoil
mass spectrum is observed (see Fig.~\ref{fig:fit}~(left)). The
measured mass and width of the structure are $(4026.3\pm 2.6\pm
3.7)$~MeV/$c^2$ and $(24.8\pm 5.6\pm 7.7)$~MeV, respectively, from
the fit with a constant-width BW function for the signal, and the
statistical significance is $13\sigma$. From the fit results,
$401\pm 47$ $\zcp$ signal events are obtained, and the associated
ratio of the production rates $\frac{\sigma(\EE\to \zcp^\pm
\pi^\mp \to (D^{*} \bar{D}^{*})^{\pm} \pi^\mp)}{\sigma(e^+e^-\to
(D^{*} \bar{D}^{*})^{\pm} \pi^\mp)}$ is determined to be $0.65\pm
0.09\pm 0.06$.

The processes $\EE \to (D^{*0} \bar{D}^{*0})\piz$ and
$(D^{*+}{D}^{*-}) \piz$ are also studied at BESIII to search for
the neutral partner of the $\zcp$ state~\cite{zc0_4025}. In this
analysis, two $D$ mesons are reconstructed together with the
bachelor $\piz$ with data at $\sqrt{s}=4.23$ and $4.26$~GeV. the
$\zcp^0$ is observed in the $\piz$ recoil mass spectrum (see
Fig.~\ref{fig:fit}~(right)). The mass spectrum is fitted with the
incoherent sum of a BW function with a mass-dependent width and MC
simulated background shape. The mass and width of its pole
position are determined to be $(4025.5^{+2.0}_{-4.7}\pm
3.1)$~MeV/$c^2$ and $(23.0\pm6.0 \pm 1.0)$~MeV, respectively.

From the simultaneous fit, $69.5\pm 9.2$ and $46.1\pm 8.5$
$\zcp^0$ signal events are obtained at 4.23 and 4.26~GeV,
respectively, with a statistical significance of $5.9\sigma$. The
Born cross section $\sigma(\EE\to \zcp^0 \piz \to(D^{*0}
\bar{D}^{*0} + D^{*+}{D}^{*-})\piz)$ is measured to be $(61.6\pm
8.2\pm 9.0)$~pb at 4.23~GeV and $(43.4\pm8.0 \pm 5.4)$~pb at
4.26~GeV. The ratio $\frac{\sigma(\EE \to \zcp^0 \piz \to(D^{*}
\bar{D}^{*})^0\piz)}{\sigma(\EE\to \zcp^+ \pi^-\to(D^{*}
\bar{D}^{*})^+\pi^-)}$ is compatible with unity at
$\sqrt{s}=4.26$~GeV, which is expected from isospin symmetry. This
also confirms the isospin of the $\zcp$ is one.

\subsubsection{Search for $\zcp\to \pi\jpsi$ }

As has been described in Sec.~\ref{Sec:jpc}, using the two data
sets at $\sqrt{s}=$4.23 and 4.26~GeV, BESIII also searches for the
process $\EE\to \zcp^+\pi^-+c.c.\to \pp\jpsi$ in the PWA, with the
$\zcp^\pm$ assumed to be a $1^+$ state added in the global fit. In
the PWA, its mass is taken from Ref.~\refcite{zc4020}, and its
width is taken as the observed value, which includes the detector
resolution. The statistical significance for $\zcp^\pm \to
\jpsi\pi^\pm$ is found to be 3$\sigma$ in the combined data. The
Born cross sections are measured to be $(0.2\pm 0.1)$~pb at
$\sqrt{s}=4.23$~GeV and $(0.8\pm 0.4)$~pb at $\sqrt{s}=4.26$~GeV,
and the corresponding upper limits at the 90\% C.L. are estimated
to be $0.9$ and $1.4$~pb, respectively.

\subsubsection{Search for $\zcp\to \rho\etac$ }

As has been shown in Fig.~\ref{fit_Zc_data_4230}, no statistically
significant $\zcp^\pm$ signal is observed in the $\rho\etac$ decay
mode at $\sqrt{s}=4.23$~GeV~\cite{ycz_charm2018}. The fit to the
invariant mass distribution yields $21\pm 13$ signal $\zcp^\pm$
events with a statistical significance of only $1.0\sigma$. The
upper limit of the production cross section at the 90\% C.L. is
determined to be 16~pb. The $\zcp^\pm$ signal at $\sqrt{s}=$4.26
and 4.36~GeV are not statistically significant either, and the
upper limits of the production cross section are determined to be
9 and 11~pb, respectively, at the 90\% C.L.

By comparing with the cross sections of $\EE\to \pi^\mp\zcp^\pm\to
\pphc$, one measures the upper limit on the ratio of the $\zcp$
decay branching fractions, $R_{\zcp}=\BR(\zcp^\pm\to \rho^{\pm}
\etac)/\BR(\zcp^\pm\to \pi^{\pm} \hc)<$1.9, 1.2, and 1.0 at c.m.
energies of 4.23, 4.26, and 4.36~GeV, respectively, at the 90\%
C.L. The ratio is smaller than the calculations based on
tetraquark model while not in contradiction with the molecule
model calculation which is about two orders of magnitude smaller
than the current upper limit~\cite{Esposito:2014hsa}.

\subsubsection{Summary on the $\zcp$}

From the above studies, we conclude that the $\zcp$ is an
isovector state with positive $G$-parity, similar to the $\zc$
state. The spin-parity quantum numbers of the $\zcp$ are not
measured, but $J^P=1^+$ are assumed in all the analyses. $\zcp$
decays into $\pi\hc$ and $D^*\dstrbar$. It may also decay into
$\pi\jpsi$ final state but the decay into $\rho\etac$ is not
observed. The neutral $\zcp$ has a negative $C$-parity.

Although the mass and width are reported in both $\pphc$ and
$D^*\dstrbar$ final states, the comparison of the numbers is not
straightforward. The reported values depend on different
assumptions on the signal shape in different analysis and the
interference between $\zcp$ and other amplitudes is neglected. In
the $\pi D^*\dstrbar$ analyses~\cite{zc4025,zc0_4025}, the
$\zcp^\pm$ and $\zcp^0$ are parametrized with different line
shapes and the pole mass and width are reported in the latter
case. In addition, the fractions of non-$\zcp$ events in $\EE\to
\pi D^*\dstrbar$ are quite different in charged and neutral modes.
All these suggest that improved measurements of $\EE\to \pi
D^*\dstrbar$, both charged and neutral modes, using more $D$-tag
modes and data at other c.m. energies are necessary.

The most reliable measurement is probably from $\EE\to \pphc$
mode~\cite{zc4020}. Although the resonance is parametrized with a
constant-width BW function and the interference with non-$\zcp$
amplitudes is also neglected, the fact that the width is very
narrow makes the line shape distortion due to these above effects
not very significant. The mass and width are measured to be
$(4022.9\pm 0.8\pm 2.7)~{\rm MeV}/c^2$ and $(7.9\pm 2.7\pm
2.6)$~MeV, respectively, in this mode. Of course, a coupled
channel analysis of $\pphc$ and $\pi D^*\dstrbar$ modes will give
more reliable measurements of the resonant parameters.

The production cross section for $\EE\to \pi\zcp$ suffers from the
same problems mentioned in the $\zcp$ mass and width
determination. It would be very important to measure the $\zcp$
production cross sections as a function of the c.m. energy with
PWA, to check if $\EE\to \pi\zcp$ is from continuum production or
from decays of some resonant structures, such as the $Y(4220)$ and
$Y(4390)$ observed in $\EE\to \pphc$~\cite{bes3_pipihc_lineshape},
this will be an important piece of information in understanding
the nature of the $\zcp$ and the production mechanism.

\subsection{The $Z_c$ structures in $\pi\psp$ system}

\subsubsection{The $Z_c(4430)$}

The Belle collaboration first reported evidence for a narrow
$Z_c(4430)^-$ peak, with mass $M=(4433\pm 4\pm 2)$~MeV/$c^2$ and
width $\Gamma=(45^{+18+30}_{-13-13})$~MeV, in the $\pi^-\psp$
invariant mass distribution in $B\to K \pi^-\psp$
decays~\cite{Belle_zc4430}. The BaBar collaboration did the same
analyses~\cite{Aubert:2008aa}, but did not confirm its existence.
On the other hand, the BaBar's results did not contradict the
Belle observation due to low statistics. This has been an open
question for a very long time since there were no new data
available until recently.

To take into account the interference effect between the
$Z_c(4430)^-$ and the $K^*$ intermediate states in $B\to K
\pi^-\psp$ decays, the Belle collaboration updated their
$Z_c(4430)^-$ results with a four-dimensional (4D) amplitude
analysis~\cite{Belle_zc4430pwa}. The $Z_c(4430)^-$ is observed
with a significance of $5.2\sigma$, a much larger mass of
$(4485\pm 22^{+28}_{-11})$~MeV/$c^2$, and a large width of
$(200^{+41+26}_{-46-35})$~MeV. The product branching fractions are
measured to be \( \BR(B^0\to Z_c(4430)^-K^+)\times
\BR(Z_c(4430)^-\to \pi^-\psp) = (6.0^{+1.7+2.5}_{-2.0-1.4})\times
10^{-5}\), and spin-parity $J^P=1^+$ is favored over the other
assignments by more than $3.4\sigma$. This was confirmed recently
by the LHCb experiment~\cite{LHCb_zc4430}.

LHCb reported a 4D model-dependent amplitude fit to a sample of
$25,176\pm 174$ $B^0\to K^+\pi^-\psp$, $\psp\to \MM$ events
reconstructed in a 3~fb$^{-1}$ data sample collected at
$\sqrt{s}=7$ and $8$~TeV. The order-of-magnitude increase in
signal yield over the Belle measurement~\cite{Belle_zc4430pwa}
improves sensitivity to the quantum numbers of the $Z_c(4430)^-$
and allows a measurement of the Argand plot.

In the amplitude fit, the $Z_c^-$ amplitude is represented by a BW
function, the measured mass of $(4475\pm
7\,{_{-25}^{+15}})$~MeV/$c^2$ and width of $(172\pm
13\,{_{-34}^{+37}})$~MeV are consistent with, but more precise
than, the Belle results~\cite{Belle_zc4430pwa}. Relative to
$J^P=1^+$, LHCb data rule out the $0^-$, $1^-$, $2^+$ and $2^-$
hypotheses by at least $9\sigma$, and establishes the spin-parity
of the $Z_c(4430)$ to be $1^+$.

\begin{figure}[hbt]
\centering
    \includegraphics*[width=0.55\textwidth]{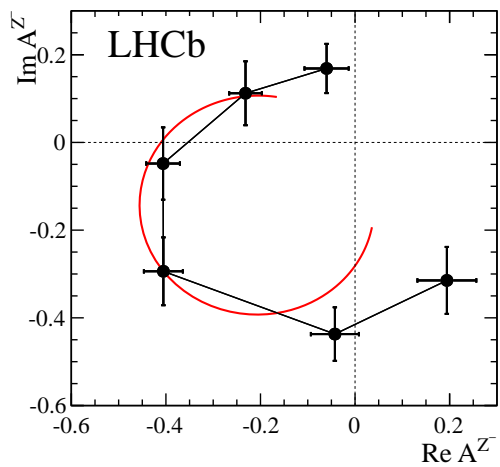}
\caption{The Argand plot of the $Z_c(4430)$ amplitudes (points
with error bars, $M_{\pi\psp}^2$ increases counterclockwise). The
curve is the prediction from a BW function with mass and width of
4475~MeV/$c^2$ and 172~MeV, respectively. \label{fig:argand}}
\end{figure}

In addition, LHCb measures Argand plot of the $Z_c^-$ amplitude as
a function of $M_{\pi\psp}$ (shown in Fig.~\ref{fig:argand}),
which is consistent with a rapid change of the $Z_c(4430)^-$ phase
when its magnitude reaches the maximum, an expected behavior of a
resonance. This is the first time an Argand plot is measured for
an exotic charmonium-like state.

\subsubsection{The $Z_c$ structure in $\EE\to \pp\psp$}

In a study of $\EE\to \pp\psp$, charged structure at
4.03~GeV/$c^2$ is observed at Belle and BESIII experiments.

\begin{figure}[htbp]
\centering
 \psfig{file=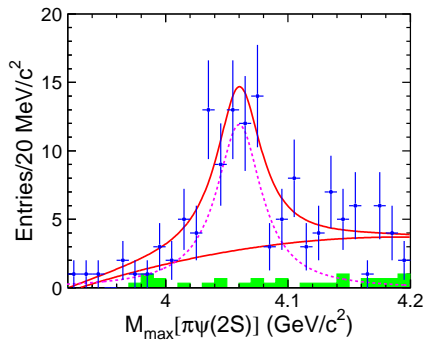,angle=-90,width=5.5cm}
\caption{The distribution of $M_{\rm max}(\pi^{\pm}\psp)$ from
$Y(4360)$ decays in $\EE\to \pp\psp$ at Belle. The points with
error bars represent the data; the histogram is from the sidebands
and normalized to the signal region; the solid curve is the best
fit and the dashed curve is the signal parametrized with a BW
function. } \label{mppsp-fit}
\end{figure}

Using the 980~fb$^{-1}$ full data sample, Belle updated the
measurement of $\EE\to \pp\psp$ using ISR technique with two
$\psip$ decay modes~\cite{belle_y4660_new}, namely, $\pp\jpsi$ and
$\MM$. Possible charged charmonium-like structures in
$\pi^{\pm}\psp$ final states from the $Y(4360)$ or $Y(4660)$
decays are searched for with the selected candidate events.
Figure~\ref{mppsp-fit} shows the sum of $M_{\pim\psip}$ and
$M_{\pip\psip}$ distributions in $Y(4360)$ decays ($4.0 <
M_{\pp\psp} < 4.5$~GeV/$c^2$) from both the $\ppjpsi$ and the
$\MM$ modes. An unbinned maximum-likelihood fit is performed on
the distribution of $M_{\rm max}(\pi^{\pm}\psp)$, the maximum of
$M(\pi^+\psp)$ and $M(\pi^-\psp)$, simultaneously with both modes.
The excess is parameterized with a BW function and the
non-resonant non-interfering background with a second-order
polynomial function. The fit yields a mass of $(4054\pm 3\pm
1)$~MeV/$c^2$ and a width of $(45\pm 11\pm 6)$~MeV. The
statistical significance of the signal is $3.5\sigma$.

\begin{figure*}[htbp]
\centering
\includegraphics*[width=3.8cm,angle=0]{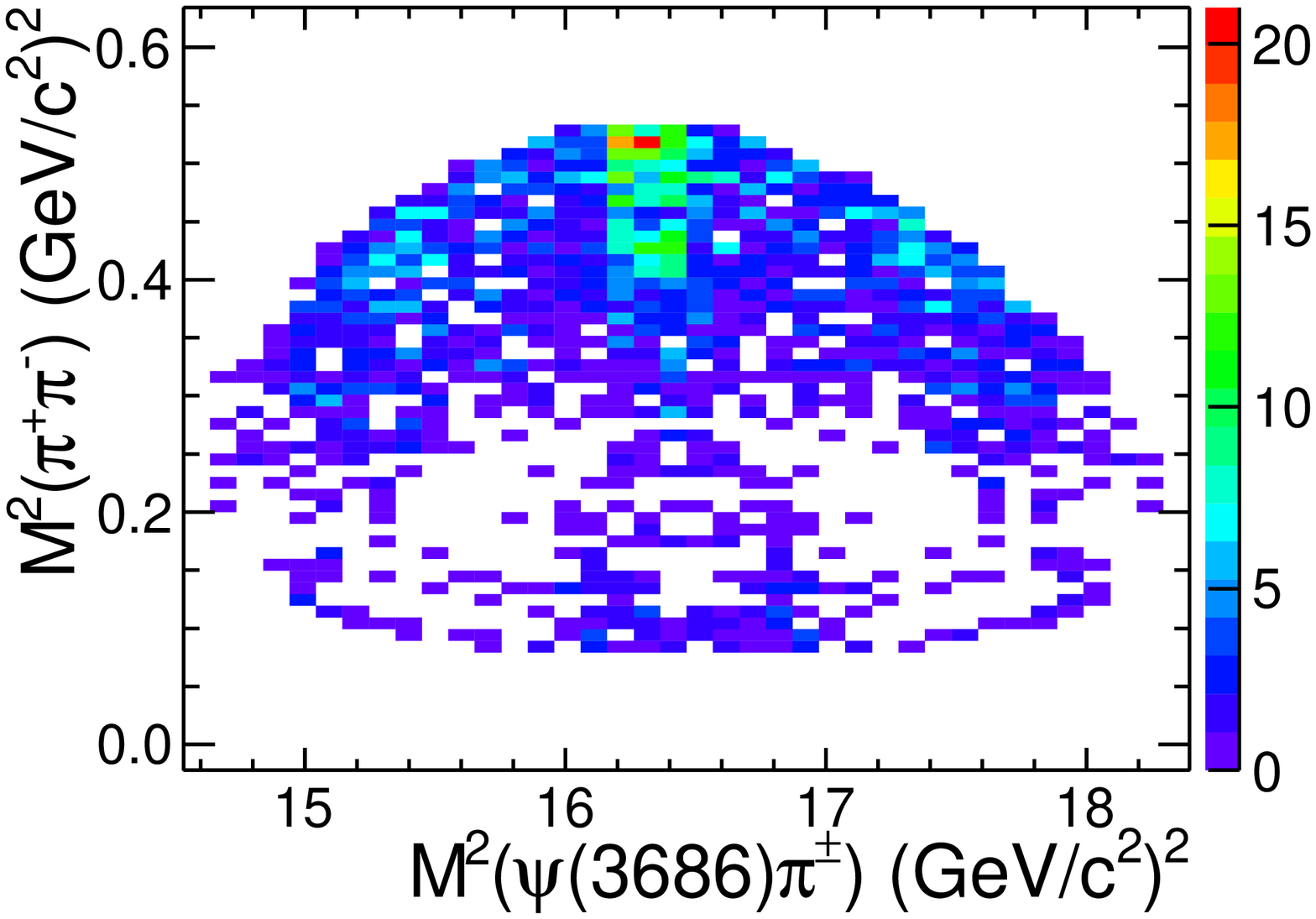}
\includegraphics*[width=3.8cm,angle=0]{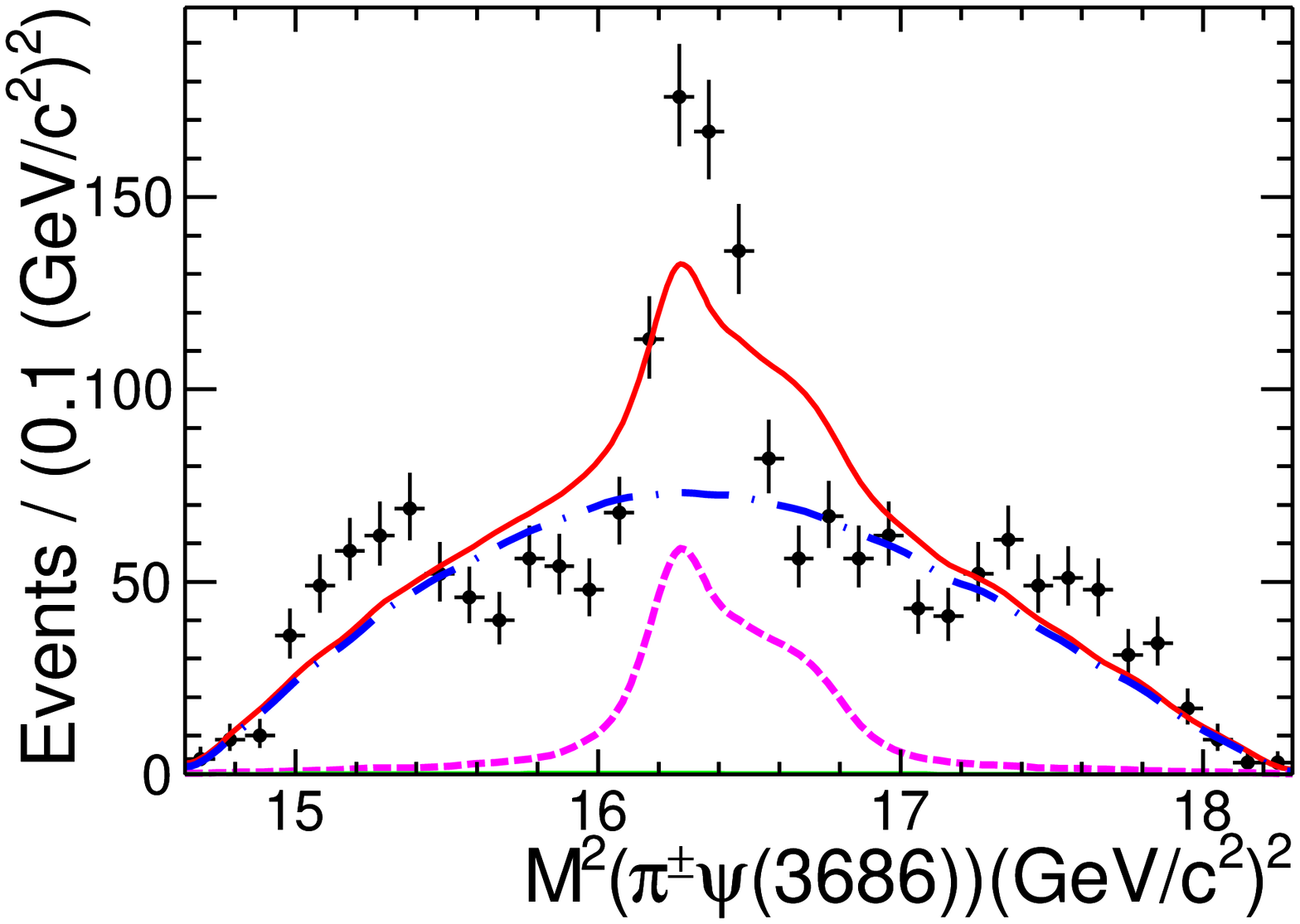}
\includegraphics*[width=3.8cm,angle=0]{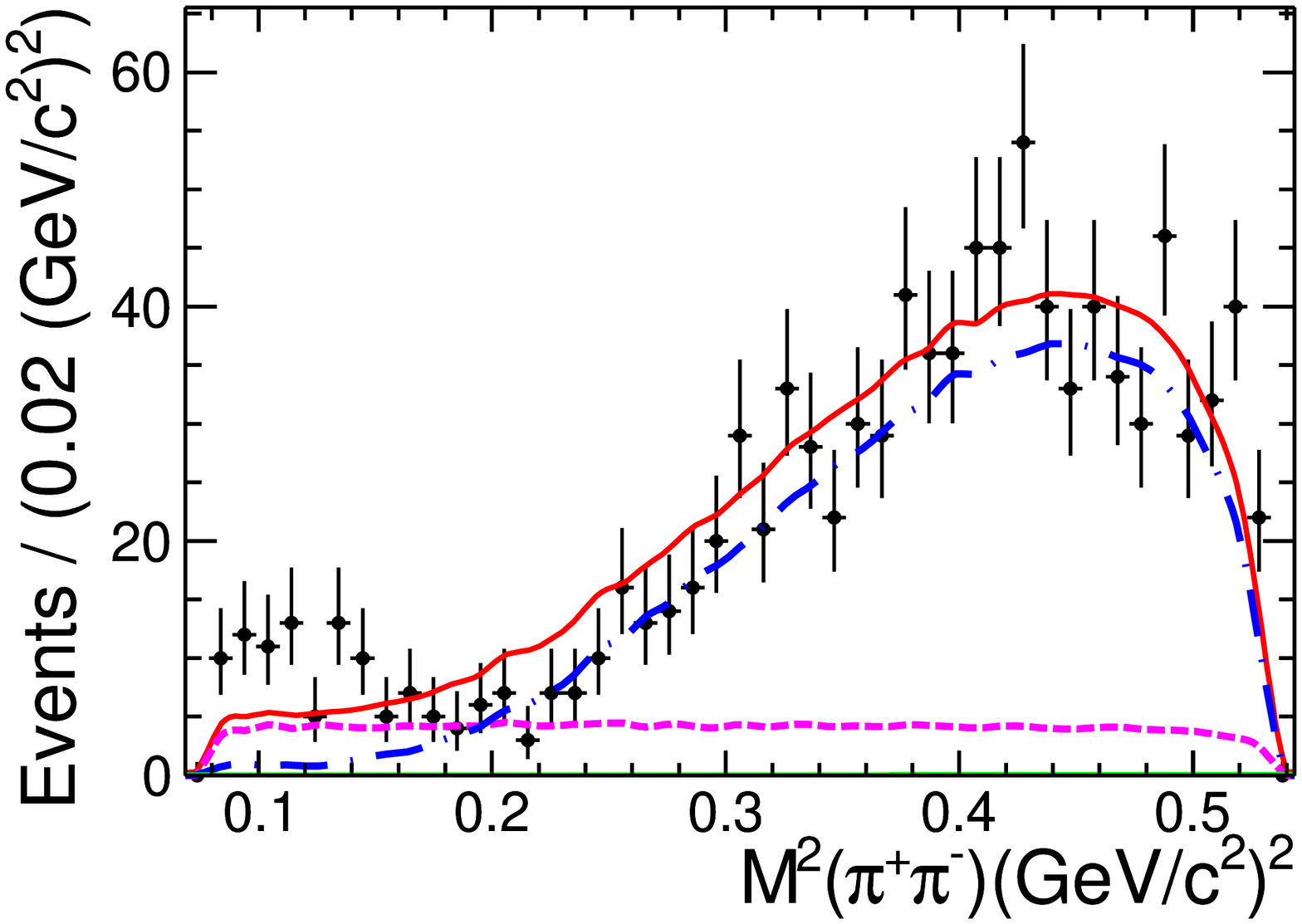}
\caption{Dalitz plot of $\EE\to \pp\psp$ at $\sqrt{s}= 4.416$~GeV
(two entries per event) from BESIII. Dots with errors are data.
The solid curves (red) are projections from the fit; the dashed
curves (pink) show the shape of the intermediate state; the
dash-dotted curves (blue) show the shape from the direct process
$\EE\to \pp\psip$ obtained from MC model. \label{intermediate}}
\end{figure*}

BESIII studies the process $\EE\to \pp\psip$ using 5.1~fb$^{-1}$
of data at c.m. energies from 4.0 to 4.6~GeV~\cite{bes3_pppsp}.
Intermediate states are investigated in the data samples that have
large integrated luminosity. The Dalitz plot and the corresponding
one-dimensional projections are shown in Fig.~\ref{intermediate}
for data at $\sqrt{s} = 4.416$~GeV, a prominent narrow structure
is observed around $4.03$~GeV/$c^2$ in the $M(\pi^\pm\psip)$
spectrum.

An unbinned maximum likelihood fit to the Dalitz plot is applied
at BESIII. Assuming an intermediate $Z_c$ state in $\pi\psp$
system with spin-parity $J^P=1^{+}$, the Dalitz plot is
parameterized by the incoherent sum of the process $\pi Z_c\to
\pp\psp$ and the direct process $\EE\to \pp\psp$. The fit yields a
mass of $M = (4032.1\pm 2.4)$~MeV/$c^2$ and a width of $\Gamma$ =
$(26.1\pm 5.3)$~MeV for the intermediate state with a significance
of $9.2\sigma$. The fit projections on $M^{2}(\pi^{\pm}\psip)$ and
$M^{2}(\pp)$ are shown in Fig.~\ref{intermediate}. It is observed
that the fit curve does not match data well, thus BESIII reported
statistical errors only for the parameters of the $Z_c$ structure.

The BESIII data are refitted with a similar model as is used in
the description of the $Z_b$ states in $\Upsilon(5S)$
decays~\cite{bondar_charm2018}. This fit considers the
interference effect between the $\pp$ amplitude and the $Z_c$
amplitude properly. It is found that the structure can be
described well with a charged state with a mass of $(4019.0\pm
1.9)$~MeV/$c^2$ and width of $(29\pm 4)$~MeV, or the same state
observed in $\pphc$ final state, $\zcp$~\cite{zc4020}, and the fit
quality is much improved. The fraction of $\pi \zcp\to \pp\psp$ is
$(12.0\pm 3.7)\%$ of the total $\EE\to \pp\psp$ cross section,
that is, $\sigma(\EE\to \pi^+\zcp^-+c.c.\to \pp\psp)=(5.1\pm
1.6)$~pb. Unfortunately there is no corresponding cross section of
$\EE\to \pi^+\zcp^-+c.c.\to \pphc$ at the same energy to compare
with. If we take the cross section of $\EE\to \pi^+\zcp^-+c.c.\to
\pphc$ at 4.36~GeV for comparison, we obtain $\frac{\BR(\zcp\to
\pi\psp)}{\BR(\zcp\to \pi\hc)}\sim 0.5\pm 0.3$. It is obvious both
$\pp\psp$ and $\pphc$ final states need to be further investigated
to understand the intermediate structures.

\subsection{Summary on the $Z_c$ states}

Although many measurements have been performed on the $\zc$,
$\zcp$, and $Z_c(4430)$, the experimental information is still not
precise enough. From the studies of the $Z_c(4430)$ we know that
the resonant parameters may change significantly by doing simple
1D fit to the invariant mass distribution~\cite{Belle_zc4430} or
by doing full amplitude analysis with the interference effects
between different amplitudes considered
properly~\cite{Belle_zc4430pwa}. The same thing may happen to the
$\zc$ and $\zcp$ cases too. Amplitude analysis of the relevant
final states to extract the resonant parameters as well as the
couplings to different modes is essential to extract correct
information for understanding the nature of these states. In
addition, the PWA also allows a measurement of the Argand plot of
the $Z_c$ amplitudes, which may be used to discriminate models of
the $Z_c$ states.

Comparison of the $Z_c$ with the $Z_b$ states~\cite{belle_zb} may
help understand these structures too. The decays of $Z_b(10610)$
and $Z_b(10650)$ to $\pi\Upsilon(nS)$ ($n=1$, $2$, $3$), $\pi
h_b(mP)$ ($m=1$, $2$), $B\bar{B}^*+c.c.$, and $B^*\bar{B}^*$ have
been measured while only significant $\zc\to \pi\jpsi$,
$D\bar{D}^*+c.c.$ and $\zcp\to \pi\hc$, $D^*\bar{D}^*$ are
observed; on the other hand, evidence for $\zc\to \rho\etac$ is
observed while no measurement of $Z_b\to \rho\eta_b$ is reported
yet. Many of these final states can be reached at BESIII and Belle
II~\cite{belle2} in the near future. Needless to say, there are
other decay modes to search for, such as final state with D-wave
quarkonium state like $\pi\psi(1D)$ and $\pi\Upsilon(1D)$.

The production of the $Z_c$ states as a function of c.m. energy
should also be measured in a similar way as has been presented for
$\zc$ and $\zcp$ at a few c.m. energies, this may reveal whether
these states are from resonance decays or continuum production. So
far $\zc$ and $\zcp$ are only observed in $\EE$
annihilation~\cite{D0-zc3900} while $Z_c(4430)$ only in $B$
decays. Search for these states in different production modes is
of great importance.

\section{The $X(3872)$}
\label{sec:x3872}

The $\xx$ was observed in $B^\pm\to K^\pm \ppjpsi$ decays 15 years
ago at the Belle experiment~\cite{bellex}. It was confirmed
subsequently by several other experiments~\cite{CDFx,D0x,babarx}.
Since its discovery, the $\xx$ state has stimulated special
interest for its nature. Both BaBar, Belle, and LHCb reported
evidence for $\xx\to \gamma\jpsi$ decay, which supports $\xx$
being a $C$-even state~\cite{babar-jpc,belle-jpc,LHCbgammapsp}.
The CDF and LHCb experiments determine the spin-parity of the
$\xx$ to be $J^P=1^+$, and the $\pp$ system is dominated by a
$\rho^0(770)$ resonance in $\xx\to
\ppjpsi$~\cite{CDF-pp,CDF-jpc,x3872_JPC_LHCb_2013,x3872_JPC_LHCb_2015}.

The $\xx$ is only observed in $B$ meson decays and hadron
collisions before. Since the quantum numbers of $\xx$ are
$J^{PC}=1^{++}$, it can be produced through the radiative
transition of excited vector charmonium or charmonium-like states
such as the $\psi$s and the $Y$s and the evidence for the $\y\to
\gamma\xx$ is reported by the BESIII experiment~\cite{BES3x}.

\subsection{The spin-parity of the $X(3872)$}

The spin-parity quantum numbers $J^P$ of the $\xx$ are restricted
to two possibilities, $1^{+}$ or $2^{-}$, by the CDF experiment,
via an analysis of the angular correlations in $\xx\to
\pi^+\pi^-\jpsi$ and $\jpsi\to \MM$~\cite{CDF-jpc}. Using
1.0~fb$^{-1}$ of $pp$ collision data, LHCb rules out $J^{P}=2^{-}$
by analyzing the angular correlations in the same decay chain,
with the $\xx$ state produced in $B^+\to \xx K^+$
decays~\cite{x3872_JPC_LHCb_2013}. However, these above analyses
assumed that the lowest orbital angular momentum between the $\xx$
decay products dominates the matrix element.

\begin{figure}[htbp]
\centering
\includegraphics*[height=6cm]{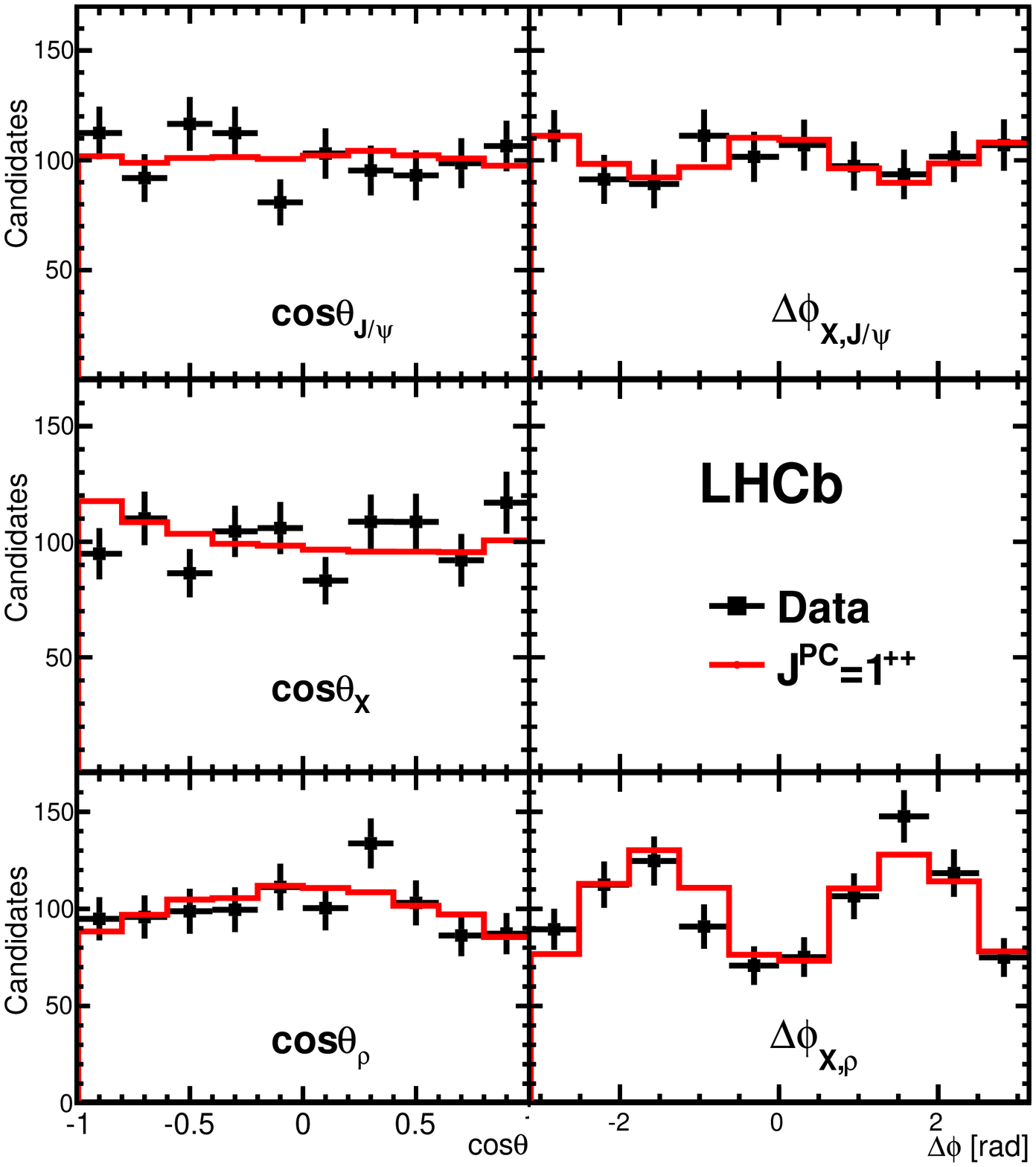}
\includegraphics*[height=6cm]{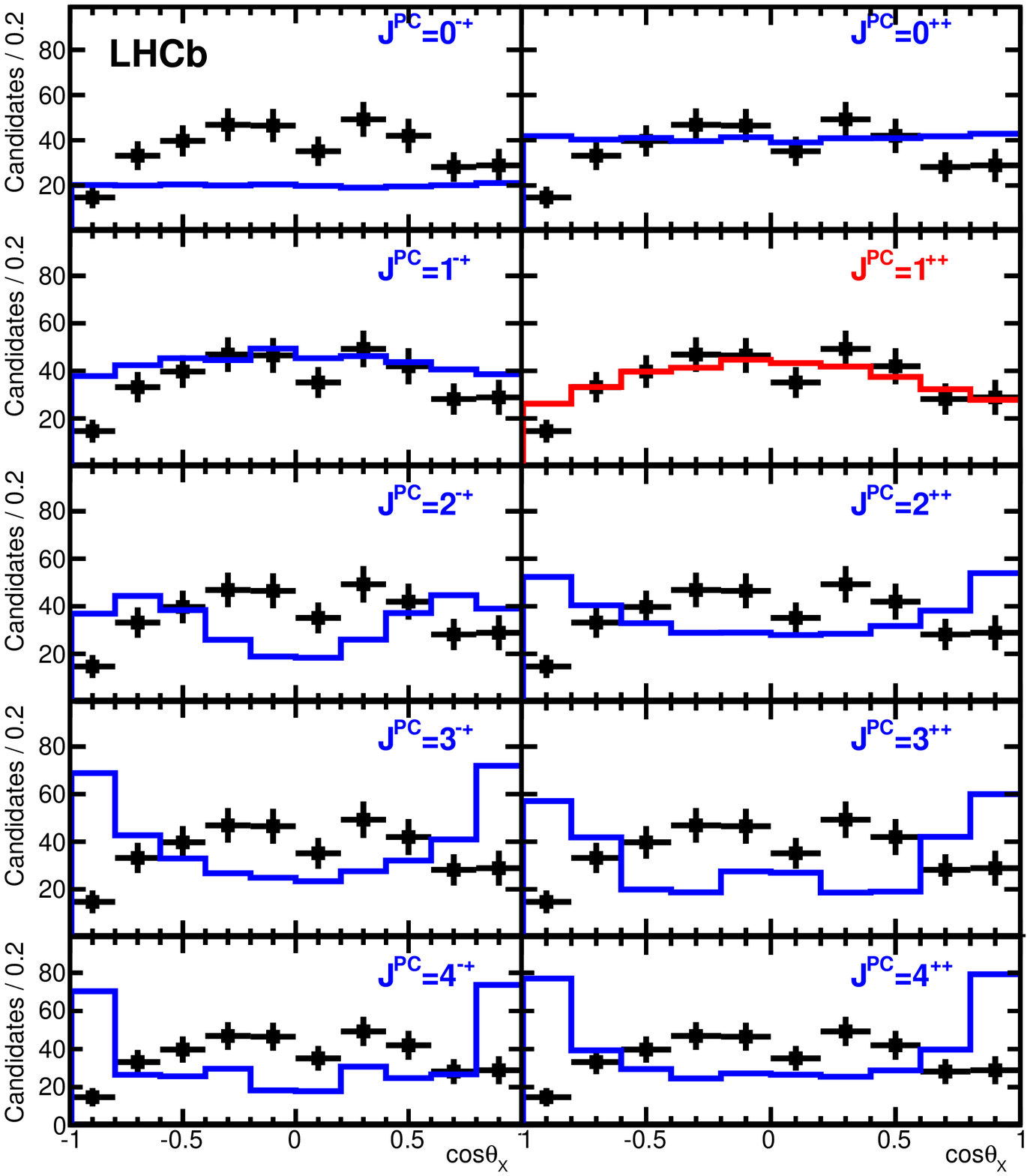}
\caption{Background-subtracted distributions of all angles for the
data (left panel) and background-subtracted distribution of
$\cos\theta_X$ for candidates with $|\cos\theta_{\rho}|>0.6$ for
the data compared with the expected distributions for various
$\xx$ $J^{PC}$ assignments (right panel). Points with error bars
are LHCb data, and the solid histograms are the fit projections
(left panel) and expected distributions (right panel).
\label{fig:angles1P}}
\end{figure}

This assumption is removed later by LHCb in a five-dimensional
(5D) angular correlation analysis in $B^+\to \xx K^+$, $\xx\to
\rho^0\jpsi$, $\rho^0\to \pi^+\pi^-$, $\jpsi\to \MM$ decays with
$1011\pm 38$ events selected from 3~fb$^{-1}$ data at $7$ and
$8$~TeV~\cite{x3872_JPC_LHCb_2015}. This analysis confirms that
the spin, parity, and charge-conjugation of the $X(3872)$ state
are $1^{++}$, the same quantum numbers as one of the P-wave
spin-triplet charmonium state $\chi_{c1}(2P)$. The comparison
between data and the fit projections for $J^{PC}=1^{++}$ is shown
in Fig.~\ref{fig:angles1P}~(left), and the comparison between data
and expectation from other spin-parity is shown in
Fig.~\ref{fig:angles1P}~(right). We can see that $J^{PC}=1^{++}$
describe data very well and other $J^{PC}$ have very different
angular distributions.

\subsection{The puzzling radiative transitions $\xx\to \gamma \jpsi$
and $\gamma \psp$}

Radiative decays of the $\xx$ to lower mass charmonium states
provide crucial information to understand its nature, especially
to check if it is a conventional charmonium or an exotic state.

The transition $\xx\to \gamma \jpsi$ was measured by
BaBar~\cite{babar_gammapsp} ($28\pm 7$ signal events with a
statistical significance of $3.6\sigma$), Belle~\cite{belle-jpc}
($36^{+9}_{-8}$ signal events with a statistical significance of
$5.5\sigma$), and LHCb experiments~\cite{LHCbgammapsp} ($591\pm
48$ signal events). So $\xx\to \gamma \jpsi$ is well established.

For $\xx\to \gamma \psip$, evidence was first reported by
BaBar~\cite{babar_gammapsp} with $33\pm 8$ observed signal events
and a statistical significance of $3.5\sigma$. The ratio of the
branching fractions is measured to be
\[
R = \frac{\BR(\xx\to \gamma\psp)}{\BR(\xx\to \gamma\jpsi)} = 3.4
\pm 1.4.
\]

In a study at the LHCb experiment based on a 3~fb$^{-1}$ data
sample at $\sqrt{s}=7$ and $8$~TeV, strong evidence for the decay
$\xx\to\gamma \psp$ was reported together with a measurement of
$R$~\cite{LHCbgammapsp}. In the data sample, $36.4\pm 9.0$ $B^\pm
\to \xx K^\pm$, $\xx\to \gamma \psp$ events are observed with a
statistical significance of $4.4\sigma$. LHCb measures
\[
R = \frac{\BR(\xx\to \gamma\psip)}{\BR(\xx\to \gamma\jpsi)} =
2.46\pm 0.64\pm 0.29.
\]
This result is in good agreement with BaBar's measurement.

In contrast, no significant signal was observed at Belle and an
upper limit of $R < 2.1$ was reported at the 90\%
C.L.~\cite{belle-jpc} (using the information from
Ref.~\refcite{belle-jpc}, we obtain \( R = \frac{\BR(\xx\to
\gamma\psp)}{\BR(\xx\to \gamma\jpsi)} = 0.6 \pm 1.4 \) as a good
estimation of the central value and uncertainty). Although not in
disagreement, BaBar, LHCb, and Belle results do show some tension
on the decay rate of $\xx\to \gamma\psp$.

As the measurements of all the above three experiments agree with
each other, we can make a weighted average to give the best
estimation of $R$. Neglecting the small correlated errors in the
measurements, we obtain
\[
\overline{R} = \frac{\BR(\xx\to \gamma\psp)}{\BR(\xx\to
\gamma\jpsi)} = 2.31\pm 0.57.
\]
This value does not support a pure $\bar{D^0}D^{*0}+c.c.$ molecule
interpretation of the $\xx$ state, but agrees with expectations if
the $\xx$ is a pure charmonium or a mixture of a molecule and a
charmonium~\cite{reviews,LHCbgammapsp}. Of course many of the
calculations have model-dependent parameters, adjustment of the
parameters may still reproduce the experimental data.

As there is still no experiment reports $\xx\to \gamma\psp$ signal
with statistical significance larger than $5\sigma$, the existence
of this transition still needs further experimental investigation.

\subsection{Decay branching fractions of the $\xx$}

So far, all the $\xx$ related measurements are product branching
fractions or relative branching ratios since the absolute
production rate of the $\xx$ is unknown in any of the
measurements. The only attempt to measure the production of the
$\xx$ is via inclusive $B$ decays into an $\xx$ and a kaon at
BaBar and Belle experiments.

BaBar set an upper limit of the $X(3872)$ production rate in the
$B$-meson decays by measuring the momentum distribution of the
inclusive kaon from $B$-meson decays with 210~fb$^{-1}$
$\Upsilon(4S)$ data~\cite{babar_Xcc}: \( \BR(B^-\to K^-
X(3872))<3.2\times 10^{-4} \) at the 90\% C.L.

A recent update is from the Belle experiment with the full sample
of $772\times 10^6$ $B\bar{B}$ pairs (711~fb$^{-1}$ $\Upsilon(4S)$
data)~\cite{belle_Xcc}. No significant $\xx$ signal is observed,
and Belle sets a more stringent upper limit
\[
\BR(B^-\to K^- X(3872))<2.7\times 10^{-4}
\]
at the 90\% C.L., and the central value is \( \BR(B^-\to K^-
X(3872)) = (1.2\pm 1.1\pm 0.1)\times 10^{-4} \).

Together with all the other measurements on the product branching
fractions $\BR(B^-\to K^- X(3872))\cdot \BR(X(3872)\to {\rm
exclusive})$ (${\rm exclusive}=\pp\jpsi$, $\pp\piz\jpsi$,
$\gamma\jpsi$, $\gamma\psp$, $D^0\bar{D}^{*0}+c.c.$)~\cite{pdg},
one obtains
\[
2.9\%<\BR(X(3872)\to \pi^+\pi^-J/\psi)<10\%,
\]
\[
0.9\times 10^{-4}<\BR(B^-\to K^- X(3872))<2.7\times 10^{-4},
\]
at the 90\% C.L.~\cite{bnote} We find that the decay width of the
$X(3872)$ to $\pi^+\pi^-J/\psi$ is larger and the production rate
of the $X(3872)$ is smaller than conventional charmonium states
such as $\eta_c$, $\psp$, and $\chi_{c1}$~\cite{pdg}. The
branching fraction of other $\xx$ decay modes can be estimated in
a similar way, while it decays dominantly into open charm final
state $D^0\bar{D}^{*0}+c.c.$, its decays into each final state
with a charmonium is at a few per cent level.

\subsection{Observation of $\EE\to \gamma X(3872)$}
\label{subsec:x3872}

BESIII observed $\EE\to \gamma\xx\to \gamma \ppjpsi$, with data
collected at 4.23, 4.26, and 4.36~GeV~\cite{BES3x}.

\begin{figure}
\begin{center}
\includegraphics[height=4cm]{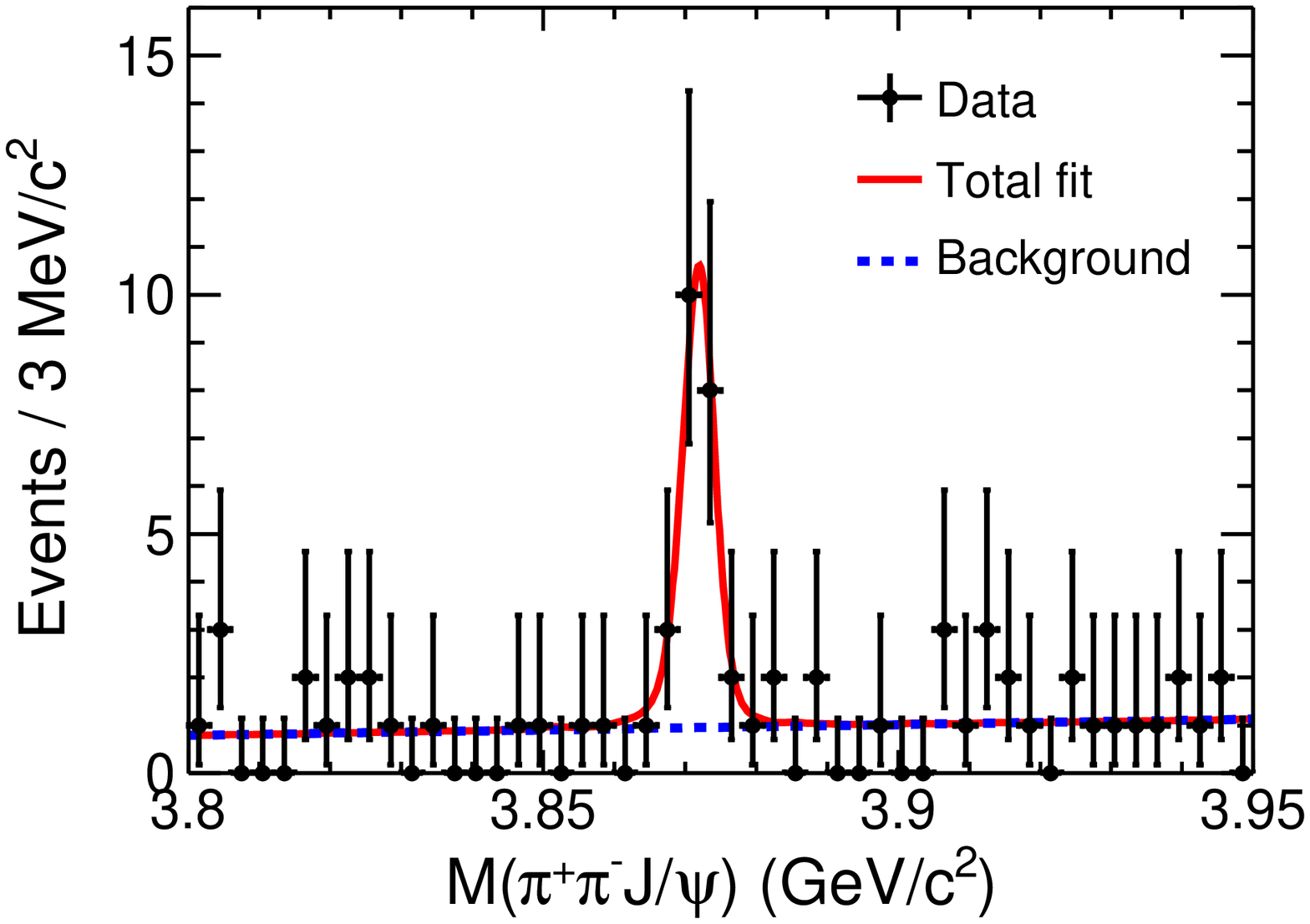}
\includegraphics[height=4cm]{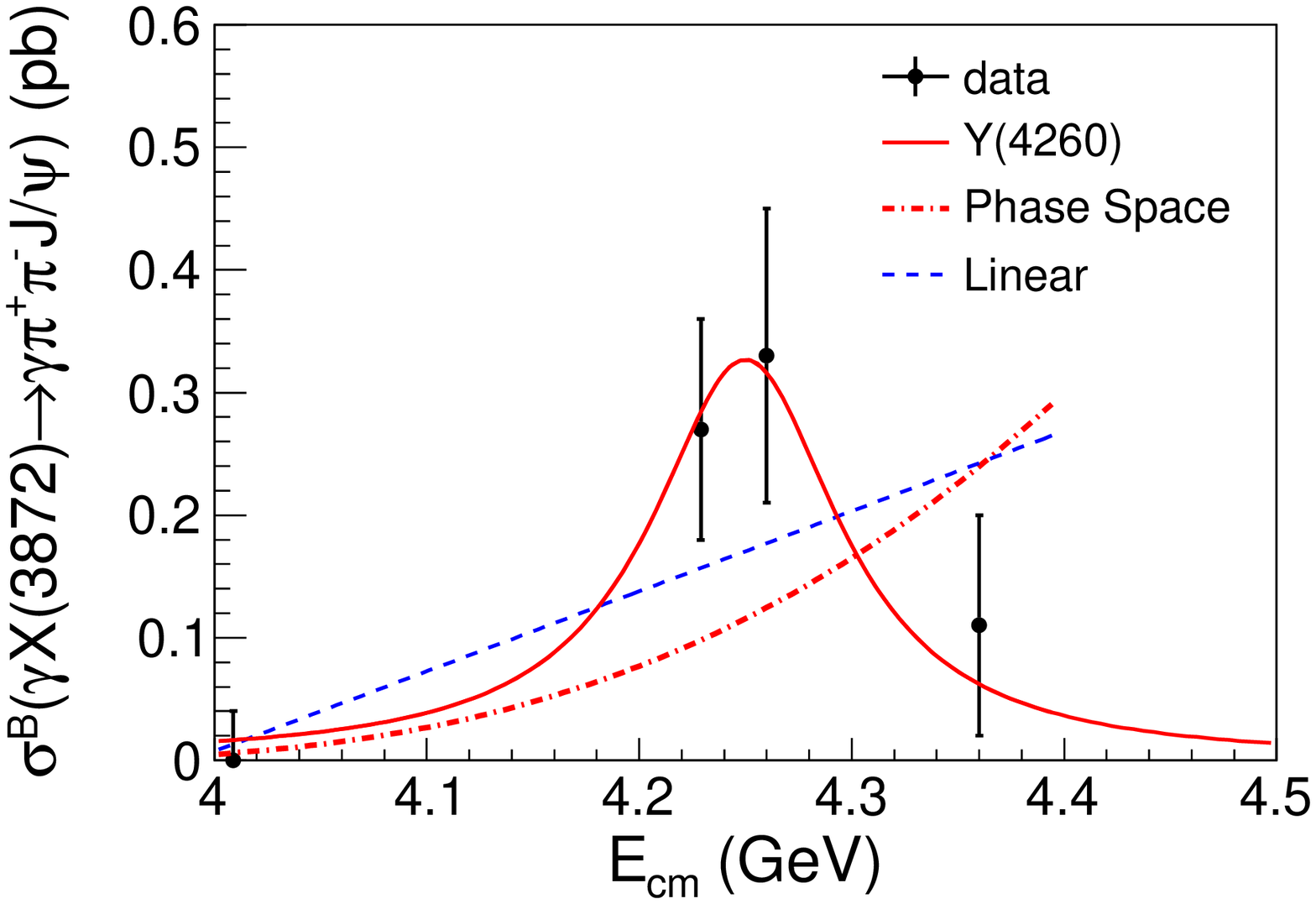}
\caption{Left panel: fit the $M(\ppjpsi)$ distribution observed at
BESIII. Dots with error bars are data, the curves are the best
fit. Right panel: fit to $\sigma^B[\EE\to \gamma\xx]\times
\mathcal{B}[\xx\to \ppjpsi]$ measured by BESIII (dots with error
bars) with a $\y$ resonance (red solid curve), a linear continuum
(blue dashed curve), or an $E1$-transition phase space term (red
dotted-dashed curve). } \label{fit-mx}
\end{center}
\end{figure}

The $M(\ppjpsi)$ distribution (summed over all energy points), as
shown in Fig.~\ref{fit-mx}~(left), is fitted to extract the mass
and signal yield of $\xx$ (BESIII experiment is insensitive to the
width of the $\xx$). The ISR $\psp$ signal is used to calibrate
the absolute mass scale and to extract the resolution difference
between data and MC simulation. Figure~\ref{fit-mx} shows the
fitted result: the measured mass of $\xx$ is $(3871.9\pm 0.7\pm
0.2)$~MeV/$c^2$ and the width is found to be less than 2.4~MeV at
the 90\% C.L. The statistical significance of $\xx$ is
$6.3\sigma$.

The Born cross section is measured, and the results are listed in
Table~\ref{sec} and plotted in Fig.~\ref{fit-mx}~(right). The
energy-dependent cross sections are compared with a $\y$ resonance
(parameters fixed to PDG~\cite{pdg} values), linear continuum, or
$E1$-transition phase space ($\propto E^3_\gamma$) term. The $\y$
resonance describes the data better than the other two options.

\begin{table}[htbp]
\tbl{The product of the Born cross section $\sigma^{B}(\EE\to
\gamma \xx)$ and $\BR(\xx\to \ppjpsi)$ at different energy points.
The upper limits are given at 90\% C.L.}
{\begin{tabular}{@{}cc@{}} \toprule
  $\sqrt{s}$~(GeV)  & $\sigma^B[\EE\to \gamma\xx]
           \cdot\mathcal{B}(\xx\to \ppjpsi)$~(pb) \\  \colrule
  4.009 &  $0.00\pm 0.04\pm 0.01$ or $<0.11$  \\
  4.229 &  $0.27\pm 0.09\pm 0.02$  \\
  4.260 &  $0.33\pm 0.12\pm 0.02$  \\
  4.360 &  $0.11\pm 0.09\pm 0.01$ or $<0.36$ \\ \botrule
\end{tabular}\label{sec}}
\end{table}

These observations strongly support the existence of the radiative
transition $\y\to \gamma\xx$. Together with the hadronic
transition to the charged charmonium-like state
$\zc$~\cite{zc3900,belley_new,seth_zc}, this suggests that there
might be some commonality in the nature of $\xx$, $\y$, and $\zc$,
and so the model developed to interpret any one of them should
also consider the other two.

Combining the measurements at c.m. energies 4.23 and 4.26~GeV with
the $\EE\to \ppjpsi$ cross sections at the same energies from
BESIII~\cite{bes3_pipijpsi_lineshape}, we obtain $\sigma^B[\EE\to
\gamma\xx]\cdot \BR[\xx\to \ppjpsi]/\sigma^B(\EE\to \ppjpsi) =
(3.7\pm 0.9)\times 10^{-3}$, under the assumption that $\xx$ and
$\ppjpsi$ are only produced from $\y$ decays. If we take
$\BR[\xx\to \ppjpsi] = 5\%$ as estimated in previous section, then
$\mathcal{R} = \frac{\BR[\y\to \gamma\xx]}{\BR(\y\to \ppjpsi)}\sim
0.1$.

\subsection{More studies at BESIII and LHCb?}

After 15 years of study, the knowledge on the $\xx$ is still very
limited, we know it is an isoscalar with $J^{PC}=1^{++}$, a very
precise mass close to the $D^0\bar{D}^{*0}$ threshold, and a very
narrow width. We only have significant observations of its decays
into $D^0\bar{D}^{*0}$, $\ppjpsi$, and $\gamma\jpsi$, while the
significances of other modes $\pp\piz\jpsi$ and $\gamma\psp$ are
still less than $5\sigma$.

Since more data have been accumulated at LHCb, some of the
measurements can be improved, such as the study of $\xx\to
\gamma\psp$ and possibly search for $\xx\to \pp\piz\jpsi$ if the
$\piz$ background can be handled properly. The BESIII accumulated
more data close to the $\y$ peak, which can be used to measure all
the final states since the background level is very low as has
been shown in the $\ppjpsi$ case~\cite{BES3x}, but the real
limiting factor is still the statistics. From the discussion in
previous section, we know the production cross section of $\EE\to
\gamma\xx$ is at a few pb level, and the sample at BESIII has only
about $10^4$ produced $\xx$ events. This only allows measurement
of final states with branching fractions at percent level.

\section{The $Y$s: vector structures in $\EE$ annihilation}

Among the charmonium-like states, there are many vectors with
quantum numbers $J^{PC} = 1^{--}$ that are usually called $Y$
states, like the $Y(4260)$~\cite{babar_y4260}, the
$Y(4360)$~\cite{babar_y4360,belle_y4660}, and the
$Y(4660)$~\cite{belle_y4660}. The $Y$ states show strong coupling
to hidden-charm final states in contrast to the vector charmonium
states in the same energy region ($\psi(4040)$, $\psi(4160)$,
$\psi(4415)$) which couples dominantly to open-charm meson
pairs~\cite{pdg}. These $Y$ states are good candidates for new
types of exotic particles and stimulated many theoretical
interpretations, including tetraquarks, molecules, hybrids, or
hadrocharmonia~\cite{reviews}.

These $Y$ states were observed at $B$ factories with limited
statistics since they are produced from ISR processes with data
collected at around 10.6~GeV in the bottomonium energy
region~\cite{babar_y4260_new,belley_new,babar_y4360_new,belle_y4660_new}
and the reconstruction efficiency is low. The high precision cross
section measurements and the study of these states in different
final states in direct $\EE$ annihilation in the charmonium energy
region at BESIII experiment~\cite{bes3} supply new insight into
their nature.

\subsection{$\EE\to \pp\jpsi$}
\label{Sec:ppjpsi}

The process $\EE\to \pi^+\pi^- J/\psi$ at c.m. energies up to
5.0~GeV was first studied by the BaBar experiment, where the
$Y(4260)$ was observed~\cite{babar_y4260}. Belle measured the
cross sections of $\EE\to \pi^+\pi^- J/\psi$ at c.m. energies
between 3.8 and 5.0~GeV and reported that $Y(4260)$ alone cannot
describe the line shape satisfactorily~\cite{belley}. Improved
measurements with both BaBar~\cite{babar_y4260_new} and
Belle~\cite{belley_new} full data samples confirmed the existence
of non-$Y(4260)$ component in $\EE\to \pi^+\pi^- J/\psi$ but the
line shape was parametrized with different models.

\begin{figure*}
\begin{center}
\includegraphics[width=0.95\textwidth]{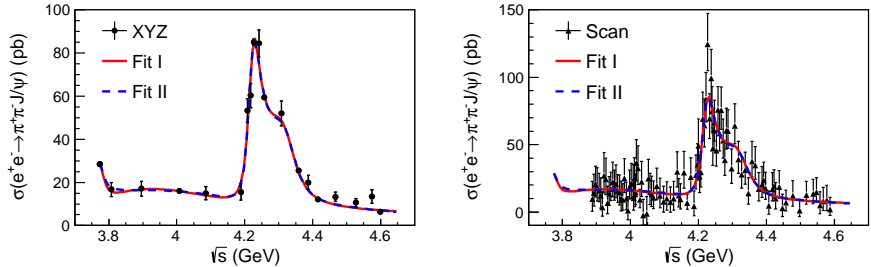}
\caption{Measured cross section $\sigma(\EE\to \ppjpsi)$ and the
simultaneous fit to the ``$\xyz$ data" (left) and ``$R$-scan data"
(right) with the coherent sum of three BW functions (Fit I, red
solid curves) and the coherent sum of an exponential continuum and
two BW functions (Fit II, blue dashed curves). Dots with error
bars are data.} \label{xsec-fit}
\end{center}
\end{figure*}

The cross sections of $\EE\to \ppjpsi$ are measured precisely at
c.m. energies from 3.77 to 4.60~GeV using 9~fb$^{-1}$ of BESIII
data~\cite{bes3_pipijpsi_lineshape}. The data sample used in this
measurement includes the ``$\xyz$ data" and the ``$R$-scan data".
Figure~\ref{xsec-fit} shows the measured cross sections, one can
see clearly the $Y(4260)$ structure observed by BaBar and Belle
experiments, but it is peaked at around 4.22~GeV rather than
4.26~GeV from the previous fits~\cite{babar_y4260_new,belley_new}.

Two resonant structures in the $Y(4260)$ peak region are needed in
a fit to the cross sections. The first one has a mass of
$(4222.0\pm 3.1\pm 1.4)$~MeV/$c^2$ and a width of $(44.1\pm 4.3\pm
2.0)$~MeV, while the second one has a mass of $(4320.0\pm 10.4 \pm
7.0)$~MeV/$c^2$ and a width of $(101.4^{+25.3}_{-19.7}\pm
10.2)$~MeV. The mass of first resonance is lower than that of the
$Y(4260)$ and it is much narrower. The second resonance is
observed in $\EE\to \ppjpsi$ for the first time, with a
statistical significance larger than $7.6\sigma$, it could be the
$Y(4360)$ resonance observed in $\EE\to
\pp\psp$~\cite{belle_y4660_new,babar_y4360_new} reported by the
BaBar and Belle experiments.

It is worth pointing out that the lower mass structure (called
$Y(4220)$ hereafter) is the main component of the $Y(4260)$
structure but with improved measurement of the resonant parameters
thanks to the high luminosity data from BESIII. We will use
$Y(4220)$ rather than $Y(4260)$ in the discussion below.

From the BESIII data, one can see that besides the $Y(4260)$ peak,
the $\pp\jpsi$ cross section is at 10--15~pb level, whether the
cross section is due to pure continuum process or from the decays
of other charmonium or charmonium-like states is still not clear.
In fact, there are not many high luminosity data points in the
$\psi(4040)$ and $\psi(4160)$ energy region, this prevents us from
measuring the coupling of these states to $\pp\jpsi$ mode, and the
interference between these amplitudes and the $Y(4260)$ structures
may affect the fit results significantly.

\subsection{$\EE\to \pp\hc$}

In 2013, BESIII reported the cross section measurement of $\EE\to
\pphc$ at 13 c.m. energies from 3.9 to 4.4~GeV and found the
magnitude of the cross sections is at the same order as that of
$\EE\to \ppjpsi$ but with a different line shape~\cite{pipihc-1}.
In this study, the $\hc$ is reconstructed via its electric-dipole
transition $\hc\to \gamma\etac$ with $\etac$ to 16 exclusive
hadronic final states.

Although no quantitative results were given in interpreting the
$\pphc$ line shape, the resonant structure at around
4.22~GeV/$c^2$ is obvious~\cite{pipihc-1}. A combined fit to the
BESIII data together with the CLEO-c measurement at
4.17~GeV~\cite{cleoc_pipihc} results in a resonant structure with
a mass of $(4216\pm 18)~{\rm MeV}/c^2$ and a width of $(39\pm
32)$~MeV~\cite{y4220_ycz}, different from any of the known $Y$ and
excited $\psi$ states in this mass region at that time~\cite{pdg}
but in agreement with the evidence of the structure observed in
$\EE\to \omega\chi_{c0}$~\cite{BES3_omegachic0}.

\begin{figure*}[htbp]
\centering
\includegraphics[width=0.8\textwidth]{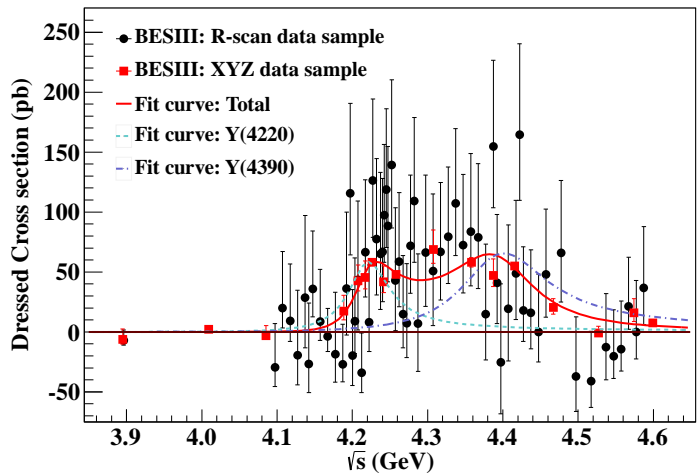}
\caption{Fit to the cross section of $\EE\to \pp\hc$ with the
coherent sum of two BW functions (solid curve). The dash
(dash-dot) curve shows the contribution from the two structures
$Y(4220)$ ($Y(4390)$). The dots with error bars are the cross
sections for $R$-scan data sample, the squares with error bars are
the cross sections for $\xyz$ data sample. Here the error bars are
statistical uncertainty only.} \label{cs}
\end{figure*}

In 2017, BESIII presented a follow-up study of $\EE\to \pp\hc$ at
c.m. energies from 3.9 to 4.6~GeV using the ``$\xyz$ data" and the
``$R$-scan data" as used in $\pp\jpsi$
analysis~\cite{bes3_pipihc_lineshape}. The cross sections are
shown in Fig.~\ref{cs} with dots and squares for $R$-scan and
$\xyz$ data samples, respectively. The measurements indicate that
the cross section does decrease as c.m. energy increases to
4.6~GeV, and there are two resonant structures in the full energy
range.

Assuming the $\pp\hc$ events come from two resonances, BESIII
obtains $M=(4218.4^{+5.5}_{-4.5}\pm 0.9)$~MeV/$c^{2}$,
$\Gamma=(66.0^{+12.3}_{-8.3}\pm 0.4)$~MeV, and the product of the
electronic partial width and the decay branching fraction to
$\pphc$ $\Gamma_{\EE}\BR(Y(4220)\to \pphc) = (4.6^{+2.9}_{-1.4}\pm
0.8)$~eV for $Y(4220)$, and $M=(4391.5^{+6.3}_{-6.8}\pm
1.0)$~MeV/$c^{2}$, $\Gamma=(139.5^{+16.2}_{-20.6}\pm 0.6)$~MeV,
and $\Gamma_{\EE}\BR(Y(4390)\to \pphc)
=(11.6^{+5.0}_{-4.4}\pm1.9)$~eV for $Y(4390)$, with a relative
phase of $\phi=(3.1^{+0.7}_{-0.9}\pm 0.2)$~rad. The parameters of
the low mass structure are consistent with those of the resonance
observed in $\EE\to \omega\chi_{c0}$~\cite{BES3_omegachic0} and in
$\EE\to \ppjpsi$~\cite{bes3_pipijpsi_lineshape} (see also
Sec.~\ref{Sec:ppjpsi}). The high mass structure is different from
the $Y(4360)$~\cite{babar_y4360_new,belle_y4660_new} and
$\psi(4415)$~\cite{pdg}.

\subsection{$\EE\to \pp\psp$}

The BaBar experiment reported the update of the study of $\EE\to
\pp\psp$ with ISR events~\cite{babar_y4360_new} with the full data
sample recorded at and near the $\Upsilon(nS)$ ($n$=2,~3,~4)
resonances and corresponds to an integrated luminosity of
520~fb$^{-1}$. The cross sections for $\EE\to \pp\psp$ from 3.95
to 5.95~GeV are measured. A fit to the $\pp\psp$ mass distribution
yields a mass of $(4340\pm 16\pm 9)$~MeV/$c^2$ and a width of
$(94\pm 32\pm 13)$~MeV for the $Y(4360)$, and a mass of $(4669\pm
21\pm 3)$~MeV/$c^2$ and a width of $(104\pm 48\pm 10)$~MeV for the
$Y(4660)$~\cite{babar_y4360_new}. The results are in good
agreement with the Belle measurement and confirm the $Y(4660)$
observed by the Belle experiment~\cite{belle_y4660}.

\begin{figure}[htbp]
\centering
 \psfig{file=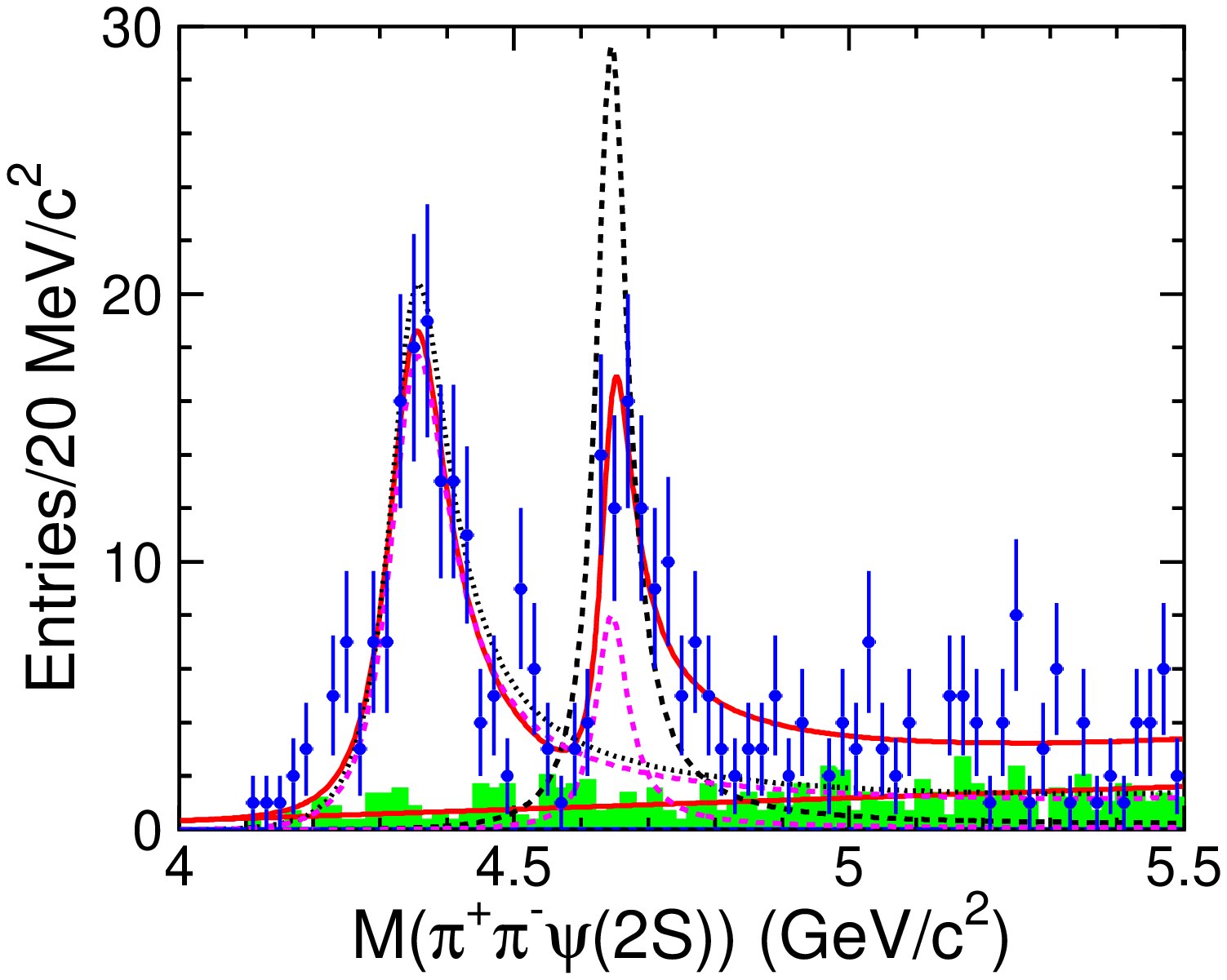, width=4.0cm, angle=-90}
 \psfig{file=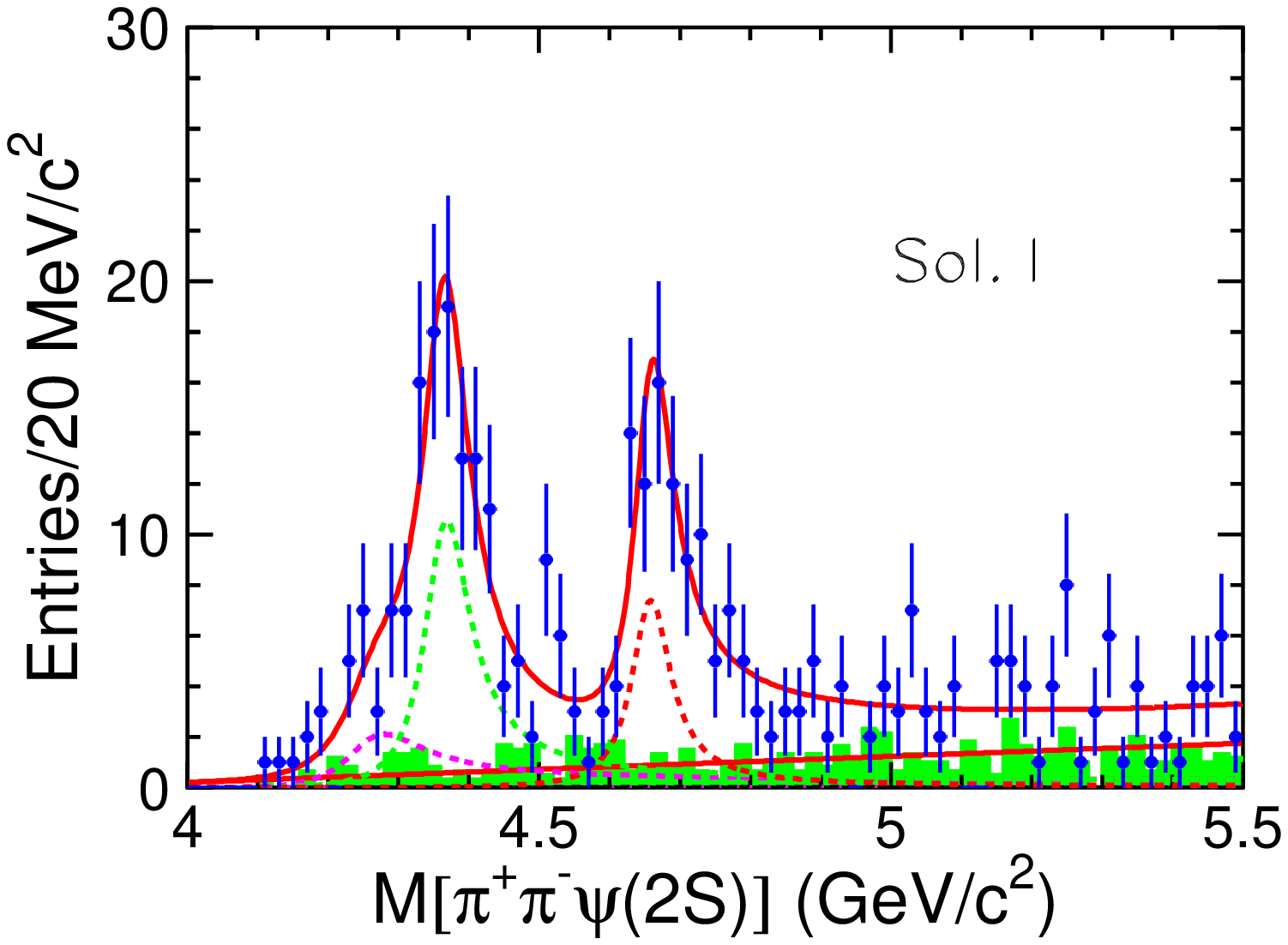, width=4.0cm, angle=-90}
\caption{The $\pp\psp$ invariant mass distribution from the Belle
experiment and the fit results with the coherent sum of two BW
functions (left panel) and with the coherent sum of three BW
functions (right panel). The points with error bars are data while
the shaded histograms the normalized $\psp$ sideband backgrounds.
The curves show the best fit and the dashed curves, show the
contributions from different BW components. } \label{2bwfit}
\end{figure}

Using the 980~fb$^{-1}$ full data sample, Belle also updated the
analysis of $\EE\to \pp\psp$ with ISR
events~\cite{belle_y4660_new}. Fitting the mass spectrum of
$\pp\psp$ with two coherent BW functions (see
Fig.~\ref{2bwfit}~(left)), Belle obtains $M[Y(4360)] = (4347\pm
6\pm 3)$~MeV/$c^2$, $\Gamma[Y(4360)] = (103\pm 9\pm 5)$~MeV,
$M[Y(4660)] = (4652\pm 10\pm 8)$~MeV/$c^2$ and $\Gamma[Y(4660)] =
(68\pm 11\pm 1)$~MeV.

Since there are some events accumulating at the mass region of the
$Y(4260)$, the fit with the $Y(4260)$ included is also performed.
In the fit, the mass and width of the $Y(4260)$ are fixed to the
latest measured values at Belle~\cite{belley_new}. The signal
significance of the $Y(4260)$ is found to be $2.4\sigma$. The fit
results are shown in Fig.~\ref{2bwfit}~(right) for one of the
solutions. In this fit, one obtains $M[Y(4360)]=(4365\pm 7\pm
4)$~MeV/$c^2$, $\Gamma[Y(4360)]=(74\pm 14\pm 4)$~MeV,
$M[Y(4660)]=(4660\pm 9\pm 12)$~MeV/$c^2$, and
$\Gamma[Y(4660)]=(74\pm 12\pm 4)$~MeV. We can find that the
resonant parameters depends on whether there is an additional
$Y(4260)$ strongly.

\begin{figure}[htbp]
\begin{center}
\includegraphics[width=0.6\textwidth]{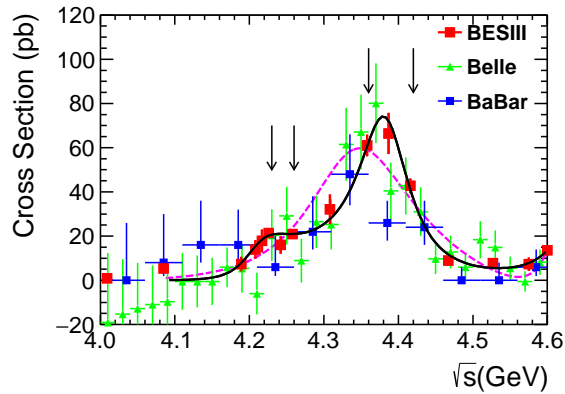}
\caption{Cross sections of $\EE\to \pp\psp$. The dots (red) are
the BESIII results, the triangles (green) and squares (blue) are
Belle and BaBar's results, respectively. The solid curve is the
fit to BESIII measurements with the coherent sum of three BW
functions. The dashed curve (pink) is the fit to BESIII
measurements with the coherent sum of two BW functions without the
$Y(4220)$. The arrows mark the locations of four energy points
with large integrated luminosities. } \label{comresult}
\end{center}
\end{figure}

BESIII measures $\EE\to \pp\psp$ using 5.1~fb$^{-1}$ of data
collected at 16 c.m. energies from 4.0 to
4.6~GeV~\cite{bes3_pppsp} with two decay modes $\psp\to \pp\jpsi$
and $\psp\to {\rm neutrals}+\jpsi$ , where ``neutrals" refers to
$\piz\piz$, $\piz$, $\eta$, and $\gamma\gamma$. The measurement is
almost background free, and BESIII measures the cross sections in
good consistency with previous BaBar and Belle
results~\cite{babar_y4360_new,belle_y4660_new}, but with much
improved precision, as shown in Fig.~\ref{comresult}.

As the BESIII data can only reach 4.6~GeV, the parameters of the
$Y(4660)$ are fixed to Belle measurement~\cite{belle_y4660_new} in
the fit to the $\EE\to \pp\psp$ cross sections with the coherent
sum of three BW amplitudes. The data require a lower-mass
resonance with a mass $M=(4209.5\pm 7.4\pm 1.4)$~MeV/$c^2$ and a
width $\Gamma=(80.1\pm 24.6\pm 2.9)$~MeV with a statistical
significance of $5.8\sigma$, this is the first observation of the
new decay mode $Y(4220)\to \pp\psp$. The fit also results in a
mass $M=(4383.8\pm 4.2\pm 0.8)$~MeV/$c^2$ and width
$\Gamma=(84.2\pm 12.5\pm 2.1)$~MeV, for the $Y(4360)$. The fit
results are also presented in Fig.~\ref{comresult}.


\subsection{$\EE\to \ddpi$}

The cross section of $\EE\to \ddpi$ was first measured by the
Belle experiment using ISR method~\cite{ddstarpi-belle}. No
evidence for the $Y(4260)$, $Y(4360)$, $\psi(4415)$, $Y(4630)$, or
$Y(4660)$ was found with limited statistics. BESIII reported
improved measurements of the production cross section of $\EE \to
\ddpi$ at c.m. energies from 4.05 to 4.60~GeV at 15 ``$\xyz$ data"
points and 69 ``$R$-scan data" points~\cite{ddstarpi-bes}. The
$D^0$ meson is reconstructed in the $D^0 \to K^-\pi^+$ decay
channel. The bachelor $\pi^+$ is also reconstructed, while the
$D^{*-}$ is inferred from energy-momentum conservation.

\begin{figure}[htbp]
\centering
\includegraphics[width=0.7\textwidth]{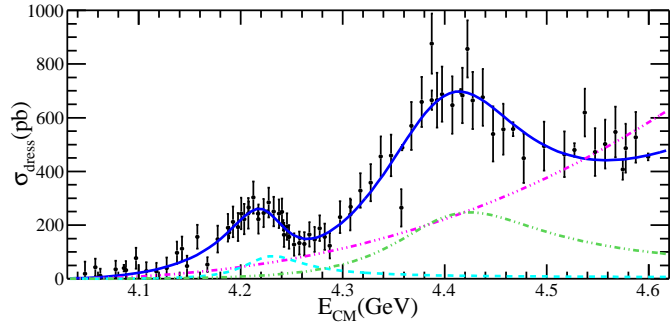}
\caption{Fit to the cross sections of $\EE\to \ddpi$ from BESIII
measurement. Dots with error bars are the data and the solid line
is the total fit. The other lines show the contribution from each
component.} \label{fig_cross}
\end{figure}

The measured cross sections are shown in Fig.~\ref{fig_cross},
which is a significant improvement over the Belle
measurement~\cite{ddstarpi-belle}. Two resonant structures around
4.23 and 4.40~GeV/$c^2$ are observed, in good agreement with the
$Y(4260)$ and $Y(4390)$ observed in $\EE\to
\pphc$~\cite{bes3_pipihc_lineshape}.

A fit to the cross section is performed to determine the
parameters of the resonant structures. The total amplitude is
described by the coherent sum of a direct three-body phase-space
term for $\EE\to \ddpi$ and two BW functions, representing the two
resonant structures. The fit yields a mass of $(4228.6\pm 4.1\pm
6.3)$~MeV/$c^2$ and a width of $(77.0\pm 6.8\pm 6.3)$~MeV for the
lower mass structure, and a mass of $(4404.7\pm 7.4)$~MeV/$c^2$
and a width of $(191.9\pm 13.0)$~MeV for the higher mass one,
where the errors are statistical only for the higher mass
structure. Since the lower mass resonance is in good agreement as
the $Y(4220)$ observed in $\pp\jpsi$ and $\pphc$
modes~\cite{bes3_pipijpsi_lineshape,bes3_pipihc_lineshape}, this
indicates the first observation of the $Y(4220)$ decays into
open-charm final state $\ddpi$

The identification of the higher mass resonance is not
straightforward. As there are known states $Y(4360)$ and
$\psi(4415)$ in this energy region, it is hard to tell whether all
the events come from a new $Y(4390)$ resonance. The above BESIII
fit neglected possible contributions from known resonances and the
complicated intermediate states in the three-body $\ddpi$ system.
As an example, there must be contribution from $\psi(4415)\to
\bar{D}D_2^*(2460)+c.c.$ in the high mass peak. Belle observed
this process in $D\bar{D}\pi$ final state~\cite{dd2_belle} and
$D_2^*(2460)$ decays into $D^*\pi$ and $D\pi$ with similar rates.
So the $\psi(4415)$ contribution in $\EE\to \ddpi$ process must be
taken into account in understanding the high mass structure.

The peak cross section of the $Y(4220)$ to this mode (consider
also charge conjugate and isospin partner final states) is much
higher than those final states with charmonium states, this
suggests stronger coupling to open charm final state than to
charmonium final state. The couplings of this state to other open
charm final states should also be further measured to understand
the decay dynamics.

\subsection{$\EE\to \omega\chi_{cJ}$ and $\phi\chi_{cJ}$}

The possible strong coupling of the $Y(4260)$ to $\omega\chi_{cJ}$
final state was proposed by a few
authors~\cite{wym_y4260,zhenghq_y4260}, the search for $\EE\to
\omega\chi_{cJ}$ are performed with the BESIII data above
4.2~GeV~\cite{BES3_omegachic0,bes3_omegachic}. The $\phi\chi_{cJ}$
thresholds are above 4.43~GeV, the study is only possible recently
with BESIII data at c.m. energy 4.6~GeV~\cite{bes3_phichic}.

\begin{figure}[htbp]
\begin{center}
\includegraphics[width=0.5\textwidth]{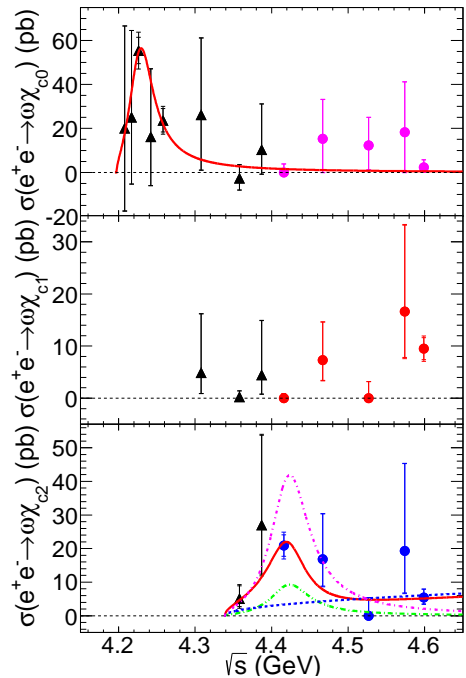}
\caption{ Measured Born cross section for $\EE \to \omega
\chi_{cJ}$ ($J=$0, 1, 2) as a function of c.m. energy. From top to
bottom are $\EE \to \omega \chi_{c0}$, $\omega \chi_{c1}$, and
$\omega \chi_{c2}$, respectively. The smaller error bars are
statistical and the larger error bars are the combined statistical
and systematic errors. The solid curves show the fit results.}
\label{fig:crosssectionall}
\end{center}
\end{figure}

The process $\EE\to \omega\chi_{c0}$ is observed at
$\sqrt{s}=4.23$ and $4.26$~GeV for the first
time~\cite{BES3_omegachic0} using $\chi_{c0}$ decays into a pair
of $\pp$ or $\kk$ (the decay rate of $\chi_{c0}\to \gamma\jpsi$ is
small and other hadronic decay modes suffer from large background
from $\EE\to {\rm light \, hadrons}$), and the Born cross sections
are measured to be $(55.4\pm 6.0\pm 5.9)$ and $(23.7\pm 5.3\pm
3.5)$~pb, respectively, which are comparable to those of $\ppjpsi$
process~\cite{bes3_pipijpsi_lineshape}. For other energy points,
no significant signals are found, and the upper limits on the
cross section at the 90\% C.L. are
determined~\cite{bes3_omegachic}. The cross sections are shown in
Fig.~\ref{fig:crosssectionall}, where a clear peaking structure
close to the threshold is observed although there are only two
statistically significant measurements available.

The processes $\EE \to \omega \chi_{c1,2}$ are observed for the
first time at high energies~\cite{bes3_omegachic}. Here the
$\chi_{c1,2}$ are reconstructed via their $\gamma\jpsi$ decays. A
significant $\omega \chi_{c2}$ signal is found at
$\sqrt{s}=$4.42~GeV, and the cross section is measured to be
$(20.9 \pm 3.2 \pm 2.5)$~pb; while at $\sqrt{s} =$4.6~GeV, a clear
$\omega \chi_{c1}$ signal is observed, and the cross section is
measured to be $(9.5 \pm 2.1 \pm 1.3)$~pb. Due to low luminosity
or low cross section at other energies, no significant signals are
observed.

The data reveal a sizable $\omega\chi_{c0}$ production at around
4.23~GeV/$c^2$, as predicted in Ref.~\refcite{zhenghq_y4260}. By
assuming the $\omega\chi_{c0}$ signals come from a single
resonance, the cross section of $\EE \to \omega \chi_{c0}$ are
fitted with a BW function, and the mass is $(4226\pm 8\pm
6)$~MeV/$c^2$, and width is $(39\pm 12\pm 2)$~MeV (shown in
Fig.~\ref{fig:crosssectionall}). The parameters are consistent
with those of the narrow structure in the $\EE\to
\pp\hc$~\cite{y4220_ycz,bes3_pipihc_lineshape},
$\pi^+\pi^-J/\psi$~\cite{bes3_pipijpsi_lineshape},
$\ddpi$~\cite{ddstarpi-bes}, and $\pp\psp$~\cite{bes3_pppsp}
processes. In fact, it is in the $\EE\to \omega\chi_{c0}$ mode
that the $Y(4220)$ (or $Y(4230)$) structure was first
reported~\cite{BES3_omegachic0}. In $\EE \to \omega \chi_{c2}$,
the cross section is higher at around 4.416~GeV than at other
energies, this may suggest $\psi(4415)\to \omega\chi_{c2}$ decays
but more measurements in the vicinity of the $\psi(4415)$ peak are
needed.

BESIII searches for the production of $\EE\to \phi\chi_{cJ}$ with
data sample at c.m. energy of $4.6$~GeV~\cite{bes3_phichic}. The
processes $\EE\to \phi\chi_{c1}$ and $\phi\chi_{c2}$ are observed
for the first time, each with a statistical significance of more
than $10\sigma$, and the Born cross sections are measured to be
$(4.2^{+1.7}_{-1.0}\pm 0.3)$ and $(6.7^{+3.4}_{-1.7}\pm 0.5)$~pb,
respectively. No significant signals are observed for $\EE\to
\phi\chi_{c0}$ and upper limit on the Born cross section is
estimated to be 5.4~pb at the $90\%$ C.L.

\subsection{$\EE\to \eta\hc$}

The process $\EE\to \eta \hc$ are searched for at c.m. energies
from 4.085 to 4.600~GeV at BESIII~\cite{bes3_etahc}. Clear signals
of $\EE\to \eta h_{c}$ are observed at $\sqrt{s}=4.226$~GeV for
the first time, $41\pm 9$ signal events with a statistical
significance of $5.8\sigma$ are observed, and the Born cross
section is measured to be $(9.5^{+2.2}_{-2.0} \pm 2.7)$~pb.
Evidence for the signal process at $\sqrt{s}=4.358$~GeV is also
observed, there are $19\pm 6$ signal events and the statistical
significance is $4.3\sigma$, the cross section is measured to be
$(10.0^{+3.1}_{-2.7}\pm 2.6)$~pb. For the other c.m. energies
considered, no significant signals are found, and upper limits on
the cross section at the 90\% C.L. are determined.

\begin{figure*}[htbp]
\centering
\includegraphics[width=0.6\textwidth]
{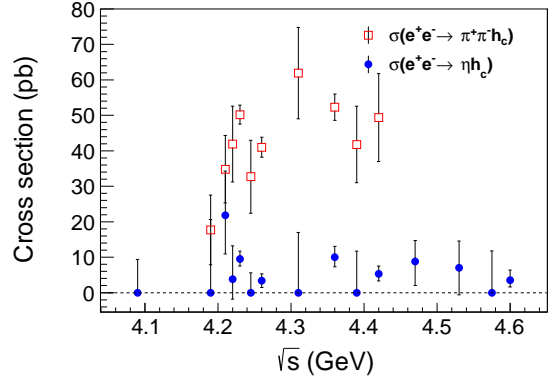}
\caption{Comparison between the cross sections of $\EE\to \eta
h_c$ (dots with error bars) and $\EE\to \pphc$ (squares with error
bars)~\cite{zc4020}. The errors are statistical only.}
\label{fig::cross_section_com}
\end{figure*}

Comparing with the cross section of $\EE\to \pphc$~\cite{zc4020},
which is shown in Fig.~\ref{fig::cross_section_com}, one finds
that the cross section at $\sqrt{s}=4.226$~GeV is larger than that
at $\sqrt{s}=4.258$~GeV in both processes. This may suggest
$\EE\to \eta\hc$ have similar line shape as $\EE\to
\pp\hc$~\cite{bes3_pipihc_lineshape} or there is $Y(4220)\to
\eta\hc$ transition. However, due to the limited statistics, one
cannot draw a solid conclusion about the production mechanism of
$\EE\to \eta h_c$.

\subsection{$\EE\to \eta\jpsi$ and $\etap\jpsi$}

\begin{figure}[htbp]
\centering
\includegraphics[height=4cm,angle=0]{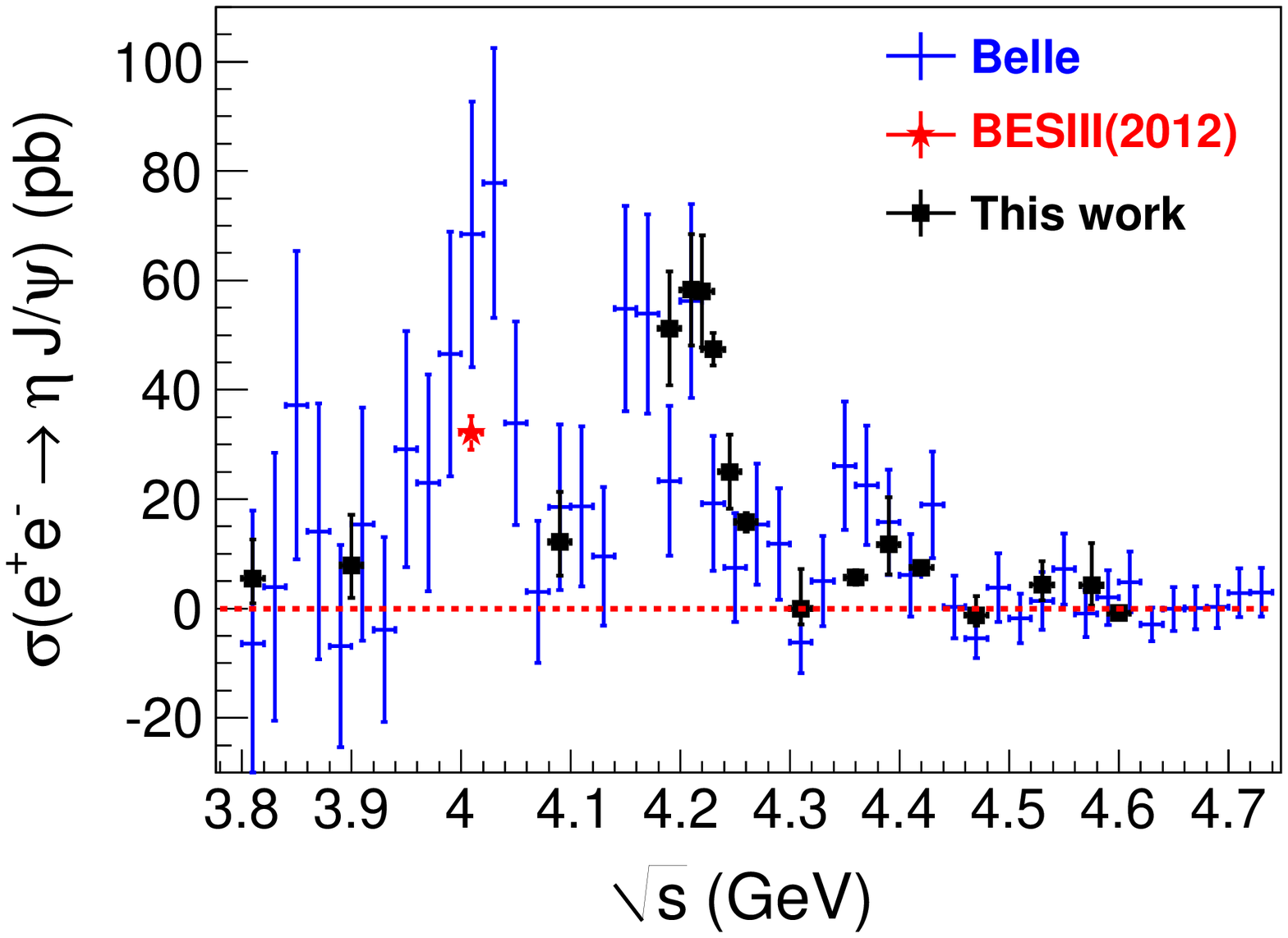}
\includegraphics[height=4cm,angle=0]{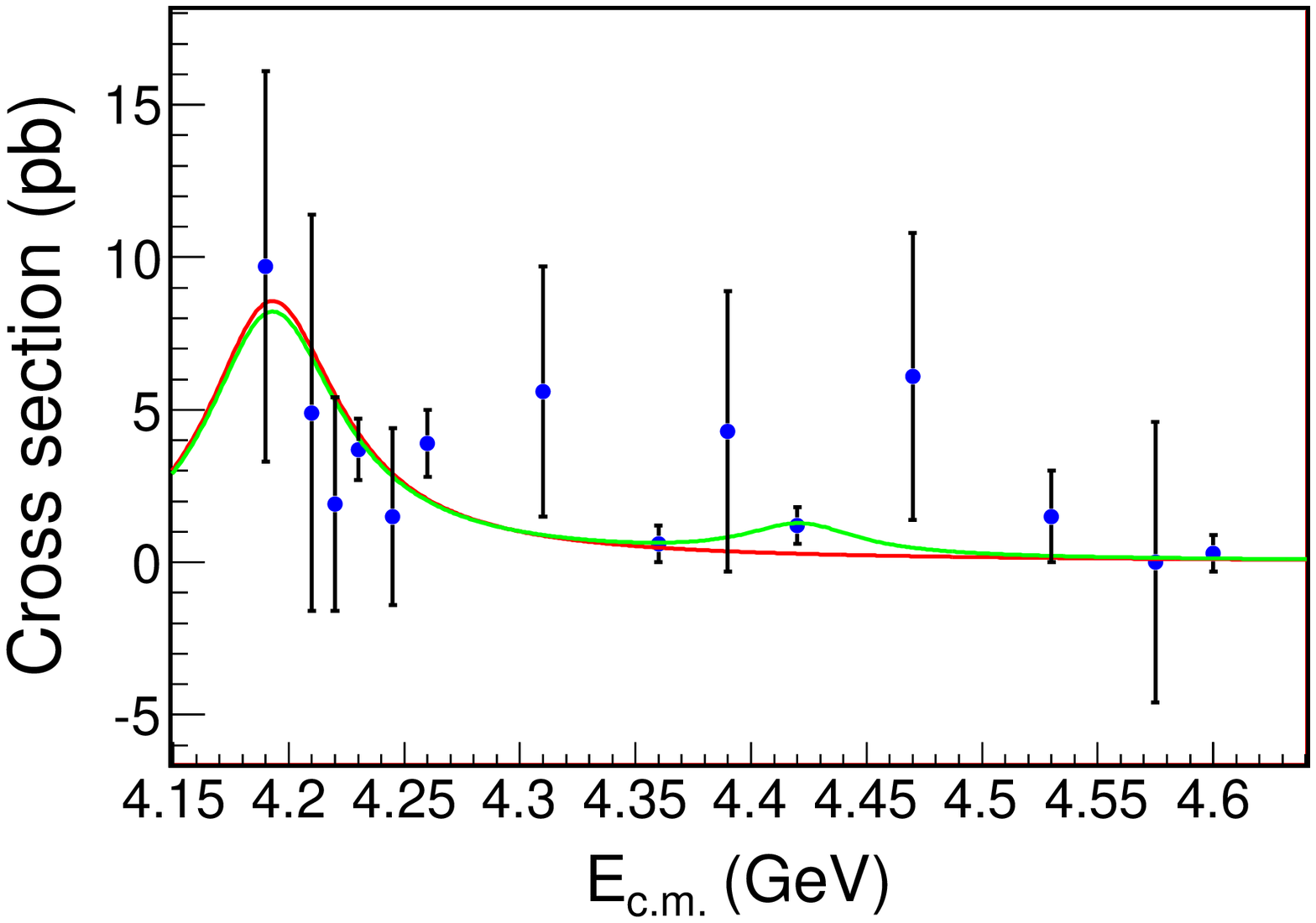}
\caption{(Left panel) Born cross sections of $\EE \to \eta \jpsi$
from BESIII~\cite{bes3-4040-etajpsi,bes3-etajpsi} (black dots and
the red star) and Belle experiments~\cite{belle-etajpsi} (blue
dots). (Right panel) Born cross sections of $\EE \to \etap \jpsi$
from BESIII~\cite{bes3-etapjpsi} (dots with error bars), and the
fit with a $\psi(4160)$ resonance (red curve), or with a
combination of $\psi(4160)$ and $\psi(4415)$ resonances (green
curve). }\label{BES_BELLE}
\end{figure}

Using data samples at c.m. energies of 3.81--4.60~GeV, BESIII
measures the Born cross sections of $\EE \to \eta
J/\psi$~\cite{bes3-4040-etajpsi,bes3-etajpsi} as shown in
Fig.~\ref{BES_BELLE}. Good agreement with the previous Belle
measurements~\cite{belle-etajpsi} is observed. As Belle measures
the cross section with ISR data, there are more data points but
the errors are large due to low statistics, the BESIII
measurements have better precision, however, are only available at
a few data points.

The process $\EE\to \eta^{\prime} J/\psi$ is investigated at 14
c.m.\ energies from 4.1 to 4.6~GeV~\cite{bes3-etapjpsi}.
Significant $\EE\to \eta^{\prime}J/\psi$ signals are observed at
$\sqrt{s} = 4.226$ and $4.258$~GeV for the first time, and the
corresponding Born cross sections are measured to be $(3.7 \pm 0.7
\pm 0.3)$ and $(3.9 \pm 0.8 \pm 0.3)$~pb, respectively. The upper
limits of Born cross sections at the 90\% C.L. are set for the
other 12 c.m. energy points where no significant signal is
observed. The results are shown in the right panel of
Fig.~\ref{BES_BELLE}.

Compared with the Born cross section of $\EE\to \eta J/\psi$, the
cross section of $\EE\to \eta^{\prime}J/\psi$ is much smaller.
BESIII have collected lots of data at a few data points around
4.2~GeV but the cross sections are not reported. With these new
data, we may have a better feeling about the line shapes of these
two modes, especially around the $\psi(4160)$--$Y(4220)$ mass
region. More data points around 4.0~GeV will be helpful to
understand the first peak observed by Belle in $\EE\to \eta\jpsi$
mode, further data around 4.4~GeV will enable a check if the small
peak observed in both Belle and BESIII data in $\EE\to \eta\jpsi$
mode is statistical fluctuation or from $Y(4360)$ or $\psi(4415)$
decays.

\subsection{Efforts in extracting resonant parameters from
combined fits}

As the cross sections of different final states have some common
features, and some of the final states have been measured by
different experiments, these data are used to do combined fit to
extract more information about the resonant
structures~\cite{gaoxy,zhangjl}.

In Ref.~\refcite{gaoxy}, the authors use the measured cross
sections of $\EE \to \omega \chi_{c0}$~\cite{bes3_omegachic},
$\pi^+\pi^-h_c$~\cite{bes3_pipihc_lineshape},
$\ppjpsi$~\cite{bes3_pipijpsi_lineshape}, and
$\ddpi$~\cite{ddstarpi-bes} processes by BESIII experiment only to
measure the resonant parameters of the $Y(4220)$.

\begin{figure*}[htbp]
\centering
 \psfig{file=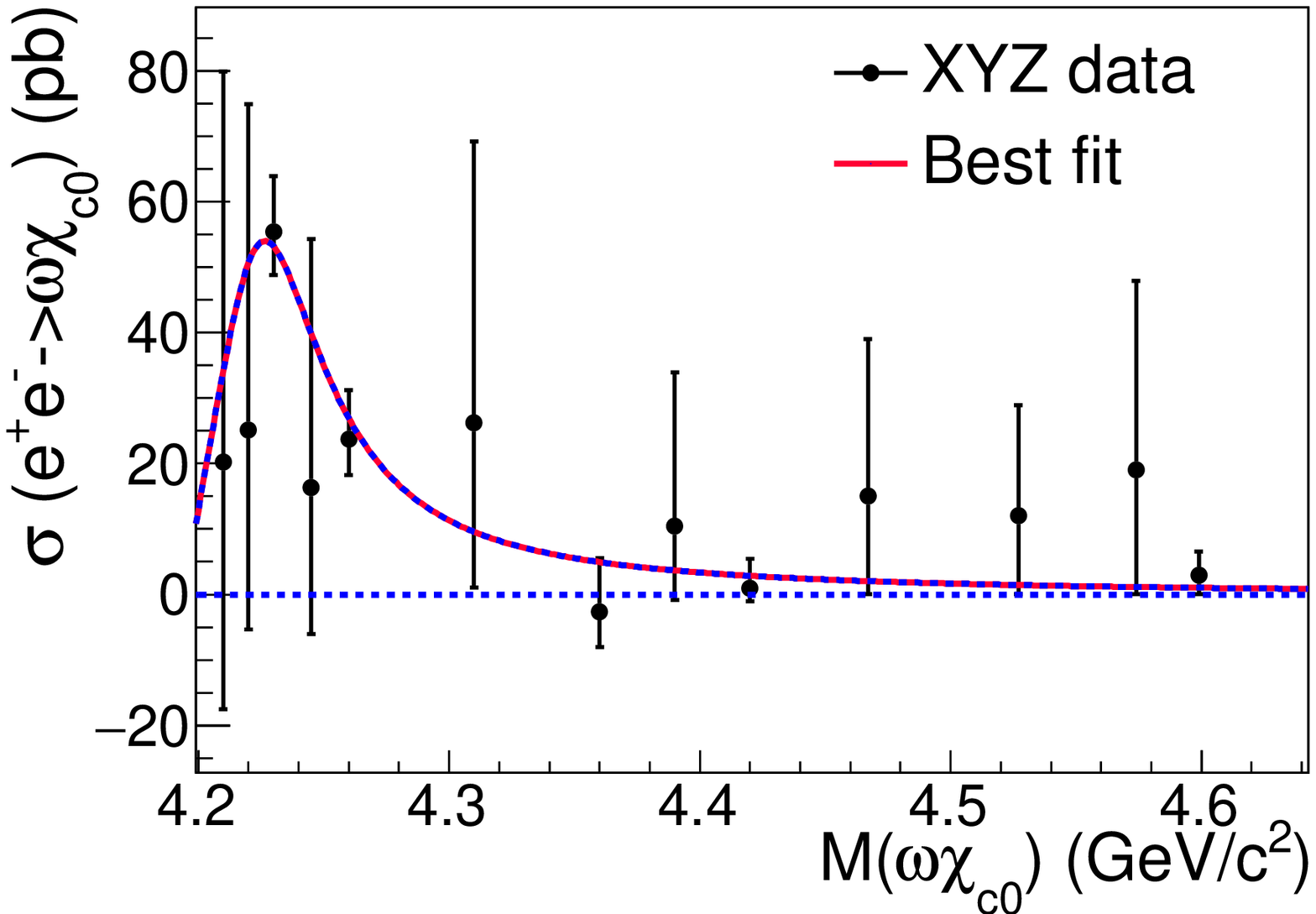,width=5cm, angle=0}
 \psfig{file=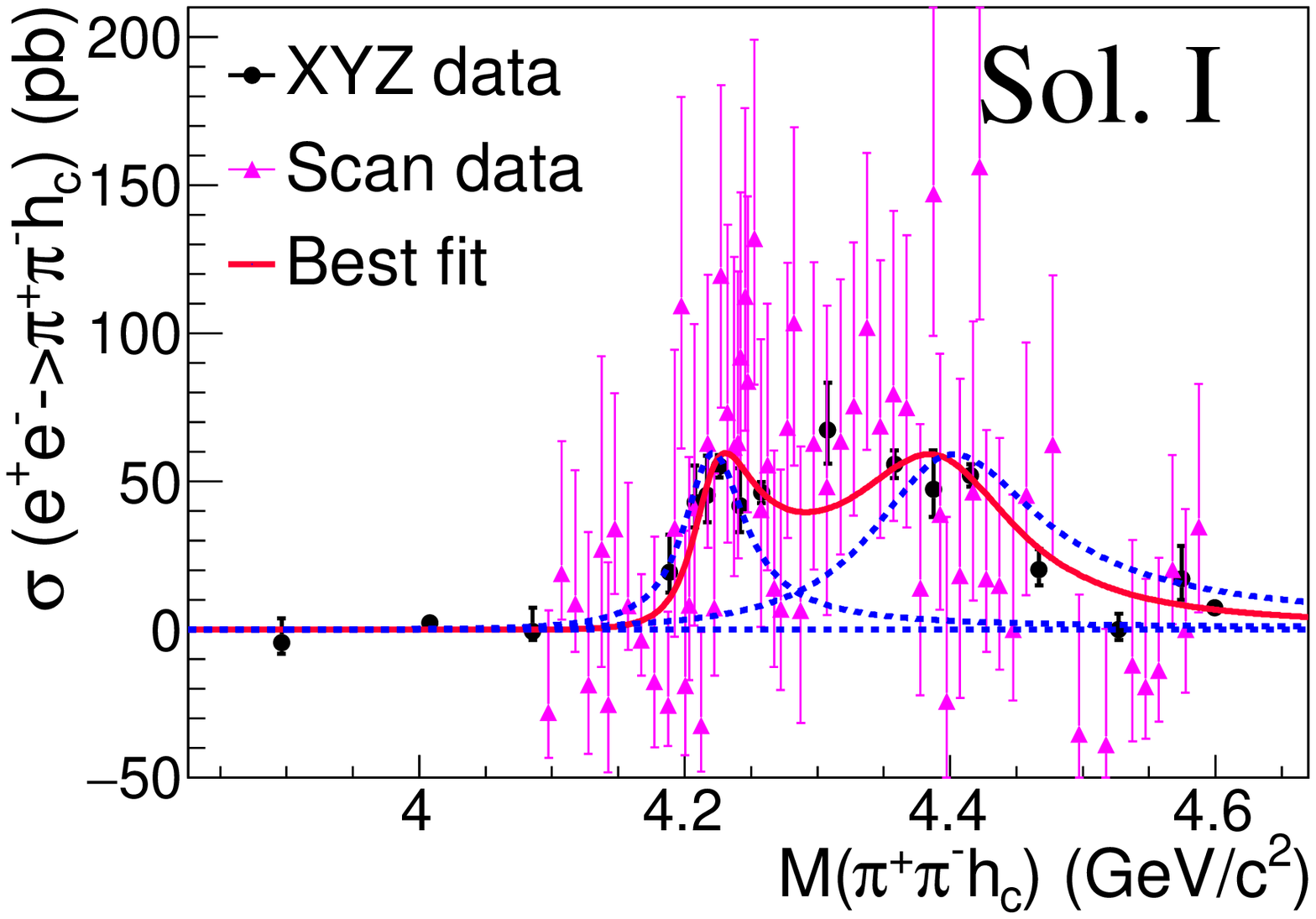,width=5cm, angle=0} \\
 \psfig{file=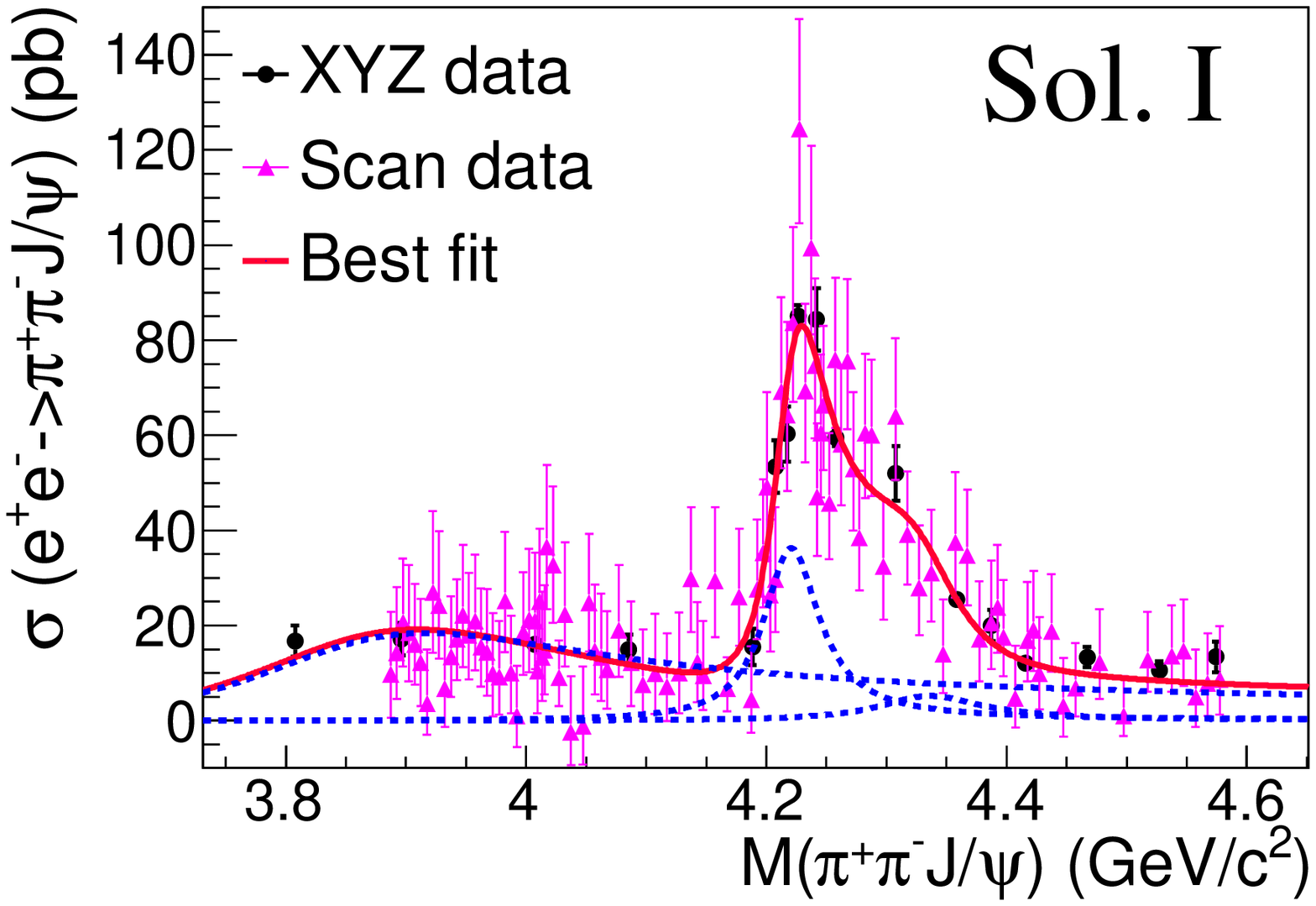,width=5cm, angle=0}
 \psfig{file=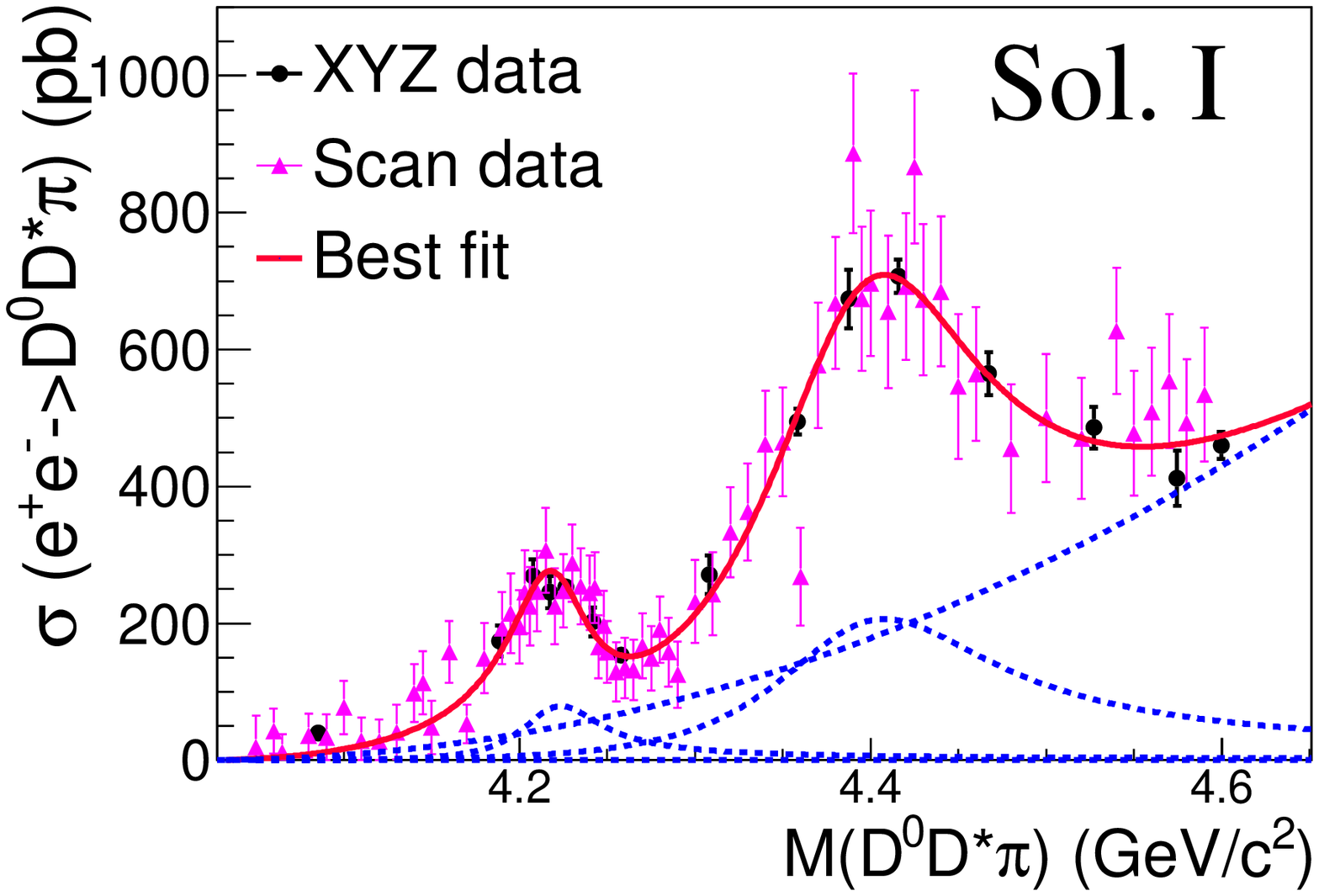,width=5cm, angle=0}
\caption{The combined fit to $\EE\to \omega\chi_{c0}$ (top left),
$\pphc$ (top right), $\pi^+ \pi^- J/\psi$ (bottom left), and
$\ddpi$ (bottom right)~\cite{gaoxy}. The dots and the triangles
with error bars are data. The solid curves are the projections
from the best fit. The dashed curves show the fitted resonance
components. Only one solution of the fit is presented here.}
\label{result-fit}
\end{figure*}

The cross sections are parametrized as the coherent sum of a few
amplitudes, either resonance represented by a BW function or
non-resonant term parametrized with a phase space distribution. In
fitting to the data shown in Fig.~\ref{result-fit}, the $Y(4220)$
is assumed to be in all the modes while the other resonances may
appear in one or more modes. The fit determines the mass of the
$Y(4220)$ as $(4219.6 \pm 3.3 \pm 5.1)$~MeV/$c^2$ and the width is
$(56.0 \pm 3.6 \pm 6.9)$~MeV. It is noticed that the mass of the
$Y(4220)$ is very close to the threshold of $D_s^*\bar{D}_s^*$
threshold (4224~MeV/$c^2$)~\cite{pdg}, the possibility of its
coupling to charmed strange meson pairs (as well as charmed
non-strange meson pairs) should be studied.

The fit also measures the product of the leptonic decay width and
the decay branching fraction to a final state. By considering the
isospin symmetric modes of the measured channels, the lower limit
on the leptonic partial width of the $Y(4220)$ is estimated to be
\( \Gamma_{e^+e^-} > (29.1\pm 2.5\pm 7.0)~\hbox{eV} \), where the
first errors are from fit and the second errors are the systematic
errors with the uncertainties from different fit scenarios. This
lower limit on the leptonic partial width of the $Y(4220)$ is
close to the prediction from LQCD for a hybrid vector charmonium
state~\cite{chenying}.

In Ref.~\refcite{zhangjl}, more modes are included in the combined
fit, and the resonant parameters of the $Y(4390)$ and $Y(4660)$
are also measured with some assumptions.

All these fits parametrize the amplitude as the coherent sum of BW
functions with constant widths. These assumptions may introduce
bias since some of the states are wide and overlap with each
other. As the states decay also into open charm final states with
thresholds not very far from the resonant peak, the width must be
energy dependent.

A coupled-channel approach is tried to perform a simultaneous fit
to the open-charm data measured by Belle in 3.7--4.7~GeV energy
region~\cite{uglov-kmatrix}. In principle, this can be extended to
all the channels to have a better understanding of the line shapes
observed. Although there are still lots of technical problems to
be solved, this is a right trend to understand the vector states
produced in $\EE$ annihilation. BESIII measurements on the
open-charm cross sections are obviously very important, more data
points in the measurements of the final states with charmonium are
also necessary.

\section{Observation of $\psi(1\,^3D_2)$}

BESIII observed $X(3823)$ in the $\EE\to \pi^+\pi^-X(3823) \to
\pi^+\pi^-\gamma\chi_{c1}$ with a statistical significance of
$6.2\sigma$ in data samples at c.m. energies of 4.23, 4.26, 4.36,
4.42, and 4.60~GeV~\cite{BES3x3823}.

\begin{figure}
\centering
\includegraphics[height=4.5cm]{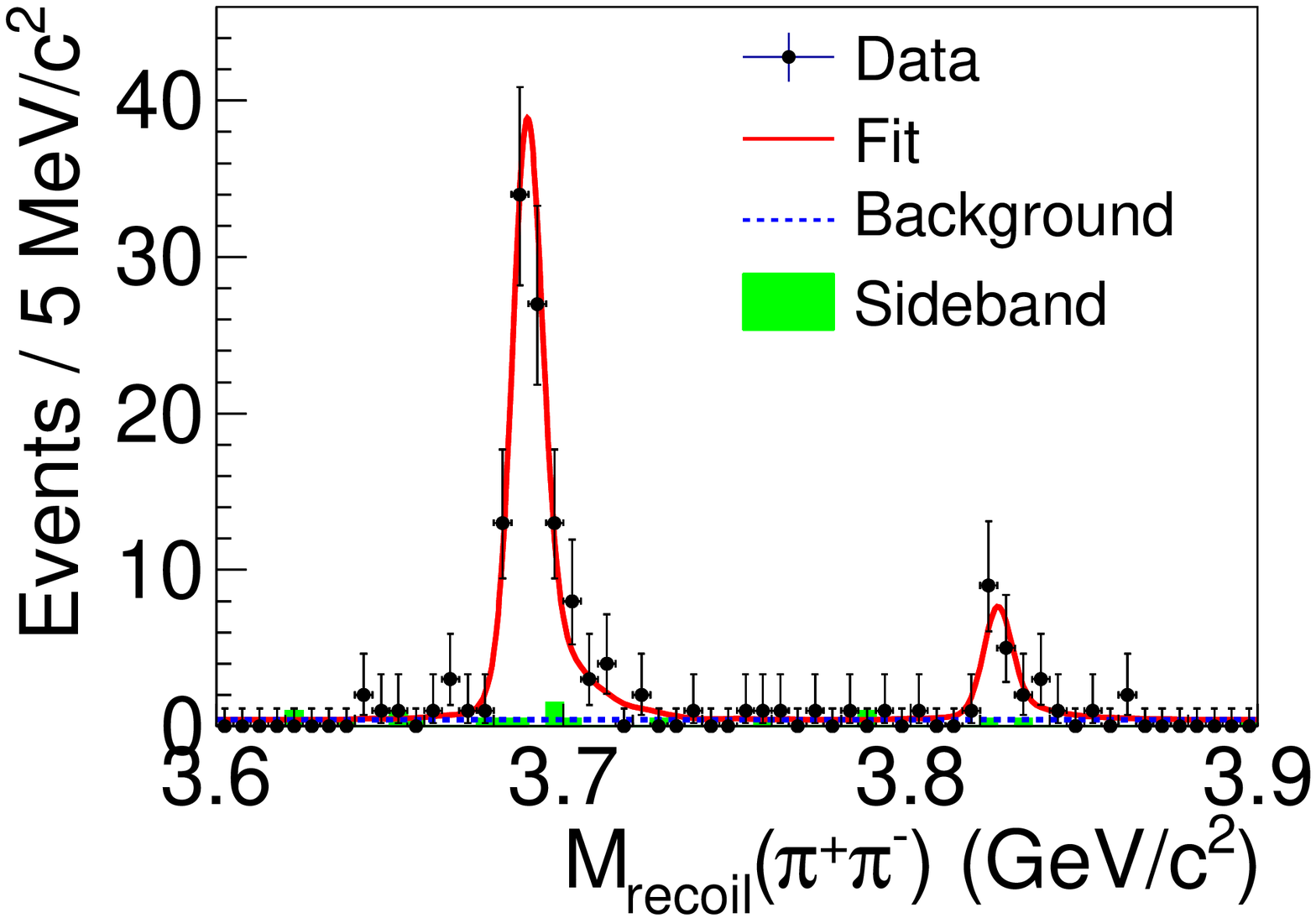}
\includegraphics[height=4.5cm]{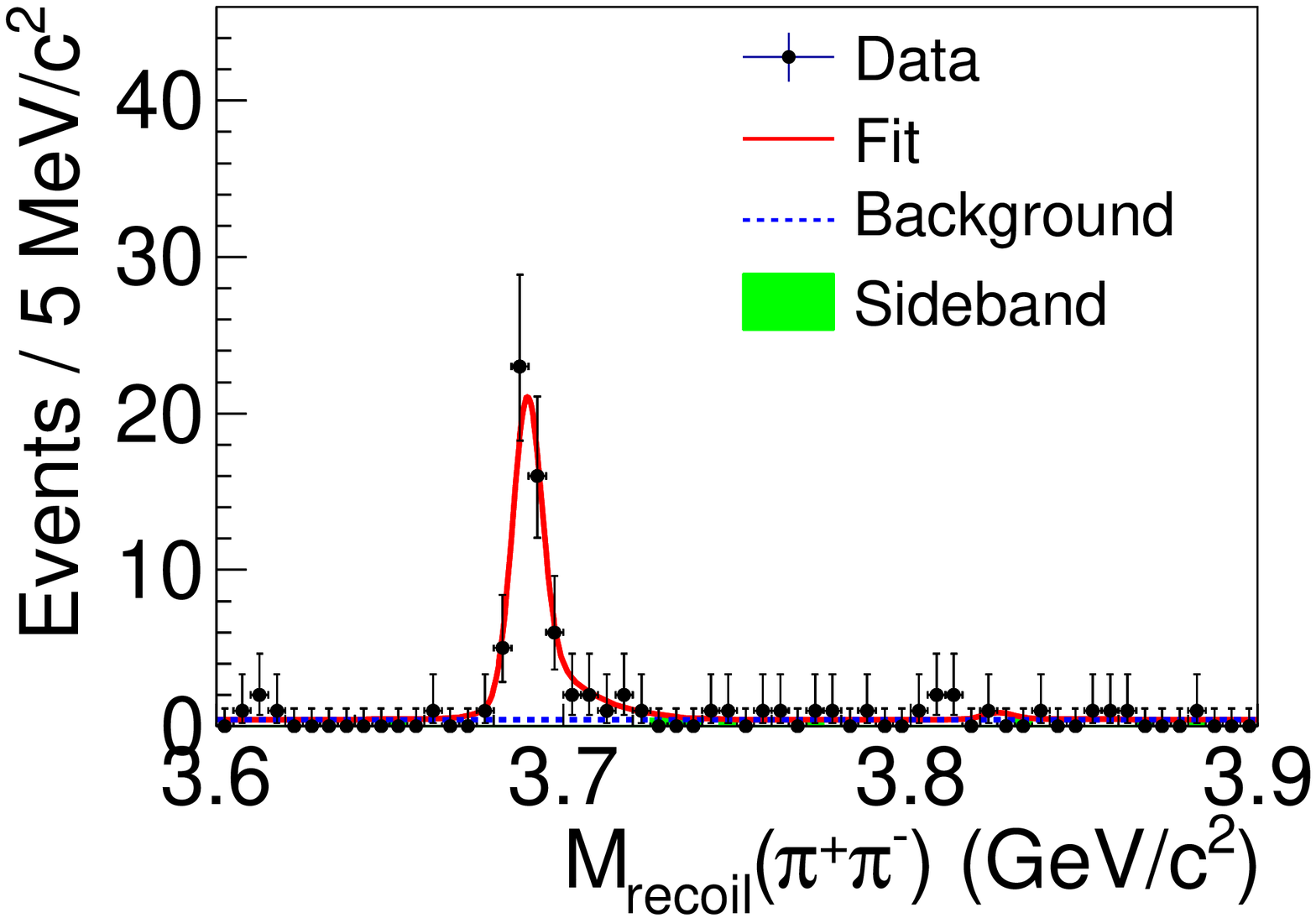}
\caption{Simultaneous fit to the $M_{\rm recoil}(\pp)$
distribution of $\gamma\chi_{c1}$ events (left) and
$\gamma\chi_{c2}$ events (right), respectively. Dots with error
bars are data, red solid curves are total fit, dashed blue curves
are background, and the green shaded histograms are $\jpsi$ mass
sideband events.} \label{X-fit}
\end{figure}

Figure~\ref{X-fit} shows the fitted results to $\pp$ recoil mass
distributions for events in the $\chi_{c1}$ and $\chi_{c2}$ signal
regions. The fit yields $19\pm 5$ $X(3823)$ signal events in the
$\gamma\chi_{c1}$ mode, with a measured mass of $X(3823)$ of
$(3821.7\pm 1.3\pm 0.7)~{\rm MeV}/c^2$. For the $\gamma\chi_{c2}$
mode, no significant $X(3823)$ signal is observed, and an upper
limit on its production rate is determined. The upper limit on the
intrinsic width of $X(3823)$ is determined as
$\Gamma[X(3823)]<16$~MeV at the 90\% C.L. This measurement agrees
with the values found by Belle~\cite{belle-3d2}.

\begin{figure}
\begin{center}
\includegraphics[height=5cm]{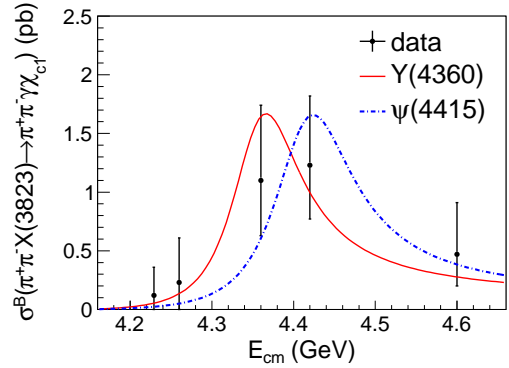}
\caption{Comparison of the energy-dependent cross sections of
$\sigma^B[\EE\to\pp X(3823)]\cdot \BR(X(3823)\to\gamma\chi_{c1})$
to the $Y(4360)$ and $\psi(4415)$ line shapes. Dots with error
bars are data. The red solid (blue dashed) curve shows a fit with
the $Y(4360)$ ($\psi(4415)$) line shape.} \label{fit-sec}
\end{center}
\end{figure}

The production cross sections $\sigma^{B}(\EE\to\pp X(3823))\cdot
\BR(X(3823)\to \gamma\chi_{c1}$, $\gamma\chi_{c2})$ are also
measured at these c.m. energies. The cross sections of $\EE\to \pp
X(3823)$ are fitted with the $Y(4360)$ shape or the $\psi(4415)$
shape, with their resonance parameters fixed to the PDG
values~\cite{pdg}. Figure~\ref{fit-sec} shows the fitted results,
we can see that both the $Y(4360)$ and $\psi(4415)$ describe the
data well.

The $X(3823)$ resonance is a good candidate for the
$\psi(1\,^3D_2)$ charmonium state. According to potential
models~\cite{potential}, the $D$-wave charmonium states are
expected to be within a mass range of 3.82--3.85~GeV/$c^2$. Among
these, the $1\,^1D_2\to \gamma\chi_{c1}$ transition is forbidden
because of $C$-parity conservation, and the amplitude for
$1\,^3D_3\to \gamma\chi_{c1}$ is expected to be
small~\cite{barnes}. The mass of $\psi(1\,^3D_2)$ is in the
$3.810\sim 3.840$~GeV/$c^2$ range predicted by several
phenomenological calculations~\cite{3d2-mass}. In this case, the
mass of $\psi(1\,^3D_2)$ is above the $D\bar{D}$ threshold but
below the $D\bar{D}^*$ threshold. Because $\psi(1\,^3D_2)\to
D\bar{D}$ violates parity, $\psi(1\,^3D_2)$ is expected to be
narrow, in agreement with the observation, and $\psi(1\,^3D_2)\to
\gamma\chi_{c1}$ is expected to be a dominant decay
mode~\cite{3d2-mass,ratio}. From the cross section measurement,
BESIII obtains the ratio $\frac{\BR[X(3823)\to \gamma\chi_{c2}]}
{\BR[X(3823)\to \gamma\chi_{c1}]}<0.42$ at the 90\% C.L., which
also agrees with expectations for the $\psi(1\,^3D_2)$
state~\cite{ratio}.

\section{The $Y(4140)$ and other states in $\phi\jpsi$ system}

Using exclusive $B^+ \to J/\psi \phi K^+$ decays, CDF
Collaboration observed a narrow structure near the $J/\psi \phi$
mass threshold with a statistical significance greater than
$5\sigma$~\cite{CDF-y4140,CDF-y4140b}. The mass and width of this
structure are fitted to be $(4143.4^{+2.9}_{-3.0}\pm
0.6)$~MeV/$c^2$ and $(15.3^{+10.4}_{-6.1}\pm 2.5)$~MeV,
respectively. There were many experimental studies in the earlier
years, but very controversial results were reported, as described
in Ref.~\refcite{yik}.

BESIII searches for $Y(4140)$ via $\EE \to \gamma \phi\jpsi$ at
$\sqrt{s}=$4.23, 4.26, 4.36, and 4.60~GeV, but no significant
$Y(4140)$ signal is observed in any of the data
samples~\cite{BES3_Y4140,bes3_phichic}. The upper limits of the
product of the cross section and branching fraction $\sigma[\EE
\to \gamma Y(4140)] \cdot \BR(Y(4140)\to \phi\jpsi)$ are
determined to be 0.35, 0.28, 0.33, and 1.2~pb at $\sqrt{s} =$4.23,
4.26, 4.36, and 4.60~GeV, respectively, at the 90\% C.L.

\begin{figure}[tbhp]
\centering
     \includegraphics[width=0.8\textwidth]{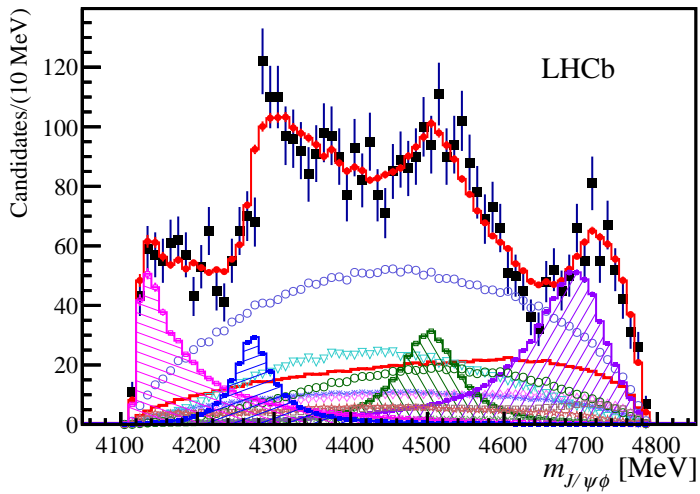}
\caption{Distributions of $\phi\jpsi$ invariant masses for the
$B^+\to\jpsi \phi K^+$ candidates (black data points) compared
with the fit containing eight $K^{*+}\to \phi K^+$ and five $X\to
\jpsi\phi$ contributions. The total fit is given by the red points
with error bars. Individual fit components are also shown.
  \label{y4140-fit}}
\end{figure}

With $3$~fb$^{-1}$ data collected at c.m. energies 7 and 8~TeV,
LHCb reconstructed $4289\pm 151$ $B^+\to \jpsi\phi K^+$
decays~\cite{y4140_LHCb}, this data sample makes the amplitude
analysis in the 6D phase space composed of invariant masses and
decay angles possible, and offers the best sensitivity to study
the resonant structures in $\phi\jpsi$ system. The data requires
not only two $J^{PC}=1^{++}$ states, the $Y(4140)$ and
$Y(4274)$~\cite{CDF-y4140b}, but also two new $J^{PC}=0^{++}$
states, $X(4500)$ and $X(4700)$. The fit results are shown in
Fig.~\ref{y4140-fit} and listed in Table~\ref{y4140}. Confirmation
from other experiments and further experimental investigation of
them are needed.

\begin{table}[tbhp]
\tbl{Results for significances, masses, and widths of the
$\phi\jpsi$ components in $B^+ \to J/\psi \phi K^+$ from LHCb
experiment.}
 {\begin{tabular}{@{}ccccc@{}} \toprule
 States  & Significance & $J^{PC}$ &  Mass (MeV/$c^2$) & Width (MeV)  \\
  \colrule
 $Y(4140)$  & $8.4\sigma$ & $1^{++}$ & $4146.5\pm 4.5^{+4.6}_{-2.8}$
                                     & $83\pm 21^{+21}_{-14}$  \\
 $Y(4274)$  & $6.0\sigma$ & $1^{++}$ & $4273.3\pm 8.3^{+17.2}_{-3.6}$
                                     & $56\pm 11^{+8}_{-11}$ \\
 $X(4500)$  & $6.1\sigma$ & $0^{++}$ & $4506\pm 11^{+12}_{-15}$
                                     & $92\pm 21^{+21}_{-20}$   \\
 $X(4700)$  & $5.6\sigma$ & $0^{++}$ & $4704\pm 10^{+14}_{-24}$
                                     & $120\pm 31^{+42}_{-33}$ \\
\botrule
\end{tabular} \label{y4140}}
\end{table}

\section{Summary and Perspectives}

With the large data samples available at LHCb and BESIII
experiments, we achieved a lot in the study of the exotic states
in the charmed sector: new charged charmonium-like states $\zc$
and $\zcp$ are discovered; the spin-parity quantum numbers of the
$\xx$, $\zc$, and $Z_c(4430)$ are determined; and the $\y$
structure is found to be dominated by the $Y(4220)$ with lower
mass, narrower width, and more decay modes; and so on. However,
there are still a lot to learn with the existing and the coming
data samples to understand these states better:
\begin{itemize}
\item In the $X$ sector:
  \begin{itemize}
  \item Search for more decay modes (including confirmation
  of $\xx\to \gamma \psp$) and measure the absolute
  branching fraction of $\xx$ decays.
  \item Measure the production cross section of $\EE\to \gamma\xx$ and
  $\pp\psi(1\,^3D_2)$, determine whether they are from resonance decays
  or continuum production.
  \item Study of the other $X$ states, such as $X(3915)$ and $X(4140)$
  ($Y(4140)$), in $\EE$ annihilation and $B$ decays.
  \end{itemize}
\item In the $Y$ sector:
  \begin{itemize}
  \item Measure more precisely the line shapes of more final
  states in $\EE$ annihilation, including open-charm and charmonium
  final states.
  \item Try coupled-channel analysis with more information.
  \item Search for the $Y$ states in $B$ or other particle decays.
  \end{itemize}
\item In the $Z$ sector:
  \begin{itemize}
  \item Measure the $Z_c$ production cross section in $\EE$
  annihilation, determine whether they are produced from resonance
  decays or continuum production.
  \item Search for $Z_c$ production in $B$ or other particle decays.
  \item Determine the quantum numbers, measure the Argand plot of
  the resonant amplitude, and search for more decay modes.
  \item Search for $Z_{cs}$ state decaying into $K^\pm\jpsi$,
  $D_s^-D^{*0}+c.c.$, or $D_s^{*-}D^0+c.c.$
  \item Search for more $Z_c$ states.
  \end{itemize}
\end{itemize}

As we have mentioned in previous sections, besides the 3~fb$^{-1}$
data at 7 and 8~TeV used in most of the LHCb analyses, there are
4~fb$^{-1}$ data at 13~TeV and will reach 5~fb$^{-1}$ by the end
of 2018. These data will allow much improved analyses of many
topics discussed above such as the $\xx$ decay properties and the
searches for the $Y$ and $\zc$ states in $B$ decays.

BESIII has achieved a lot in the study of the $\xyz$ states and
the conventional charmonium states. However, there are still some
data (see Table~\ref{ecm_lum_xyz}) not analyzed and data at more
energy points will be taken~\cite{fop}. More analyses with these
data samples will allow many improved understanding of the $\xyz$
states, especially the $\xx$, $\y$, $\zc$, and $\zcp$. BEPCII is
upgrading the maximum c.m. energy from 4.6 to 4.9~GeV in two
years, this will enable a full coverage of the
$Y(4660)$~\cite{belle_y4660} and $Y(4630)$~\cite{belle_y4630}
resonances, improved measurements of their properties are
expected.

\begin{figure}[htbp]
\begin{center}
\includegraphics[width=0.8\textwidth]{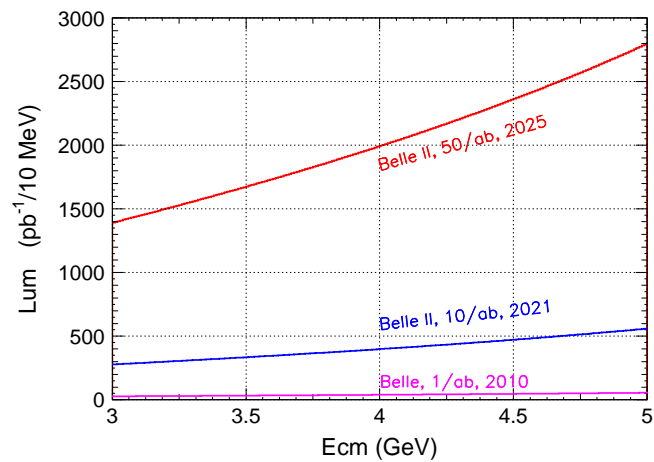}
\caption{Effective luminosity at low energy in the Belle and Belle
II $\Upsilon(4S)$ data samples.} \label{lum_belle2}
\end{center}
\end{figure}

Belle II~\cite{belle2} will start collecting data in 2019, and
will accumulate 50~ab$^{-1}$ data at the $\Upsilon(4S)$ peak by
2025. These data samples can be used to study the $\xyz$ and
charmonium states in many different ways~\cite{PBFB}, among which
ISR can produce events in the same energy range covered by BESIII.
Figure~\ref{lum_belle2} shows the effective luminosity at BEPCII
energy in the Belle II data samples. We can see that, 50~ab$^{-1}$
Belle II data correspond to 2,000--2,800~pb$^{-1}$ data for every
10~MeV from 4--5~GeV, similar statistics will be reached for modes
like $\EE\to \ppjpsi$ at Belle II and BESIII taking into account
the fact that Belle II has lower efficiency. Belle II has the
advantage that data at different energies will be accumulated at
the same time, making the analysis much simpler than at BESIII at
many data points.

There are two super $\tau$-charm factories proposed, the HIEPA in
China~\cite{HIEPA} and the SCT in Russia~\cite{SCT_charm2018}.
Both machines run at c.m. energies of up to 5~GeV or higher and a
peak luminosity of $10^{35}$~cm$^{-2}$s$^{-1}$ which is a factor
of 100 improvement over the BEPCII. This enables a systematic
study of the $\xyz$ and charmonium states with unprecedented
precision.

\section*{Acknowledgments}

This work is supported in part by National Natural Science
Foundation of China (NSFC) under contract Nos. 11235011, 11475187,
and 11521505; the Ministry of Science and Technology of China
under Contract No. 2015CB856701; Key Research Program of Frontier
Sciences, CAS, Grant No. QYZDJ-SSW-SLH011; and the CAS Center for
Excellence in Particle Physics (CCEPP).

\end{document}